\documentclass[11pt]{article}

\newsavebox{\foobox}
\newcommand{\slantbox}[2][0]{\mbox{%
        \sbox{\foobox}{#2}%
        \hskip\wd\foobox
        \pdfsave
        \pdfsetmatrix{1 0 #1 1}%
        \llap{\usebox{\foobox}}%
        \pdfrestore
}}
\newcommand\unslant[2][-.25]{\slantbox[#1]{$#2$}}

\newcommand{\mmu}{\text{\unslant\mu}}
\newcommand{\mpi}{\text{\unslant[-.18]\pi}}
\newcommand{\mdelta}{\text{\unslant[-.18]\delta}}
\newcommand{\mzeta}{\text{\unslant[-.15]\zeta}}

\usepackage[left=2cm, right=2cm, top=2.5cm, bottom=2.5cm]{geometry}
\usepackage[x11names]{xcolor}
\usepackage{fancyhdr, amssymb, cancel, amsmath, graphicx, pgfplots, tikz}
\usepackage{isomath}

\usetikzlibrary{shadows}

\newcommand{\stylecolor}{blue!50!black}

\usepackage[labelfont={bf,sf, color=\stylecolor}, margin={1.5cm,0cm}]{caption}

\usepackage[colorlinks=true, urlcolor=\stylecolor!70!white, linkcolor=\stylecolor, citecolor=\stylecolor!70!white, hyperindex=true, linktocpage=true]{hyperref}

\usepackage[explicit]{titlesec}

\newcommand*\sectionlabel{}
\titleformat{\section}
  {\gdef\sectionlabel{}
   \Large\bfseries\scshape}
  {\gdef\sectionlabel{\thesection }}{0pt}
  {\begin{tikzpicture}[remember picture,overlay]
       \end{tikzpicture}
  }
\titlespacing*{\section}{0pt}{0pt}{0pt}

\newcommand*\subsectionlabel{}
\titleformat{\subsection}
  {\gdef\subsectionlabel{}
   \large\bfseries\scshape}
  {\gdef\subsectionlabel{\thesubsection  }}{0pt}
  {\begin{tikzpicture}[remember picture]
    	\draw (-0.15, 0) node[left] {\color{\stylecolor} \textsf{\subsectionlabel}};
	\draw (0.15, 0) node[right] {\color{\stylecolor} \textsf{#1}};
	\fill[color=\stylecolor] (-0.05, -0.23) rectangle (0.05, 0.23);
       \end{tikzpicture}
  }
\titlespacing*{\subsection}{-4pt}{10pt}{0pt}

\newcommand*\subsubsectionlabel{}
\titleformat{\subsubsection}
  {\gdef\subsubsectionlabel{}
   \bfseries\scshape}
  {\gdef\subsubsectionlabel{\thesubsubsection.\ \  }}{0pt}
  {\begin{tikzpicture}[remember picture]
    	\draw (0, 0) node[left] {\color{\stylecolor} \textsf{\subsubsectionlabel}};
	\draw (0, 0) node[right] {\color{\stylecolor} \textsf{#1}};
       \end{tikzpicture}
  }
\titlespacing*{\subsubsection}{-4pt}{7pt}{0pt}

\pgfplotsset{every axis legend/.append style={at={(1.02,1)},anchor=north west}}

\begin{document}

\allowdisplaybreaks

\pagestyle{fancy}
\renewcommand{\headrulewidth}{0pt}
\fancyhead{}

\fancyfoot{}
\fancyfoot[C] {\textsf{\textbf{\thepage}}}

\begin{equation*}
\begin{tikzpicture}
\draw (\textwidth, 0) node[text width = \textwidth, right] {\color{white} easter egg};
\end{tikzpicture}
\end{equation*}

\begin{equation*}
\begin{tikzpicture}
\draw (0.5\textwidth, -3) node[text width = \textwidth] {\huge  \textsf{\textbf{Hydrodynamics of electrons in graphene}} };
\end{tikzpicture}
\end{equation*}
\begin{equation*}
\begin{tikzpicture}
\draw (0.5\textwidth, 0.1) node[text width=\textwidth] {\large \color{black} \textsf{Andrew Lucas}$^{\color{\stylecolor} \mathsf{a}}$ \textsf{and Kin Chung Fong}$^{\color{\stylecolor} \mathsf{b}}$};
\draw (0.5\textwidth, -0.5) node[text width=\textwidth] {$^{\color{\stylecolor} \mathsf{a}}$ \small\textsf{Department of Physics, Stanford University, Stanford, CA 94305, USA}};
\draw (0.5\textwidth, -1) node[text width=\textwidth] {$^{\color{\stylecolor} \mathsf{b}}$ \small\textsf{Raytheon BBN Technologies, Quantum Information Processing Group, Cambridge, MA 02138, USA}};
\end{tikzpicture}
\end{equation*}
\begin{equation*}
\begin{tikzpicture}
\draw (0, -13.1) node[right, text width=0.5\paperwidth] {\texttt{ajlucas@stanford.edu,  kc.fong@raytheon.com}};
\draw (\textwidth, -13.1) node[left] {\textsf{\today}};
\end{tikzpicture}
\end{equation*}
\begin{equation*}
\begin{tikzpicture}
\draw[very thick, color=\stylecolor] (0.0\textwidth, -5.75) -- (0.99\textwidth, -5.75);
\draw (0.12\textwidth, -6.25) node[left] {\color{\stylecolor}  \textsf{\textbf{Abstract:}}};
\draw (0.53\textwidth, -6) node[below, text width=0.8\textwidth, text justified] {\small Generic interacting many-body quantum systems are believed to behave as classical fluids on long time and length scales.   Due to rapid progress in growing exceptionally pure crystals,  we are now able to experimentally observe this collective motion of electrons in solid-state systems, including  graphene.   We present a review of recent progress in understanding the hydrodynamic limit of electronic motion in graphene,  written for physicists from diverse communities.  We begin by discussing the ``phase diagram" of graphene, and the inevitable presence of impurities and phonons in experimental systems.  We  derive hydrodynamics,  both from a phenomenological perspective and using kinetic theory.   We then describe how hydrodynamic electron flow is visible in electronic transport measurements.   Although we focus on graphene in this review, the broader framework naturally generalizes to other materials.  We assume only basic knowledge of condensed matter physics, and no prior knowledge of hydrodynamics.};
\end{tikzpicture}
\end{equation*}

\tableofcontents

\begin{equation*}
\begin{tikzpicture}
\draw[very thick, color=\stylecolor] (0.0\textwidth, -5.75) -- (0.99\textwidth, -5.75);
\end{tikzpicture}
\end{equation*}

\titleformat{\section}
  {\gdef\sectionlabel{}
   \Large\bfseries\scshape}
  {\gdef\sectionlabel{\thesection }}{0pt}
  {\begin{tikzpicture}[remember picture]
	\draw (0.2, 0) node[right] {\color{\stylecolor} \textsf{#1}};
	\draw (0.0, 0) node[left, fill=\stylecolor,minimum height=0.27in, minimum width=0.27in] {\color{white} \textsf{\sectionlabel}};
       \end{tikzpicture}
  }
\titlespacing*{\section}{0pt}{20pt}{5pt}

\section{Introduction}
The study of the hydrodynamic behavior of electrons in solid-state systems is attracting a diverse community of physicists from a broad variety of backgrounds.   It is important to have a clear and pedagogical introduction to the fundamentals of electronic hydrodynamics, emphasizing the particular challenges required to observe this regime in experiment.  This topical review aims to fill this gap in the literature.  To make this review accessible to physicists from diverse fields including condensed matter physics, high energy physics, hydrodynamics and plasma physics,  we assume minimal knowledge of solid-state physics, at the level of \cite{ashcroft}, and no knowledge of fluid dynamics.   

We have chosen to focus on electronic hydrodynamics in a particular material: graphene.   Graphene is both a particularly rich playground for electronic hydrodynamics and well suited for connecting theory with experimental data due to the simplicity of its band structure.  As we will discuss, hydrodynamics is a universal description of interacting, thermalizing physical systems.  Thus, many of the topics discussed here will immediately be relevant for other materials as well.   However, the onset of the hydrodynamic limit of electron fluids exhibits material-specific peculiarities too.  The challenge of electron hydrodynamics is to tease out the universal physics from non-universal, often material-specific phenomena.  We hope that our focus on graphene, with brief forays into other materials, conveys this theme throughout the review.

\subsection{``Relativistic" Electrons in Graphene}
\begin{figure}
\centering
\includegraphics[width=2in]{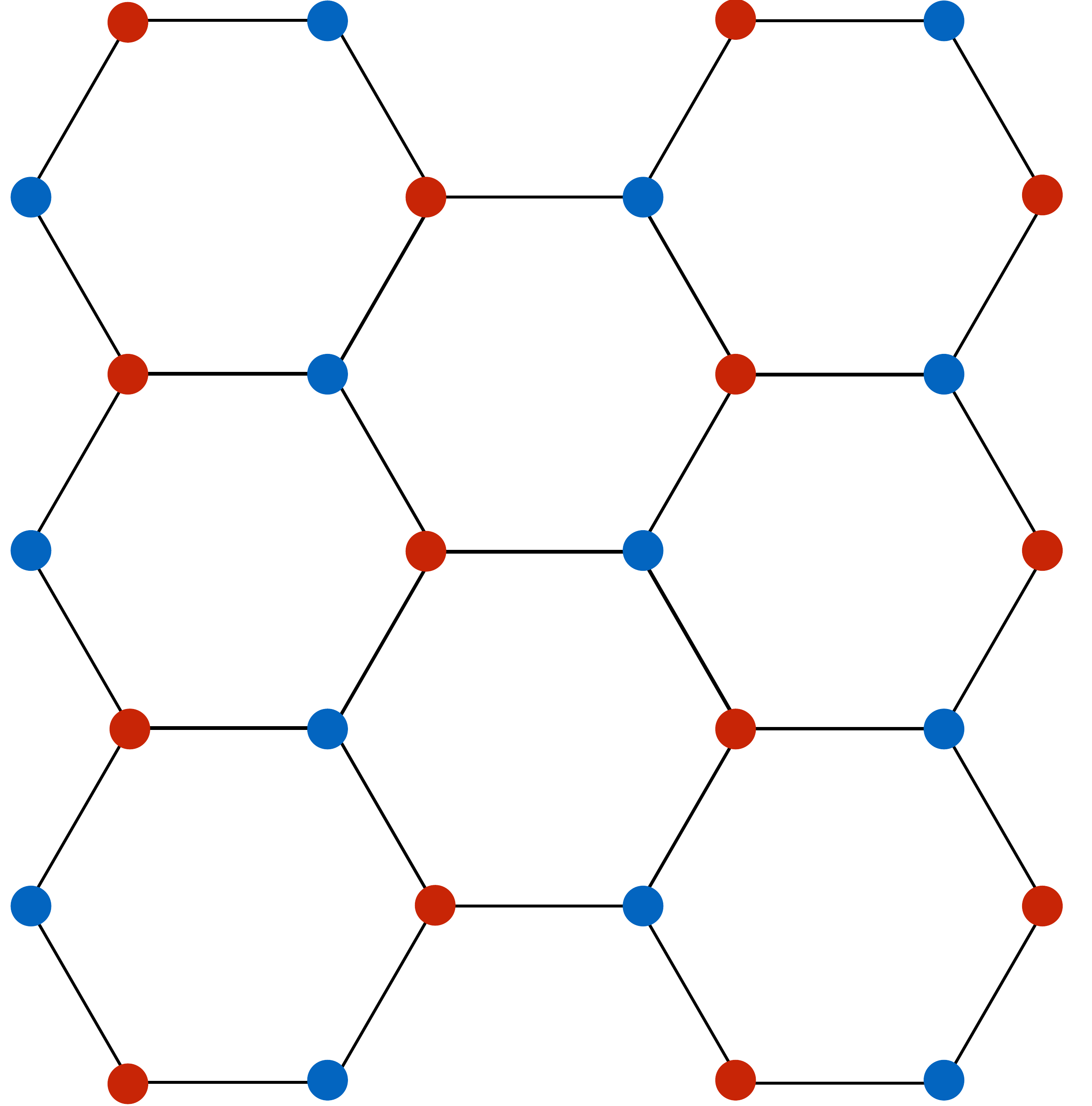}
\caption{The honeycomb lattice in two spatial dimensions.   The red and blue atoms both represent carbon in graphene -- the colors denote the bipartite sublattices.}
\label{fig:honeycomb}
\end{figure}

We begin with a brief introduction to monolayer graphene.    Graphene is a honeycomb lattice of carbon atoms in two spatial dimensions:  see Figure \ref{fig:honeycomb}.    A simple calculation, shown in Section \ref{sec:band}, reveals that the quasiparticles in the honeycomb lattice have a ``relativistic" dispersion relation \cite{wallace}:  \begin{equation}
\epsilon(\mathbf{k}) = \pm  \hbar v_{\mathrm{F}}|\mathbf{k}|, \label{eq:disprel}
\end{equation}
with $v_{\mathrm{F}}$ the Fermi velocity of graphene.  The dispersion relation (\ref{eq:disprel}), describing \emph{massless} fermions, was confirmed experimentally in a pair of seminal papers \cite{geim2005, kim2005}.

The electrons in graphene are charged particles, and interact with one another via the standard electrostatic Coulomb force.   How strong are these interactions?   A useful way to answer this question is to compute the dimensionless parameter \begin{equation}
\alpha = \frac{\text{typical potential energy}}{\text{typical kinetic energy}} = \dfrac{\displaystyle \frac{e^2}{4\mpi \epsilon r}}{\displaystyle \frac{\hbar v_{\mathrm{F}}}{r}} = \frac{e^2}{4\mpi \epsilon \hbar v_{\mathrm{F}}}.  \label{eq:alpha1}
\end{equation}
We have denoted $r$ as the typical ``distance" between electrons, and used (\ref{eq:disprel}), together with the estimate $k\sim 1/r$ for the typical quasiparticle wavenumber.   For realistic samples of graphene, we should take $v_{\mathrm{F}} \approx 1.1\times 10^6$ m/s, and  $1\lesssim \epsilon/\epsilon_0 \lesssim 5$ \cite{geim2005, kim2005}.    It is useful to rewrite (\ref{eq:alpha1}) as \begin{equation}
\alpha = \frac{e^2}{4\mpi \epsilon_0 \hbar c} \times \frac{c}{v_{\mathrm{F}}}\times \frac{\epsilon_0}{\epsilon} \approx \alpha_{\mathrm{QED}} \frac{c}{v_{\mathrm{F}}} \frac{\epsilon_0}{\epsilon}  \approx   \frac{1}{137} \times 300  \frac{\epsilon_0}{\epsilon} \sim 1.  \label{eq:alphaest}
\end{equation}
Here $c\approx 3\times 10^8$ m/s is the speed of light, and $\alpha_{\mathrm{QED}}$ is the bare coupling strength of quantum electrodynamics (QED) in  $3+1$ spacetime dimensions.   Unlike QED, the Coulomb interactions between electrons in graphene are not always weak.

The effective theory of the interacting electrons in graphene depends dramatically on the number density $n$ of electrons.   We will always measure this density relative to the Dirac point, or charge neutrality point,   where the Fermi energy in (\ref{eq:disprel}) is 0.     When $n$ is large (relative to the density of thermal excitations),  graphene behaves as a conventional Fermi liquid, albeit one in two spatial dimensions.   In this Fermi liquid, one can find long-lived quasiparticles and understand the dynamics of the electrons via the gas dynamics of the quasiparticles.   When $n=0$, there is no Fermi surface.    The resulting theory of the electron fluid is -- at reasonable temperatures $T$ -- strongly interacting;  we call this state the Dirac fluid.   We will discuss these regimes in Section \ref{sec:phases}.

\subsection{Hydrodynamics of Quantum Systems}
At finite temperature, both the Fermi liquid and the Dirac fluid are interacting many-body quantum systems.    The dimension of the many-body Hilbert space grows exponentially quickly with system size.   A direct solution to the many-body dynamics problem is intractable.   Nonetheless, there are many fundamental open questions.  For example:  how does the microscopic Hamiltonian affect the measurable properties of a (finite temperature)  quantum system, on short and long time scales?   What does it even mean for a closed quantum system to thermalize?

In some respects, these questions have very similar classical counterparts.   A macroscopic body of water consists of $\sim 10^{23}$ molecules, strongly interacting with each other.    Nonetheless, we have a reasonable understanding of the dynamics of water (at least in simple settings).   The reason is that we only care about the dynamics of water on long length scales.  The only degrees of freedom which we can reasonably measure are the \emph{conserved} quantities:   the number of molecules, and their energy and momentum.   The resulting effective theory is called \emph{hydrodynamics} \cite{landau}.   

Because the assumptions of our theory of statistical mechanics do not depend in a fundamental way on whether the microscopic degrees of freedom are classical or quantum,   it is natural to postulate that an interacting many-body quantum system also has a hydrodynamic description, on long length scales and at finite temperature \cite{kadanoff}.       The resulting equations of motion are classical differential equations, describing the rearrangement of the conserved quantities in space and time.    They are valid on time scales $t \gg \tau_{\mathrm{ee}}$, the electron-electron interaction time,  and length scales $\ell \gg \ell_{\mathrm{ee}}$, the mean free path for electron-electron collisions.      These classical equations can be derived by symmetry considerations and thermodynamic postulates alone.  We will do so in Section \ref{sec:hydro}.    While it may not be obvious how to define $\tau_{\mathrm{ee}}$ and $\ell_{\mathrm{ee}}$ in a quantum theory, we will provide a partial answer in Section \ref{sec:kinetic}.    These hydrodynamic descriptions have been used to describe spin waves \cite{forster} and superfluids \cite{putterman} for decades;  this review describes  the new systems where the electrons themselves behave as a charged fluid, and how the motion of this charged fluid leads to novel electronic phenomena.

Let us emphasize from the start that there is a confusing but popular jargon that the hydrodynamics of superfluids is ``quantum hydrodynamics".   A superfluid is a quantum state which spontaneously breaks a global symmetry.  There are associated gapless modes, Goldstone bosons, associated with this broken symmetry, and the hydrodynamics of these Goldstone bosons is called ``quantum".  There is a long literature of both theory and experiment on  superfluid hydrodynamics \cite{putterman}.   In this review, we focus on non-superfluid quantum systems.   We again emphasize that the hydrodynamics of these systems is \emph{classical}, in spite of the quantum dynamics on the smallest length scales.

When $t\lesssim \tau_{\mathrm{ee}}$ and $\ell \lesssim \ell_{\mathrm{ee}}$, a simple hydrodynamic description of many-body dynamics fails.    If quasiparticles are long-lived, and these quasiparticles interact weakly with one another,  then one can instead build a ``quantum" kinetic description of the dynamics \cite{kamenev}.   A kinetic description is valid on length scales where quasiparticles can be approximated as point-like:  $\ell \gg \lambda_{\mathrm{F}}$,  where $\lambda_{\mathrm{F}}$ is the Fermi wavelength of quasiparticles.   In Section \ref{sec:kinetic}, we will discuss the kinetic theory approach to many-body dynamics.

So far, our discussion of hydrodynamics has made no reference to a specific quantum system.  Indeed, in the absence of disorder,  we believe that interacting many-body quantum systems will generically exhibit a hydrodynamic limit.   Experiments have uncovered evidence for hydrodynamics of finite temperature quantum systems in a broad variety of settings, including cold atomic gases \cite{cao} and the quark-gluon plasma \cite{shuryak}.      So it may seem surprising that despite the long history of solid-state physics, we have only relatively recently found experimental evidence for electronic hydrodynamics in GaAs \cite{molenkamp2, molenkamp},  graphene \cite{bandurin, crossno, levitov1703}, $\mathrm{PdCoO}_2$ \cite{mackenzie} and $\mathrm{WP}_2$ \cite{felser}.    The main challenge for observing the hydrodynamics of the electronic fluid alone is that metals contain impurities and phonons.  The scattering of electrons off of impurities and phonons destroys the collective hydrodynamic flow of the electrons alone.   Hence, to see hydrodynamic electron flow, one must ensure that the electron-electron scattering, which occurs at a rate \begin{equation}
\frac{1}{\tau_{\mathrm{ee}}} \sim \alpha^2  \frac{k_{\mathrm{B}}T}{\hbar} \times \min\left(1,\frac{k_{\mathrm{B}}T}{E_{\mathrm{F}}}\right) \sim \frac{1}{0.1 \; \mathrm{ps}} \times \frac{T}{100 \; \mathrm{K}} \min\left(1,\frac{k_{\mathrm{B}}T}{E_{\mathrm{F}}}\right)  \label{eq:taueeintro}
\end{equation}
in graphene, occurs sufficiently quickly.   This scattering rate may seem quite fast;  however, electron-impurity scattering rates in typical metals can easily be $(0.01 \; \mathrm{ps})^{-1}$.    Not surprisingly then,  one of the major obstacles to observing electronic hydrodynamics in graphene has been the growth of exceptionally clean crystals.  Indeed, the possibility of hydrodynamic electron flow was realized in the 1960s \cite{gurzhi}, but was mostly forgotten because at the time it was not feasible to see experimentally.  

Hydrodynamics thus provides a partial answer to the question of \emph{how} a (quantum) system reaches global thermal equilibrium.  The hydrodynamic limit exists \emph{after} the system has decohered, yet it does tell us directly about the long time response of a quantum system.   As we will see, hydrodynamics tells us that certain response functions have a universal functional form, but will not tell us how to compute overall prefactors.   Despite these limitations, it is still valuable  to understand the hydrodynamic limit well.   The hydrodynamic regime is precisely when the collective nature of the many-body dynamics becomes most pronounced.   The usual regime that we study in condensed matter physics is the opposite:  when the dynamics of electrons can be understood in a non-interacting approximation \cite{ashcroft}.   A complete solution of the many-body dynamics problem will necessarily involve an understanding of both  limits.

\subsection{The Challenge of Strange Metals and Non-quasiparticle Physics}
For historical reasons, our conventional theory of condensed matter physics relies almost entirely on the assumption that electronic correlations can be neglected \cite{ashcroft}.    In most ``typical" metals, this assumption is sensible.    However, we have known for a long time that there are non-Fermi liquid ``strange metals" with strongly correlated electrons \cite{SSBK11}, which cannot be described in our conventional framework.  There are no quasiparticles in these systems, and it is likely that they must be described through a many-body framework, non-perturbative in interaction strength.    Hydrodynamics and gauge-gravity duality \cite{lucasrmp} are two such descriptions, yet only the former is directly relevant to experimental systems at present.  The universality of hydrodynamics is further appealing, because experiments suggest universal behavior of many non-Fermi liquids.  For example:  why do so many of these strange metals have  a very high electrical resistivity $\rho$, which scales linearly with temperature $T$ \cite{mackenzie2013}?   

As we will see in Sections \ref{sec:FLtrans} and \ref{sec:DFtrans}, the electrical resistivity $\rho$ (and transport phenomena more broadly) depend critically on the nature of electronic scattering.  Hence, properly accounting for electron-electron scattering is crucial.   Furthermore, because of the strong electronic correlations, the apparent mean free path $\ell_{\mathrm{ee}}$ can become comparable to interatomic distances.   In graphene, $\ell_{\mathrm{ee}} \sim v_{\mathrm{F}}\tau_{\mathrm{ee}} \sim 100$ nm.   As hydrodynamic electron flow in graphene has been observed, it is not crazy to postulate that hydrodynamic phenomena could be relevant for understanding the behavior of correlated electrons in strange metals.     Indeed, in recent years there have been various proposals that the ubiquitous observation of $\rho \propto T$ has a hydrodynamic origin \cite{dsz, hartnoll1, hartnoll1704, hartnoll1705}.   

As we will see in Section \ref{sec:DFtrans}, the consequences of hydrodynamic flow on electronic transport can be quite subtle in unconventional non-Fermi liquid phases.   As argued in \cite{hartnoll1704, hartnoll1705}, if typical strange metals are in a hydrodynamic regime, it may be a rather unconventional one.   Thus, the Dirac fluid of graphene is an important experimental test case:  it is a non-Fermi liquid that we understand (at least qualitatively) theoretically.   As we will show,  conventional electrical transport experiments are not usually sufficient for discovering collective hydrodynamic motion of electrons in the Dirac fluid (or other exotic hydrodynamic regimes).   We will propose ways of detecting hydrodynamic electron flow in the unconventional Dirac fluid in Sections \ref{sec:DFtrans} and \ref{sec:drag}, but we emphasize that the search for crisp signatures of electronic hydrodynamics is an important open problem for both theory and experiment.


\section{Graphene}\label{sec:graphene}
With a broad view as to why the study of electronic dynamics in graphene is exciting, we now turn to a  review of the physics of electrons and phonons in graphene.  Many more details can be found the reviews \cite{geimrmp,Sarma:2011br}.   In this section, we will assume the validity of the textbook theory, and not carefully treat  electron-electron interactions.   In the Fermi liquid in graphene, many (but not all) experiments can be understood in such a limit.
\subsection{Band Structure of the Honeycomb Lattice}
\label{sec:band}

Neglecting electron-electron interactions, phonons and impurities, the mobile electrons in graphene are approximately described by a tight-binding model.   For the moment, we ignore electronic spin. Letting $c_i^\dagger$ and $c_i$ be creation and annihilation operators for a fermion on site $i$ of the honeycomb lattice, shown in Figure \ref{fig:honeycomb}, the Hamiltonian is \begin{equation}
H_0 = -t\sum_{\langle ij\rangle} c^\dagger_i c_j +  \mathrm{H.c.}  \label{eq:hopping}
\end{equation}
with $t\approx2.8$  eV \cite{geimrmp}, and the sum over $\langle ij\rangle$ running over nearest neighbor sites.  For completeness, we remind the reader that\begin{equation}
c^\dagger_i c_j  + c_j c^\dagger_i = \mdelta_{ij}.
\end{equation}
The displacement between two atoms in the honeycomb lattice is given $\pm a \mathbf{e}_1$ or $\pm a\mathbf{e}_2$ or $\pm a \mathbf{e}_3$, where \begin{subequations}\begin{align}
\mathbf{e}_1 &=  \hat{\mathbf{x}}, \\ 
\mathbf{e}_2 &=  \frac{-\hat{\mathbf{x}} - \sqrt{3}\hat{\mathbf{y}}}{2}, \\
\mathbf{e}_3 &=  \frac{-\hat{\mathbf{x}} + \sqrt{3}\hat{\mathbf{y}}}{2},
\end{align}\end{subequations}
and $a\approx 0.14$ nm is the spacing between carbon atoms in the honeycomb lattice.  The honeycomb lattice is not a Bravais lattice.   The unit cell consists of one atom on each sublattice (denoted with red vs. blue in Figure \ref{fig:honeycomb}), and they form a triangular lattice (which is fundamental) with unit vectors $a (\mathbf{e}_1 - \mathbf{e}_2)$ and $a (\mathbf{e}_1 - \mathbf{e}_3)$.   The reciprocal lattice consists of vectors $\mathbf{k}$ for which \begin{subequations}\begin{align}
\mathbf{k} \cdot a (\mathbf{e}_1 - \mathbf{e}_2) &= 2\mpi m_1, \\
\mathbf{k} \cdot a (\mathbf{e}_1 - \mathbf{e}_3) &= 2\mpi m_2, \\
\end{align}\end{subequations} 
for $m_1$ and $m_2$ integers.   One finds \begin{equation}
\mathbf{k} = \frac{4\mpi}{3a}\left((m_1+m_2)\hat{\mathbf{x}} + \sqrt{3}(m_1-m_2)\hat{\mathbf{y}}\right).  \label{eq:reciplattice}
\end{equation}

It is well-known how to find the eigenvalues of such a Hamiltonian \cite{ashcroft}.   For (\ref{eq:hopping}),  this was first done in \cite{wallace}, and a pedagogical discussion can be found in \cite{sachdevhubbard}.     We write \begin{equation}
c_A(\mathbf{k}) \equiv  \sum_{i \in A} \mathrm{e}^{\mathrm{i}\mathbf{k}\cdot \mathbf{r}_i} c_i.
\end{equation}
with the index $A$ denoting one of the two sublattices (red vs. blue).   (\ref{eq:hopping}) becomes \begin{equation}
H_0 = -t\int \frac{\mathrm{d}^2\mathbf{k}}{(2\mpi)^2} \left(\mathrm{e}^{\mathrm{i}\mathbf{k} \cdot a\mathbf{e}_1} + \mathrm{e}^{\mathrm{i}\mathbf{k} \cdot a\mathbf{e}_2} + \mathrm{e}^{\mathrm{i}\mathbf{k} \cdot a\mathbf{e}_3}\right)   c^\dagger_A(-\mathbf{k}) c_B(\mathbf{k})   + \mathrm{H.c.}
\end{equation}
with the wave number integral over the Brillouin zone only.   The eigenvalues of the global Hamiltonian $H_0$ are simply given by the energies of all occupied electronic states.   The energy levels of the electronic states are given by \begin{equation}
\epsilon(\mathbf{k}) = \pm \left|\mathrm{e}^{\mathrm{i}\mathbf{k} \cdot a\mathbf{e}_1} + \mathrm{e}^{\mathrm{i}\mathbf{k} \cdot a\mathbf{e}_2} + \mathrm{e}^{\mathrm{i}\mathbf{k} \cdot a\mathbf{e}_3}\right|.  \label{eq:epsmicro}
\end{equation}
Note that there is an additional degeneracy due to the electronic spin, which we have neglected.

Of interest to us will be the charge neutrality point, where there is exactly one electron per site.    This means that exactly half of the energy levels will be occupied, and from the symmetry of (\ref{eq:epsmicro}) it is clear that the charge neutrality point has Fermi energy, or chemical potential, $\mu=0$.   The Fermi surface consists of points $\mathbf{k}$, in the Brillouin zone, where (\ref{eq:epsmicro}) vanishes.   There are exactly two such points: \begin{equation}
\mathbf{k} =  \pm \frac{4\mpi}{3\sqrt{3}a}\left(\frac{\sqrt{3}}{2} \hat{\mathbf{x}} + \frac{1}{2}\hat{\mathbf{y}}\right) .
\end{equation}
All other $\mathbf{k}$ at which $\epsilon=0$ are related to these two by a reciprocal lattice vector given in (\ref{eq:reciplattice}).    For $\epsilon \ll t$, we may Taylor expand (\ref{eq:epsmicro}) and obtain (\ref{eq:disprel}) 
with \begin{equation}
v_{\mathrm{F}} = \frac{3ta}{2\hbar} \approx 1.1\times 10^6\;\frac{\mathrm{m}}{\mathrm{s}}.
\end{equation}

This dispersion relation is identical to that of a massless Dirac fermion, with Hamiltonian \begin{equation}
H \approx \hbar v_{\mathrm{F}} \left(\sigma^x k_x + \sigma^y k_y\right),   \label{eq:HDirac}
\end{equation}
where $\sigma^x$ and $\sigma^y$ are the usual Pauli matrices.   For energies $\epsilon \ll t$, one may just as well use this Hamiltonian.    Of course, because there are inequivalent two Dirac points,  as well as two electronic spin degrees of freedom, there are a total of $N=4$ Dirac fermions in this effective description of graphene.      

The vanishing of a gap is related to the symmetry between the red and blue sublattices in Figure \ref{fig:honeycomb}.   There are perturbations to graphene that can break this symmetry, such as uniaxial strain, which can open a gap \cite{zhni}.  This is equivalent to perturbing (\ref{eq:HDirac}) with a term $\sigma^z\Delta$, with $\Delta$ the energy of the gap.    Indeed, there are many microscopic  Hamiltonians which can give rise to emergent massless Dirac fermions;  such points are generically symmetry-protected.

The energy scale at which the curvature of the band structure becomes significant is quite high.  In units of temperature:  \begin{equation}
T_\Lambda \equiv \frac{t}{k_{\mathrm{B}}} \sim 10^5\; \mathrm{K}.    \label{eq:TLambda}
\end{equation}
For practical purposes, we can think of the electrons in graphene as quasirelativistic.

\subsection{Charge Puddles}
\label{sec:puddle}
\begin{figure}
\centering
\includegraphics[width=4in]{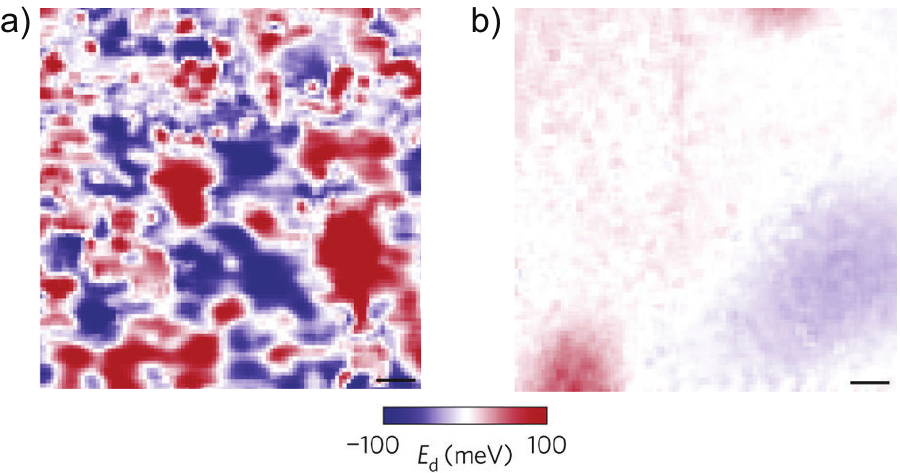}
\caption{Spatial maps of the charge puddles measured with scanning probe microscopy for graphene on (a) silicon dioxide and (b) hexagonal-boron nitride. Scale bars are 10 nm. Figure adapted from \cite{xue} with permission.}
\label{fig:ChargePuddles}
\end{figure}

The electronic properties of any material, including graphene, will be strongly affected by the phonons and impurities/imperfections which are always present in any experimentally realized system. Even for ``defect-free" graphene at low temperatures with negligible phonon excitations,  impurities modify electronic transport.  In graphene, the dominant source of impurities are believed to be  charged impurities, located out of the plane of graphene, in the substrates on which a monolayer of graphene is inevitably placed \cite{Nomura:2007ed, sarma1, Chen:2008hp}.  These impurities pin spatial fluctuations in the Fermi energy, or chemical potential.  Near the Dirac point, this impurity-induced fluctuation can be larger than the average chemical potential. The charge density will then be positive in some regions of space, while negative in others;  the resulting regions are often called charge puddles \cite{adamchargepuddles}.  Both the magnitude of these fluctuating potentials, and their spatial size, can be mapped using scanning probes \cite{xue, crommie, yacoby2007}. In Figure \ref{fig:ChargePuddles}a, the scanning tunneling microscope shows that these disorder puddles can lead to local fluctuations of 56 meV in the chemical potential, over a length scale $\sim 10$ nm, when graphene is put on top of a silicon dioxide ($\mathrm{SiO}_2$) surface.   In units of temperature, this is $\sim 650$ K, and it implies that until the electronic temperature is larger than 650 K, experiments will measure the physics of doped graphene, and not charge neutral graphene.

A major improvement comes from suspending the graphene monolayer \cite{bolotin}, or replacing the amorphous $\mathrm{SiO}_2$ by planar hexagonal-boron nitride (hBN) \cite{Dean:2010jy}.  As the electronic wave function is  mostly confined in the plane between boron and nitrogen atoms,  hBN is an exceptional dielectric with band gap $\sim 6$ eV.   With a band structure remarkably similar to graphene, along with a comparable interatomic spacing,  it provides a clean electrical environment which encapsulates the graphene without significantly modifying the band structure (\ref{eq:disprel}) in most cases (though see \cite{ortix, yankowitz}). Figure \ref{fig:ChargePuddles}b shows that both the size of  charge puddles now increase up $\gtrsim$ 100 nm, and the chemical potential fluctuations reduce to $\sim 5$ meV.    More improvement can come from a graphite back gate to further screen away any remote static potential.

We can also see the impact of charge puddles from electrical transport measurements \cite{sarma1}.  In textbook transport theory \cite{ashcroft}, the electrical conductivity $\sigma$ is proportional to the carrier density $n$ in the presence of the long-range impurity scattering.  However, this linearity will break down when the carrier density is comparable to the density induced by the charge puddles.  Schematically \begin{equation}
\sigma \sim \sigma_{\mathrm{min}} \sqrt{1+\frac{n^2}{n_{\mathrm{imp}}^2}}
\end{equation}
where $n_{\mathrm{imp}}$ is (very crudely) the local carrier density in the presence of charge puddles.  For temperatures below $T_{\mathrm{imp}} = \hbar v_{\mathrm{F}} \sqrt{\pi n_{\mathrm{imp}}}/k_{\mathrm{B}}$, the charge puddles are essentially themselves tiny patches of Fermi liquid, and we find that $n_{\mathrm{imp}}$ is approximately $T$-independent.   For temperatures above $T_{\mathrm{imp}}$, we observe $n_{\mathrm{imp}}$ to be an increasing function of $T$ \cite{crossno}.  This signifies the presence of thermally excited electrons and holes near the Dirac point.

\subsection{Phonons}
\label{sec:ephonon}
\begin{figure}
\centering
\includegraphics[width=6in]{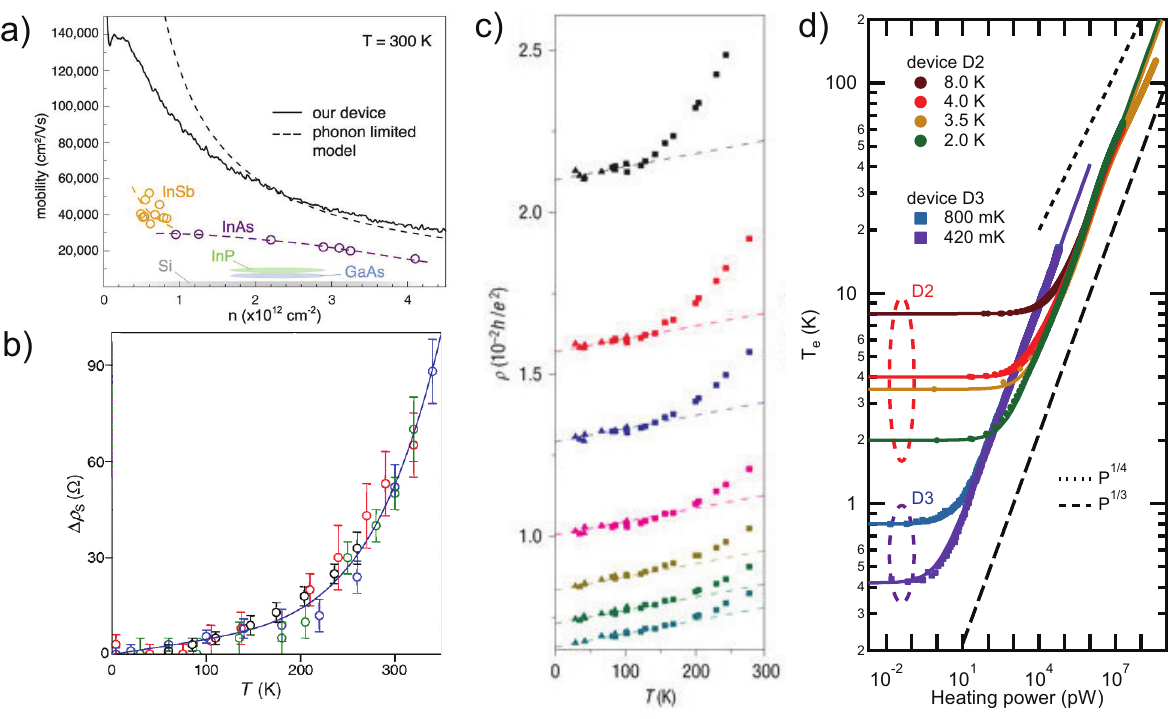}
\caption{(a) Phonon-limited electronic mobility at high density and temperature \cite{wang13}. (b) Resistivity after subtracting a temperature-independent component \cite{Morozov:2008uf}. (c) Temperature dependent resistivity,  demonstrating the contributions of various phonon scattering processes to the resistivity with gate voltages at 10, 15, 20, 30, 40, 50, 60 V from top to bottom (increasing carrier density) \cite{Chen:2008by}. (d) Measuring the electron-phonon coupling in graphene by heat transfer \cite{Fong:2013hl}. The scaling $\rho \sim T^3$ is a hallmark of supercollisions. Figures adapted from \cite{wang13, Chen:2008by, Morozov:2008uf, Fong:2013hl} with permission.}
\label{fig:ElectronPhonon}
\end{figure}

While the electronic properties of graphene are limited by impurities at low temperatures, it is lattice vibrations, or phonons, that constrain its electronic mobility at higher temperatures \cite{hwang07, Chen:2008by, Morozov:2008uf, wang13, sarma13}.  In experiments, a clean graphene monolayer can reach a mobility of about $4\times 10^4$ cm$^{2}$/V$\cdot$s at the carrier density of 4.5 $\times 10^{12}$ cm$^{-2}$ \cite{wang13}, corresponding to a mean-free-path of almost 1 $\mmu$m (see Figure \ref{fig:ElectronPhonon}a).  This mobility is higher than the two-dimensional III-V semiconductor heterostructure \cite{ancona, ashley}.   It is only limited by  scattering off acoustic phonons in the graphene lattice at high density.  In addition to acoustic phonons, the electrons in graphene can also be scattered by the surface optical phonons on the substrate, or by ripples/strain in the single atomic layer of graphene.  The scattering rates of acoustic vs. optical phonons can be disentangled by their carrier density and temperature dependence in the electrical conductivity (see Figures \ref{fig:ElectronPhonon}b and \ref{fig:ElectronPhonon}c). For instance, the resistivity $\rho$ due to  longitudinal acoustic phonons is directly proportional to temperature ($\rho \sim T$) whereas $\rho(T)$ highly non-linear when it is dominated by activated surface phonons \cite{Chen:2008by}. Unlike conventional metals, graphene has a small Fermi surface but a large Debye temperature ($\sim$2800 K) due to the stiffness of the chemical bonds.  The characteristic temperature of the electron-phonon coupling in  graphene is thus set by the Bloch-Gr\"{u}neisen temperature $T_{\mathrm{BG}}$ instead of Debye temperature \cite{Efetov:2010fu}. Below (above) $T_{\mathrm{BG}}$, the phonon system is (non-)degenerate, giving a different resistivity temperature power law, i.e. $\rho\sim T^{4}$ ($\rho\sim T$), that can be measured in experiments.

Electron-phonon coupling can also be studied using Raman spectroscopy \cite{Yan:2007ie, Ferrari:2007fba} and heat transfer \cite{Tse:2009il, Bistritzer:2009ht, Kubakaddi:2009br, viljas, Chen:2012et}.  Since the electron-phonon coupling is weak in graphene, it is possible for the electrons in graphene  to maintain a higher temperature than the phonon background for some time. The resulting heat transfer process can be measured by the time response of a photocurrent \cite{Graham:2013vl} or in thermal conductivity measurements \cite{Betz:2012wya, Fong:2012ut, Betz:2013up, Fong:2013hl}.  By measuring the dependence of the heat transfer rate on temperature, these experiments can identify the dominant energy relaxation process (see Fig. \ref{fig:ElectronPhonon}c), such as ``supercollisions" \cite{song2011}: a three-body collision between an   electron, phonon and impurity that leads to $\rho \sim T^3$ scaling. Heat transfer measurements also confirm the importance of the substrate as a source of optical phonons which strongly couple to the electrons in graphene at higher temperatures \cite{crossno2}. 

\subsection{Is Electron-Electron Scattering Negligible?}
\label{sec:negligible}
Due to recent advances in pump-probe spectroscopy \cite{ma1, song2, song, johannsen, ma2}, we can rather directly measure the rate of electron-electron scattering in graphene, at least in the limit of high energy excitations.  The experiments are usually performed by observing the transmission or reflection properties of the graphene sample with a probe beam shortly after a pump beam promotes interband absorption.  Experiments find a very fast decay signal, typically on the order of 10-100 fs, followed by a slow decay, on the order of 10 ps. They are attributed to the electron-electron and electron-phonon interaction time, respectively.   More precisely, the initially excited electron-hole pairs have a very high energy relative to the ambient temperature.   On fs time scales, electron-electron scattering causes rapid thermalization of the electrons,  while on ps time scales electron-phonon scattering brings the combined electron-phonon system to thermal equilibrium.   
Remarkably,  the electronic fluid appears to thermalize so quickly that on all time scales accessible experimentally, the distribution of electrons is of the equilibrium Fermi-Dirac form.   The temperature of the electrons (found from the fitted form of the distribution) lowers as energy is lost to the phonon bath over time. 
  This is strong evidence that electronic interactions cannot be neglected in any study of the quantum many-body dynamics of graphene.

Because the Fermi temperature is much smaller in graphene than in conventional metals, the electron-electron scattering time is expected to be relatively fast even in the doped regime. Near the Dirac point, it is almost as fast as possible.   The rest of this review is about the measurable consequences of electron-electron interactions.

\section{Electronic ``Phase Diagram"}
\label{sec:phases}
As we saw in Section \ref{sec:negligible}, electron-electron interactions can be quite fast  in graphene, especially near the neutrality point.   We must now include them in our model.  The most important electron-electron interactions are simply Coulomb interactions:   \begin{equation}
H = H_0 +  \sum_{i\ne j}  \frac{\alpha}{|\mathbf{x}_i - \mathbf{x}_j|} c^\dagger_i c_i c^\dagger_j c_j, \label{eq:HDF}
\end{equation}
with $H_0$ given in (\ref{eq:hopping}).
We emphasize that these Coulomb interactions have a $1/r$ tail, despite the fact that graphene is a ``two-dimensional metal", and in two dimensions Coulomb potentials are $\log r$.  This is because the Coulomb interactions are mediated by out-of-plane electromagnetism, in three spatial dimensions.

There are other types of interactions one could, in principle, include.  For example, one could add a Hubbard interaction, penalizing electrons of opposite spin on the same lattice site \cite{sachdevhubbard}:  \begin{equation}
H = H_0 + U\sum_i  c^\dagger_{i\uparrow} c_{i\uparrow} c^\dagger_{i\downarrow} c_{i\downarrow}.
\end{equation}
This interaction arises from the same Coulomb interaction as (\ref{eq:HDF}), but between the orbitals on a single atom \cite{hubbard}.   When $U$ is very large, such interactions lead to an antiferromagnetic insulating phase \cite{herbut1, herbut2, herbut3}.  Experimentally, graphene is observed to be a conductor -- evidently, $U$ is not large enough in real graphene.    It may be possible to experimentally drive an insulating phase upon applying strain \cite{adam1505}.   Our focus in this review is on the conducting limit of ordinary graphene.

\subsection{Fermi Liquid}
When the Fermi energy is large compared to the temperature, graphene behaves like a ``conventional" two-dimensional metal with long-lived quasiparticles \cite{Sarma:2007ej}.     The standard arguments suggest the ``robustness" of this Fermi liquid \cite{pines, shankarRMP}:  since electrons can only scatter into states very close to the Fermi surface, the scattering rate for electrons is \begin{equation}
\tau_{\mathrm{ee}} \sim \frac{1}{\alpha^2} \frac{\hbar \mu}{(k_{\mathrm{B}}T)^2},   \label{eq:FLtime}
\end{equation}
where $\mu$ is the chemical potential or Fermi energy.
The thermodynamic functions will also be described by conventional Fermi liquid theory. We will discuss these results more quantitatively in Section \ref{sec:kinetic}.

A generic Fermi liquid has superconducting instabilities at low temperature \cite{shankarRMP}.  It is generally not possible to observe this in graphene due to the  low density of states, made worse by the relativistic energy spectrum \cite{uchoa}.   It may be possible to obtain (chiral) superconductivity at higher temperatures by doping graphene very far from the neutrality point \cite{mcchesney, nandkishore2011}.  For a detailed review of possible superconducting instabilities in graphene, see \cite{eegraphenermp}.   At the end of the day, when thinking about hydrodynamics we may safely neglect any superconducting instabilities of the Fermi liquid in doped graphene due to (\emph{i}) the extremely low $T_{\mathrm{c}}$ of any putative instability anywhere close to charge neutrality, and (\emph{ii}) at very low $T$, the hydrodynamic description will break down as $\ell_{\mathrm{ee}} \gg \ell_{\mathrm{imp}}$.

If graphene is an ordinary Fermi liquid at low temperature, albeit one in two spatial dimensions, why is graphene a good candidate material for observing the effects of electron-electron interactions?  The high quality with which samples of graphene may be fabricated (Section \ref{sec:puddle}) and the lack of strong electron-phonon coupling (Section \ref{sec:ephonon}) both allow us to make $\ell_{\mathrm{imp}}$ very large,  even if $\ell_{\mathrm{ee}}$ is relatively ``large" itself.  The simple quasi-relativistic band structure of graphene also has a straightforward hydrodynamic description, unlike more complicated Fermi surfaces \cite{hartnoll1704, hartnoll1705}.  However, graphene is not the unique material with favorable properties, and indeed signatures of Fermi liquid electronic hydrodynamics have been observed in GaAs \cite{molenkamp}, $\mathrm{PdCoO}_2$ \cite{mackenzie} and $\mathrm{WP}_2$ \cite{felser}.   While certain quantitative simplicities of the relativistic hydrodynamic description described in Section \ref{sec:hydro} will not apply to these materials, so long as there are no additional hydrodynamic degrees of freedom, much of the theory of Section \ref{sec:FLtrans} will apply to these more complicated materials.

\subsection{Dirac Fluid}
The reason why the Fermi liquid is weakly interacting is essentially the fact that the Fermi surface provides strong kinematic constraints on the possible scattering pathways.   If we place the chemical potential in graphene at the neutrality point, then there is no longer a Fermi surface, and so this argument no longer applies.   Furthermore, because the only low energy scale is $T$, dimensional analysis implies that \begin{equation}
\tau_{\mathrm{ee}} \sim \frac{1}{\alpha^2} \frac{\hbar }{k_{\mathrm{B}}T}.  \label{eq:DFtime}
\end{equation}
Recall the definition of $\alpha$ in (\ref{eq:alpha1});   the prefactor of $1/\alpha^2$ will be justified in Section \ref{sec:kinetic}.     As we discussed in (\ref{eq:alphaest}),  a naive estimate of $\alpha$ in graphene is quite large.   Also, at low temperatures,  this time scale grows much more slowly than (\ref{eq:FLtime}).   So we expect that electron-electron interactions ought to be much more important in charge-neutral graphene.

\subsubsection{Renormalization Group}
As is well known in quantum field theory, the fact that the bare `coupling constant' $\alpha$ in graphene is quite large is not sufficient to ensure that the low energy effective theory of graphene is strongly coupled.   Indeed,  $\alpha$ is large in the Fermi liquid phase, and yet the large Fermi surface screens out the strong interactions.   A more sophisticated renormalization group (RG) analysis \cite{wilsonRG} allows us to compute an \emph{effective} value of $\alpha$, $\alpha_{\mathrm{eff}}$ at a given temperature $T$.   This effective value of the coupling accounts for both thermal and quantum fluctuations, and serves as a diagnostic for the true coupling $\alpha$ which is in (\ref{eq:DFtime}).

This paragraph describes the RG analysis of the Dirac fluid; readers unfamiliar with this approach may wish to consult a textbook such as \cite{kardar2}.   This analysis assumes that $\alpha$ is perturbatively small.  Readers uninterested in the details can skip to (\ref{eq:alphaeffT}) for the physical result.    Let us consider the quantum field theory version of the Hamiltonian (\ref{eq:HDF}),  which replaces the electron creation/annihilation operators $c^\dagger_i /c_i$ (defined on honeycomb lattice sites) by $N_{\mathrm{f}}=4$ Dirac fermions $\Psi^A(\mathbf{x},t)$  ($A=1,\ldots,N_{\mathrm{f}}$) in the spacetime continuum.   One then writes down an effective action which depends on the energy scale $\tilde\mu$ above which our theory is (by construction) ill-defined:
\begin{equation}
S_{\mathrm{eff}}(\mu) = \int \mathrm{d}t \mathrm{d}^2\mathbf{x} \; \mathrm{i}Z(\tilde\mu) \overline{\Psi}^A \left[\gamma^t \partial_t + v_{\mathrm{F}}(\tilde\mu) \gamma^i \partial_i  \right] \Psi^A    - \int \mathrm{d}t \mathrm{d}^2\mathbf{x}  \mathrm{d}^2\mathbf{x}^\prime \frac{e^2}{8\mpi \varepsilon(\tilde\mu)} \frac{n(\mathbf{x}) n(\mathbf{x}^\prime)}{|\mathbf{x}-\mathbf{x}^\prime|}  \label{eq:SDF}
\end{equation}
with $\overline{\Psi}^A = \Psi^{\dagger A}\gamma^t$ and $n = \Psi^{\dagger A} \Psi^A$.     $\gamma^t$ and $\gamma^i$ are Dirac's gamma matrices, and $Z$ is a prefactor corresponding to the ``renormalization" of the quasiparticle weight.  One integrates out quantum fluctuations of the fermion field $\Psi$ with momentum $|\mathbf{k}| \ge \mu$,  and (schematically) looks for a $Z(\mu)$,  $\varepsilon(\mu)$ and $v_{\mathrm{F}}(\mu)$ such that the correlation functions $\langle \Psi^\dagger(\mathbf{k}_1)\cdots  \Psi(\mathbf{k}_n)\rangle_{|\mathbf{k}_i| < \mu}$ are the same, whether one evalutes the correlation function using the effective action $S_{\mathrm{eff}}(\tilde\mu)$,  or the effective action $S_{\mathrm{eff}}(\tilde\mu^\prime)$, with $\tilde\mu^\prime > \tilde\mu$.      For the theory in (\ref{eq:SDF}), one finds that $Z(\tilde\mu)$ and $\varepsilon(\tilde\mu)$ are constants.   The only parameter which varies is $v_{\mathrm{F}}(\tilde \mu)$.  It is common to write down a `flow equation' for the effective Fermi velocity as a function of energy scale.  At leading order in $\alpha$ one finds \cite{vozmediano1, vozmediano2, vafek, schmalian}  \begin{equation}
\frac{\mathrm{d}v_{\mathrm{F}}}{\mathrm{d}\log \tilde\mu} = -\frac{e^2}{16\mpi \varepsilon}.
\end{equation}
Using (\ref{eq:alpha1}), it is more instructive to write this equation as \begin{equation}
\frac{\mathrm{d}\alpha}{\mathrm{d}\log \tilde\mu} = \frac{\alpha^2}{4}.  \label{eq:alphaRG1}
\end{equation}
We now assume that at a very high energy scale of $\mu = k_{\mathrm{B}}T_\Lambda \sim 10^5$ K (the energy scale at which the dispersion relation in graphene is not relativistic: see (\ref{eq:TLambda})), the coupling constant $\alpha$ is given by its `bare' value $\alpha_0$.   Integrating this equation down to an energy scale $\mu=k_{\mathrm{B}}T$, we find 
\begin{equation}
\alpha_{\mathrm{eff}}(T) = \dfrac{\alpha_0}{\displaystyle 1 + \frac{\alpha_0}{4} \log \dfrac{T_\Lambda}{T}}.  \label{eq:alphaeffT}
\end{equation}
We say that Coulomb interactions are \emph{marginally irrelevant} because the dimensionless coupling constant is vanishing logarithmically fast at low temperature.   

If as $T\rightarrow 0$,  $\alpha_{\mathrm{eff}}(T)$ is given by (\ref{eq:alphaeffT}),  then why did we emphasize in the introduction that electron-electron interactions could still play an important role in graphene?   Suspended graphene has $\alpha_0 = 2.2$;  graphene on substrates has $\alpha_0 \approx 0.8$ due to the dielectric constants of the substrates \cite{Dean:2010jy}.    At a temperature of $T=100$ K, we estimate $\alpha_{\mathrm{eff}} \approx 0.46$ for suspended graphene, and $\alpha_{\mathrm{eff}} = 0.34$ for graphene on subsbtrates.   These are \emph{not}  small coupling constants.   Because these coupling constants are not small,  one should not take (\ref{eq:alphaeffT})  too seriously, and ultimately any coefficients (such as viscosity) which will be very sensitive to  the value of $\alpha$ can be treated as phenomenological fit parameters and experimentally  measured.

Of course, the right hand side of (\ref{eq:alphaRG1}) has corrections at $\mathrm{O}(\alpha^3)$.  Those corrections can be explicitly computed, and one finds \cite{vafek2, sarma14} \begin{equation}
\frac{\mathrm{d}\alpha}{\mathrm{d}\log \tilde\mu} = -\frac{\alpha^2}{4} \left(1 - \frac{\alpha}{\alpha_{\mathrm{c}}}\right) + \cdots,
\end{equation} 
where $\alpha_{\mathrm{c}} \approx 0.8$.   This suggests that the Dirac fluid may be unstable at strong coupling $\alpha>\alpha_{\mathrm{c}}$.  The proposed endpoint of such an instability is an excitonic insulator \cite{vafek2}.  And while early numerical studies had suggested that the true ground state of charge-neutral graphene was not the Dirac fluid, but an insulator \cite{drut1, drut2}, experiments unambiguously show that charge-neutral graphene is a conductor.  More sophisticated treatments of the RG \cite{vozmediano2, sarmaRPA, kopietz},  along with more recent numerical studies \cite{polikarpov, tupitsyn}, have confirmed that the Dirac fluid is not unstable at large values of $\alpha$ -- (\ref{eq:alphaeffT}) is qualitatively correct, although the precise numerical coefficients may be incorrect.

\begin{figure}[t]
\centering
\includegraphics[width=2.1in]{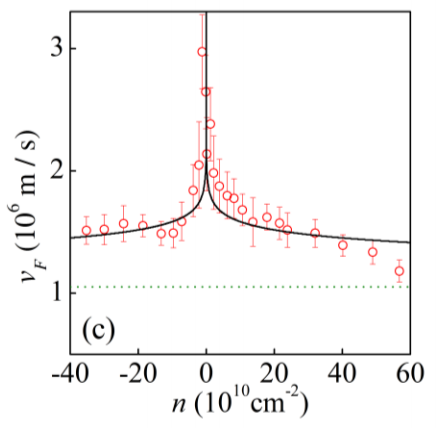}
\caption{An apparent logarithmic divergence in the effective Fermi velocity near the Dirac point in graphene, experimentally observed by measuring the decay rate of quantum oscillations.    Figure adapted from \cite{elias} with permission.}
\label{fig:elias}
\end{figure}

A direct physical prediction of the RG described above is that the effective Fermi velocity of the Dirac fermions in graphene is temperature dependent.    Combining (\ref{eq:alpha1}) and (\ref{eq:alphaeffT}),  we find \begin{equation}
v_{\mathrm{F,eff}}(T) \approx v_{\mathrm{F,0}} \left(1+\frac{\alpha_0}{4}  \log \frac{T_\Lambda}{T}\right).   \label{eq:vFeff}
\end{equation}
A similar prediction can be made for an effective density-dependent Fermi velocity at low densities \cite{vozmediano1}: \begin{equation}
v_{\mathrm{F,eff}}(n) \approx \tilde v_{\mathrm{F,0}} \left(1+\frac{\alpha_0}{8}  \log \frac{n_\Lambda}{n}\right).  \label{eq:vFlog}
\end{equation} 
The constants $v_{\mathrm{F,0}}$ and $\tilde v_{\mathrm{F,0}}$ can be different. Direct experimental evidence for this Fermi velocity renormalization as a function of density was observed in \cite{elias}:  see Figure \ref{fig:elias}.   In this experiment, a magnetic field was applied to suspended graphene, and the resulting quantum oscillations in the conductivity were measured.   Applying the standard quasiparticle-based theories \cite{sharapov}, together with the proportionality between $v_{\mathrm{F,eff}}$ and the cyclotron frequency in graphene, \cite{elias} was able to observe a slight modification of (\ref{eq:vFlog}), with relatively good experimental agreement with (\ref{eq:vFlog}).   Evidence for the enhancement of the Fermi velocity near the charge neutrality point was also observed using ARPES in \cite{siegel2011}.

\begin{figure}
\centering
\includegraphics[width=5in]{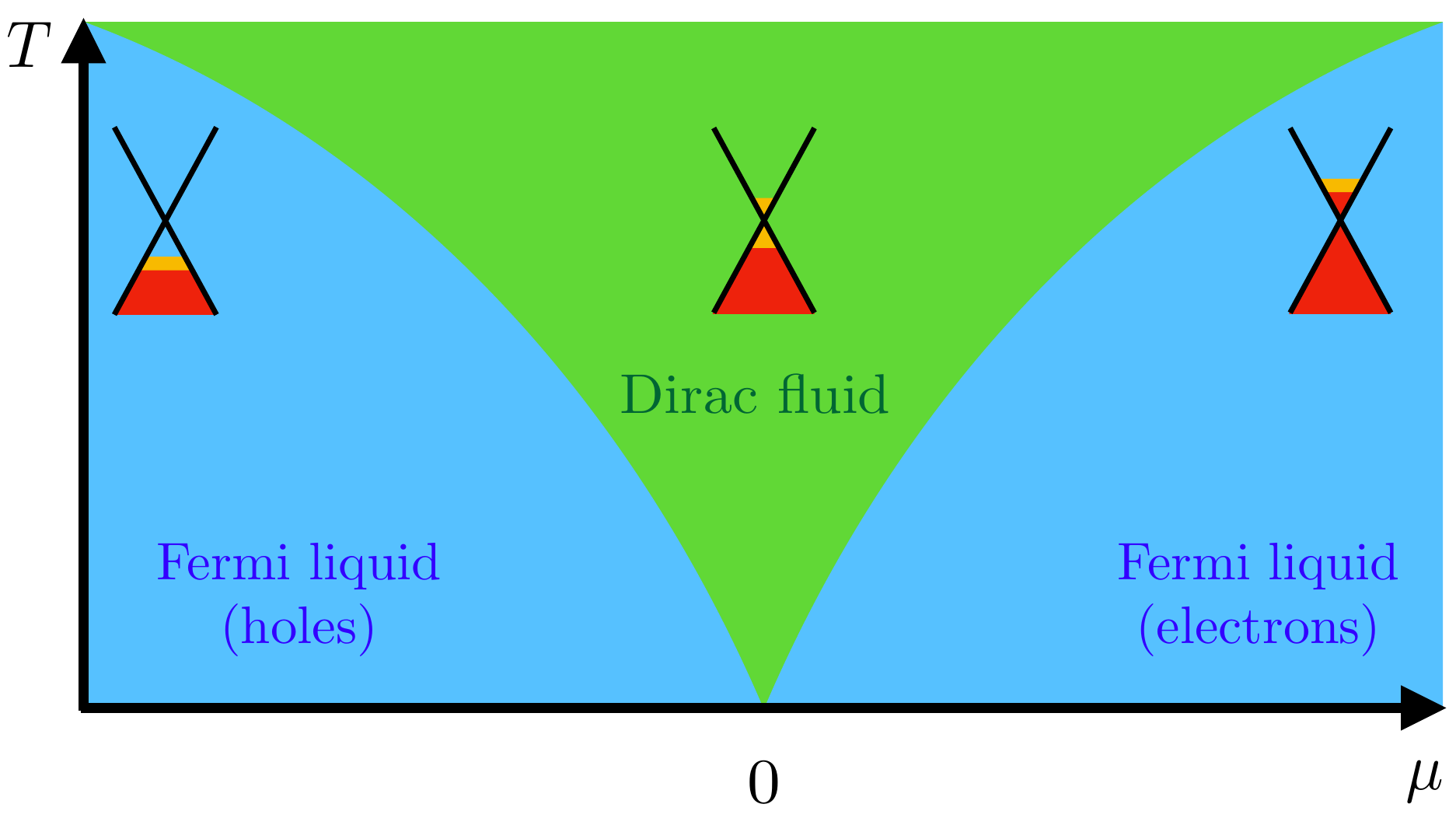}
\caption{The ``phase diagram" of electronic dynamics in graphene.  The blue regions denote the Fermi liquids, which can be either electron-like or hole-like.   The green region denotes the Dirac fluid, an electron-hole plasma with relatively strong interactions.   The boundary between blue and green regions denotes a  crossover and is not sharp.  We also depict the band structure in each part of the phase diagram:  red denotes filled electronic states with negligible thermal fluctuations, and yellow denotes where thermal fluctuations are significant.}
\label{fig:phasediagram}
\end{figure}

Our reuslts so far are summarized in Figure \ref{fig:phasediagram}.    Graphene gives rise to an ordinary Fermi liquid when $\mu \gg k_{\mathrm{B}}T$,  and a (relatively) strongly coupled Dirac fluid when $\mu \ll k_{\mathrm{B}}T$ (up to logarithmic corrections).   Electronic dynamics in ultrapure graphene is described by hydrodynamics across the phase diagram, but as we will see,  the qualitative change in the interaction rate from (\ref{eq:FLtime}) to (\ref{eq:DFtime}) will lead to profound, and experimentally measurable, changes in the hydrodynamic response of the theory:   as we will detail in Section \ref{sec:kinetic},  the hydrodynamic coefficients scale very differently with temperature across this ``phase diagram."

The qualitative phase diagram shown in Figure \ref{fig:phasediagram} is reminiscent of the theory of quantum criticality \cite{SSBK11}:  at finite temperature, the interplay of thermal and quantum fluctuations lead to a very strongly interacting quantum system.   In fact, there is a rather trivial quantum critical point between a hole-like Fermi liquid and an electron-like Fermi liquid at $T=0$.

\section{Relativistic Hydrodynamics}
\label{sec:hydro}
In this section we discuss quantitatively the hydrodynamics of a relativistic system.   The approach is that of \cite{landau}.  Here, we will assume relativistic invariance, and focus on the physics near the charge neutrality point;  we justify the relativistic assumption in Section \ref{sec:userel}.   This limit requires us to interpret the hydrodynamics in a slightly different way 
\cite{hkms} than is done in the astrophysics literature.   In a typical relativistic plasma in astrophysics, one has a fluid of heavy ions coupled to a fluid of light electrons.   As such, the number of ions and the number of electrons are \emph{separately} conserved.   This is not the case in graphene:   there are processes which create electrons and holes, conserving only the net electric charge: see Section \ref{sec:imbalance} for a detailed discussion of this issue.     So it will be important to consider a fluid which can be either positively  or negatively charged, instead of a two-fluid model.

\subsection{Thermodynamics}
\label{sec:thermo}
Before a discussion of hydrodynamics, it is important to understand the static backgrounds about which we will build our hydrodynamic theory.   These are states in thermal equilibrium.   So let us begin with some elementary thermodynamics.   Consider a system in a $d$-dimensional region of volume $V$, with a conserved charge and energy.   In graphene, $d=2$, but we might as well keep $d$ general for now.   The first law of thermodynamics states that \begin{equation}
\mathrm{d}E = T\mathrm{d}S + \mu \mathrm{d}N  - P \mathrm{d}V  \label{eq:firstlawthermo}
\end{equation}
where $E$ is the total energy,  $T$ is the temperature, $S$ the total entropy, $\mu$ the chemical potential, $N =- Q/e$ with $Q$ the net charge,  $P$ the pressure and $V$ the volume of the sample (keep in mind that for graphene in $d=2$, this volume is physically interpreted as the surface area of the sample).   Dividing through by $\mathrm{d}V$ and demanding extensivity we obtain the Gibbs-Duhem relation \begin{equation}
\epsilon +P = \mu n + Ts,  \label{eq:gibbsduhem}
\end{equation}
with $\epsilon$ the energy density,  $n$ the (relative) number density, and $s$ the entropy density.   Now suppose that we work in a fixed area, so $\mathrm{d}V = 0$, and $\mathrm{d}E = V\mathrm{d}\epsilon$, $\mathrm{d}N = V\mathrm{d}n$ and $\mathrm{d}S = V\mathrm{d}s$.     Simple manipulations give: \begin{equation}
\mathrm{d}P = \mathrm{d}( \mu n + Ts - \epsilon) = n \mathrm{d}\mu + s\mathrm{d}T.   \label{eq:dP}
\end{equation}
This thermodynamic identity implies that if we treat $\mu$ and $T$ as our tuning parameters (as is useful for theoretical purposes),  then the pressure $P$ plays the role of our thermodynamic potential.   All information about the thermodynamics of the fluid is contained in $P(\mu, T)$.

Suppose that the only two energy scales in the problem are the temperature $k_{\mathrm{B}}T$ and the chemical potential $\mu$.   Dimensional analysis requires that \begin{equation}
P(\mu, T) = \frac{(k_{\mathrm{B}}T)^{d+1}}{(\hbar v_{\mathrm{F}})^d} \mathcal{F}\left(\frac{\mu}{k_{\mathrm{B}}T}\right) 
\end{equation}
where $\mathcal{F}$ is an arbitrary function that obeys thermodynamic requirements such as $s\ge 0$.   In graphene, assuming the interaction strength $\alpha$ is small, one can find the explicit form of $\mathcal{F}$, and it is presented explicitly in (\ref{eq:54P}), later in this review.   Using (\ref{eq:dP}) we find \begin{subequations}\label{eq:nsthermo}\begin{align}
n(\mu, T) &= \left(\frac{k_{\mathrm{B}}T}{\hbar v_{\mathrm{F}}}\right)^d \mathcal{F}^\prime \left(\frac{\mu}{k_{\mathrm{B}}T}\right), \\
s(\mu, T) &=  (d+1)k_{\mathrm{B}} \left(\frac{k_{\mathrm{B}}T}{\hbar v_{\mathrm{F}}}\right)^d \mathcal{F} \left(\frac{\mu}{k_{\mathrm{B}}T}\right) - \frac{k_{\mathrm{B}}^d\mu T^{d-1}}{(\hbar v_{\mathrm{F}})^d}\mathcal{F}^\prime \left(\frac{\mu}{k_{\mathrm{B}}T}\right).
\end{align}\end{subequations}
Combining (\ref{eq:gibbsduhem}) and (\ref{eq:nsthermo}) we obtain \begin{equation}
\epsilon = dP.  \label{eq:epsdP}
\end{equation}
This relation will prove to have important consequences.

There are two points our discussion has overlooked.   Firstly, due to the weak logarithmic temperature dependence of $\alpha(T)$ in the Dirac fluid, the thermodynamics of graphene may be slightly more complicated than the above.   This is unlikely to be a qualitative effect, and so we will neglect it in what follows for simplicity.   Secondly, in a sample of graphene with charge puddles,  there is another scale $\mu_{\mathrm{rms}}$: the root-mean-square fluctuations of the chemical potential.   This implies that the thermodynamics described above is too simple.  Further discussion of both points can be found in \cite{lucas3}.
 
\subsection{The Gradient Expansion}
Our modern understanding of hydrodynamics is that it is the effective theory describing the dynamics of any many-body system relaxing to thermodynamic equilibrium  \cite{kadanoff, landau}.    The thermodynamic equilibria we described above -- with a slight generalization to allow for finite momentum density in an infinite volume -- must then be static solutions to the hydrodynamic equations.    Our key postulate is that on time scales large compared to the mean free time between electron-electron collisions: $\tau_{\mathrm{ee}}$,  and on long length scales compared to $\ell_{\mathrm{ee}} \equiv v_{\mathrm{F}}\tau_{\mathrm{ee}}$,  the only slow dynamics is associated with locally conserved quantities.   For us, this will consist of charge, energy and momentum.  

In the discussion that follows, it is convenient to choose units where $v_{\mathrm{F}} = 1$.    Factors of $v_{\mathrm{F}}$ can be restored with dimensional analysis.   We will restore such factors explicitly whenever an important physical formula is found.

In non-relativistic notation, the conservation law for a local charge density is \begin{equation}
\frac{\partial n}{\partial t} + \nabla \cdot \mathbf{J} = 0,
\end{equation}
where $\mathbf{J}$ is a spatial charge current.  In relativistic notation, we write 
\begin{equation}
J^\mu = (J^t,\mathbf{J})= (n, \mathbf{J}),\;\;\;\; \partial_\mu J^\mu = 0.  \label{eq:dJ0}
\end{equation}
The conservation law for energy and momentum reads \begin{equation}
\partial_\nu T^{\mu\nu}= 0;  \label{eq:dT0}
\end{equation}
the stress-energy tensor $T^{\mu\nu}$ describes both densities and spatial currents of energy and momentum:  $T^{tt}$ is the energy density,  $T^{ti}$ is the energy current,  $T^{it}$ is the momentum density, and $T^{ij}$ is the momentum current:  more commonly known as the stress tensor.     Relativistic invariance requires \begin{equation}
T^{\mu\nu} = T^{\nu\mu}.
\end{equation}

(\ref{eq:dJ0}) and (\ref{eq:dT0}) will form the basis for the hydrodynamic equations of motion.   We now must find expressions for $J^\mu$ and $T^{\mu\nu}$ in terms of the slow degrees of freedom.  These are the local ``charge" density \begin{equation}
n = n_{\mathrm{el}} - n_{\mathrm{hole}},
\end{equation}
with $n_{\mathrm{el}}$ and $n_{\mathrm{hole}}$ the number densities of electrons and holes respectively, the local energy density $\epsilon(\mathbf{x})$,  and the local momentum density $\mathbf{\Pi}(\mathbf{x})$.      It is conventional in hydrodynamics to not solve for $n$, $\epsilon$ and $\mathbf{\Pi}$.  Instead, one solves for thermodynamic conjugate variables:  the local chemical potential $\mu$,   the local temperature $T$, and the relativistic velocity vector \begin{equation}
u^\mu \equiv \frac{1}{\sqrt{1-|\mathbf{v}|^2}} (1,\mathbf{v}).   \label{eq:umu}
\end{equation}
Let us briefly remind the reader of relativistic index notation.   In this review, indices are raised and lowered by multiplying by the matrix  \begin{equation}
g^{tt} = -1, \;\;\; g^{ij} = \mdelta^{ij}, \;\;\; g^{ti} = g^{it} = 0;
\end{equation}
hence $u_t = -u^t$ and $u_i = u^i$.   Greek letters $\mu\nu$ will denote spacetime indices, while Latin letters $ij$ will denote spatial indices.   The simple identity \begin{equation}
u_\mu u^\mu = -1  \label{eq:umu2}
\end{equation}
follows from (\ref{eq:umu}).

We now proceed along the lines of \cite{hkms} to derive the hydrodynamic equations of motion.   As we have previously stated, hydrodynamics is an effective theory.   This means that one writes down the most complicated possible equations of motion consistent with basic principles such as symmetry,  up to a given order in a small parameter ``$\delta$".   In hydrodynamics, that small parameter is the ratio of the electronic mean free path to the size $\xi$ of perturbations:   $\delta = \ell_{\mathrm{ee}}/\xi$.    Alternatively, $\delta \sim \ell_{\mathrm{ee}}\partial$, with derivatives understood to act on the slowly varying functions such as $T$ or $u^\mu$.     We will then write down the most general $\mathrm{O}(\partial^n)$ expressions for $J^\mu$ and $T^{\mu\nu}$, for a small value of $n=0,1$.     Finally, we assume a local second law of thermodynamics, following \cite{hkms}.\footnote{Recently hydrodynamics has been understood from even more basic principles \cite{rangamani, glorioso}, but these are extremely technical and well beyond the scope of this review.}    
 
\subsubsection{Zeroth Order}
We begin at zeroth order in derivatives.   Let us imagine that we have a fluid, exactly at rest, and in global thermodynamic equilibrium.   We then know that the charge current is given by \begin{equation}
J^\mu = (n,\mathbf{0}),  \label{eq:J0}
\end{equation}
and the stress-energy tensor is given by \begin{equation}
T^{\mu\nu}= \left(\begin{array}{cc}  \epsilon &\ \mathbf{0} \\ \mathbf{0} &\  P \mdelta^{ij} \end{array}\right).  \label{eq:T0}
\end{equation}
Here $n$ is the charge density, $\epsilon$ is the energy density and $P$ is the pressure.  The reason that pressure corresponds to a momentum flux is that the force acting on a hard wall of area $A$ is given by $P \times A$ -- that is the momentum per unit time impacting the boundary.     As we have shown in Section \ref{sec:thermo}, $n$, $\epsilon$ and $P$ are not independent functions of $\mu$ and $T$.    

It now remains to consider a fluid which is moving.   This means that we must write (\ref{eq:J0}) and (\ref{eq:T0}) in terms of the vector $u^\mu = (1,\mathbf{0})$, as well as the metric $g^{\mu\nu}$.    The resulting hydrodynamics will be Lorentz-covariant:   the fluid velocity (set by the \emph{state} of the system) destroys the full relativistic Lorentz symmetry by picking a preferred reference frame (where the fluid is at rest),  even if the microscopic action is invariant under arbitrary Lorentz transformations.  It is simple to accomplish this:  one finds \begin{subequations}\label{eq:curr0}\begin{align}
J^\mu &=  nu^\mu, \\
T^{\mu\nu} &=  (\epsilon+P)u^\mu u^\nu + P g^{\mu\nu}.
\end{align}\end{subequations}

At zeroth order we find an ``accidental" conservation law for entropy.   Defining the entropy current $s^\mu \equiv  su^\mu$,  with the entropy $s = \partial P/\partial T$, as given in (\ref{eq:dP}), we claim that \begin{equation}
\partial_\mu \left(su^\mu\right) = 0.   \label{eq:scurr0}
\end{equation}
To prove this result, we note that for the zeroth order stress tensor given by (\ref{eq:curr0}), \begin{equation}
u_\mu \partial_\nu T^{\mu\nu} = 0 = - u^\mu \partial_\mu \epsilon - (\epsilon+P)\partial_\mu u^\mu   \label{eq:temp42}
\end{equation}
We have used (\ref{eq:umu2}) to simplify this result.   Combining (\ref{eq:temp42}) with  (\ref{eq:gibbsduhem}) and (\ref{eq:dP}) we find \begin{align}
0 &= u^\mu \partial_\mu (P- \mu n - Ts) - (\mu n + Ts) \partial_\mu u^\mu = -\mu u^\mu \partial_\mu n - Tu^\mu \partial_\mu s - (\mu n + Ts)\partial_\mu u^\mu \notag \\
&= -\mu \partial_\mu (nu^\mu)  - T\partial_\mu (su^\mu).   \label{eq:dmusmu0}
\end{align}
From (\ref{eq:curr0}) and (\ref{eq:dJ0}), the zeroth order charge conservation equation is $\partial_\mu (nu^\mu) = 0$.   Hence we obtain (\ref{eq:scurr0}).

\subsubsection{First Order}\label{sec:422}
We now wish to go to first order in derivatives:  this will require adding terms to $J^\mu$ and $T^{\mu\nu}$ that contain a single spatial derivative: for example,  $\partial_\mu T$ or $\partial_\nu u_\mu$.   However, there is an immediate subtlety that arises.  Strictly speaking, we defined $\mu$, $T$ and $u^\mu$ from a thermodynamic perspective.    What does it mean to discuss thermodynamic properties in the presence of spatial gradients which (as we will see) do not generally persist to infinite time?  Because the only meaningful quantities within hydrodynamics are physical objects which are the expectation values of quantum operators, like $J^\mu$ or $T^{\mu\nu}$, it is best to assert that $\partial_\mu T$, $\partial_\mu \mu$ and $\partial_\mu u_\nu$ do not have any `microscopic' definitions,  and can be chosen at will.   Importantly, the freedom to re-define our hydrodynamic degrees of freedom at first order in derivatives is not inconsistent with anything we have done so far.  For example, suppose that we shift $T \rightarrow T + K u^\nu \partial_\nu \mu $.   The expression for the charge density $J^t$ will then be modified:  $J^t  \rightarrow n +  (\partial_T n) K u^\nu \partial_\nu \mu + \cdots$;  similar statements can be made for $T^{\mu\nu}$.   The crucial point is that the re-definition of $T$ has led to first order corrections to $J^\mu$ and $T^{\mu\nu}$ -- precisely what we still have to classify.   So any field re-definitions that we make are compensated by a shift in the expressions for the conserved currents at higher orders in derivatives.    This is called the freedom to choose a fluid frame.   As we will see, it is most useful to pick a fluid frame where the equations \begin{subequations}\label{eq:landauframe}\begin{align}
u_\mu J^\mu &\equiv -n, \\
u_\mu T^{\mu\nu} &\equiv -\epsilon u^\nu,
\end{align}\end{subequations}
are exact to all orders in derivatives.   This is called the Landau frame.    Physically, we simply assert that the charge density $n$ and energy density $\epsilon$ obey their thermodynamic relations to all orders in derivatives.  $\mu(x^\mu)$ and $T(x^\mu)$ are locally defined by thermodynamic relations.   The momentum density (and energy current, by Lorentz covariance) is proportional to the spatial components of the  velocity, $u^i$.


We are now ready to study the hydrodynamic gradient expansion to first order in derivatives.   We write  \begin{subequations}\begin{align}
J^\mu &= nu^\mu + \widehat{J}^\mu, \\
T^{\mu\nu} &=  (\epsilon+P)u^\mu u^\nu + Pg^{\mu\nu} + \widehat{T}^{\mu\nu},
\end{align}\end{subequations}
with $u_\mu \widehat{J}^\mu = u_\mu \widehat{T}^{\mu\nu} = 0$ imposed by (\ref{eq:landauframe}).   We now make one more physical assumption:   the existence of a local entropy current whose divergence is non-negative:  \begin{equation}
\partial_\mu s^\mu \ge 0.
\end{equation}
This is the statement that the second law of thermodynamics holds locally.  At zeroth order in derivatives, $s^\mu = s u^\mu$ has already been defined.   At first order in derivatives, this object does not have a non-negative divergence:  
\begin{equation}
T\partial_\mu \left(su^\mu\right) = \mu \partial_\mu \widehat{J}^\mu - \widehat{T}^{\mu\nu}\partial_\mu u_\nu.  \label{eq:dmusmu1}
\end{equation}
To obtain this result, we repeat the same steps as in (\ref{eq:dmusmu0}), but now include the effects of the first order corrections $\widehat{J}^\mu$ and $\widehat{T}^{\mu\nu}$ to the constitutive relations, and note that \begin{equation}
u_\nu \partial_\mu \widehat{T}^{\mu\nu} = \partial_\mu \left(\widehat{T}^{\mu\nu}u_\nu\right) - \widehat{T}^{\mu\nu}\partial_\mu u_\nu =  - \widehat{T}^{\mu\nu}\partial_\mu u_\nu;
\end{equation}
this follows from the Landau frame choice.   We now re-write (\ref{eq:dmusmu1}) as \begin{equation}
\partial_\mu \left(su^\mu - \frac{\mu}{T}\widehat{J}^\mu \right) = - \frac{1}{T}\widehat{T}^{\mu\nu} \partial_\mu u_\nu - \widehat{J}^\mu \partial_\mu \frac{\mu}{T}.
\end{equation}
If we now define the entropy current at first order to be \begin{equation}
s^\mu \equiv s u^\mu - \frac{\mu}{T}\widehat{J}^\mu = \frac{(\epsilon+P)u^\mu - \mu J^\mu}{T},  \label{eq:entropy4}
\end{equation}
then $\widehat{J}^\mu$ and $\widehat{T}^{\mu\nu}$ can be chosen in such a way as to make $\partial_\mu s^\mu \ge 0$.   Defining the projection tensor \begin{equation}
\mathcal{P}^{\mu\nu} = \eta^{\mu\nu} + u^\mu u^\nu,
\end{equation}
we write \begin{subequations}\begin{align}
\widehat{J}^\mu &=  -\sigma_{\textsc{q}}T \mathcal{P}^{\mu\nu} \partial_\nu \frac{\mu}{T}, \\
\widehat{T}^{\mu\nu} &=  \mathcal{P}^{\mu\rho} \mathcal{P}^{\nu\sigma} \left[\eta \left(\partial_\rho u_\sigma + \partial_\sigma u_\rho - \frac{2}{d} g_{\rho\sigma} \partial_\alpha u^\alpha \right) + \zeta g_{\rho\sigma} \partial_\alpha u^\alpha\right]  \label{eq:widehatTmunu}
\end{align}\end{subequations}
with the constants $\sigma_{\textsc{q}}$,  $\eta$ and $\zeta$ all non-negative,   we see that both (\ref{eq:landauframe}) and $\partial_\mu s^\mu \ge 0$ are satisfied.   The three new coefficients we have introduced are called the intrinsic electrical conductivity\footnote{$\sigma_{\textsc{q}}$ also goes by the name of ``quantum critical" conductivity, or ``incoherent" conductivity, in some of the recent literature.}, a shear viscosity and a bulk viscosity respectively.   

In this review, we will see the practical consequences of these dissipative coefficients within hydrodynamics for experiments.  But let us mention from the outset that one immediate consequence of this formalism is that even in charge neutral plasma ($n=0$),  it is possible to have $J^\mu \ne 0$ in the presence of a chemical potential gradient.    The microscopic intuition for this is that a finite temperature plasma consists of positive and negatively charged ``particles".   Chemical potential gradients will drive oppositely charged particles in opposite directions, leading to a charge current.    We will return to this in Sections \ref{sec:imbalance} and \ref{sec:MFdisorder}.

In many papers on relativistic hydrodynamics, one works in a fluid frame where \begin{equation}
J^\mu \equiv  \left. nu^\mu \right|_{\text{historical frame}}.  \label{eq:histframe}
\end{equation}
is exact to all orders in derivatives.   We have called this a ``historical" frame because conventionally one would define (\emph{i}) the conserved charge density by the number density of a lone species of particles, and (\emph{ii}) the velocity as an ``average" velocity of each individual particle, and so find (\ref{eq:histframe}) by construction: see  the kinetic theory discussion in Section \ref{sec:hydrofromkt}.     It is not a good choice for hydrodynamics in graphene, however.  In graphene, it is quite natural to study the charge neutrality point where $n=0$.   The charge currents do \emph{not} vanish at the charge neutrality point,  and this means that the velocity $u^\mu$ becomes singular at first order in derivatives.    In contrast, within the Landau frame that we have described, the charge current can be finite even at points where $n=\mu=0$, so long as $\nabla \mu \ne 0$.  What this frame choice means in practice is that the dissipative coefficient that we have called $\sigma_{\textsc{q}}$ is often called\footnote{Most sources, except \cite{lucasrmp}, do not include an explicit subscript here.  We think it is important to do so to distinguish between the hydrodynamic coefficient $\kappa_{\textsc{q}}$ and the experimentally measured thermal conductivity $\kappa$:  see Section \ref{sec:DFtrans}.} $\kappa_{\textsc{q}}$ in textbooks:   one finds an energy current $\propto -\kappa_{\textsc{q}} \nabla T$ in the absence of charge flow.

\subsubsection{External Electromagnetic Fields}
\label{sec:extF}
In many situations, we will interested in the hydrodynamic equations in the presence of external electric and magnetic fields -- such external perturbations are natural in the solid-state laboratory.    We can combine these external electromagnetic fields into an antisymmetric tensor $F^{\mu\nu}$.  For example in $2+1$ spacetime dimensions (relevant for graphene): \begin{equation}
F^{\mu\nu} = \left(\begin{array}{ccc}  0 &\  E_x &\  E_y \\ -E_x &\ 0 &\ B \\ -E_y &\ -B &\ 0 \end{array}\right).  \label{eq:Fmunu}
\end{equation}
The hydrodynamic equations are modified in two ways under such perturbations.   Firstly, energy and momentum are  no longer  conserved  (for example,  Joule heating occurs in a background electric field).  This modifies (\ref{eq:dT0}) to \begin{equation}
\partial_\nu T^{\mu\nu} = F^{\mu\nu}J_\nu.   \label{eq:dTFJ}
\end{equation}
This equation can be derived on general grounds as a Ward identity \cite{lucasrmp}, but we will assert it here without proof.    This modifies the derivation of the entropy current in the previous subsection, and one finds that $\widehat{J}^\mu$ must be replaced by \begin{equation}
\widehat{J}^\mu = -\sigma_{\textsc{q}}T \mathcal{P}^{\mu\nu}\left[\partial_\nu \frac{\mu}{T} - \frac{1}{T}F^{\nu\rho}u_\rho\right].
\end{equation}
Some of the dissipative charge current is due to the background fields.   This can be understood as the statement that the fluid only cares about the total value of the electrochemical potential,  which has contributions both from the external $F^{\mu\nu}$ and the internal $\mu$.   In fact, such reasoning can be used to fix how $F^{\mu\nu}$ modifies the hydrodynamic equations.   Expanding around a stationary state with $u^\mu = (1,\mathbf{0})$,  one must replace $\partial_\nu \mu  \rightarrow \partial_\nu \mu - F^{\nu\rho}u_\rho$.

Let us summarize what we have learned.   The hydrodynamic equations of motion are conservation laws for charge, energy and momentum:  $\partial_\mu J^\mu = 0$ and  $\partial_\nu T^{\mu\nu} = F^{\mu\nu}J_\nu$.   The conventional hydrodynamic variables are $\mu$, $T$ and $u^\mu$, and their values point-by-point are tied to the local values of the charge, energy and momentum densities, and we found that \begin{subequations}\label{eq:genconst}\begin{align}
J^\mu &=  n u^\mu - \sigma_{\textsc{q}} \mathcal{P}^{\mu\nu} \left[\partial_\nu \mu - \frac{\mu}{T}\partial_\nu T - F_{\nu\rho}u^\rho\right], \\
T^{\mu\nu} &= (\epsilon+P)u^\mu u^\nu + P\eta^{\mu\nu} - \eta \mathcal{P}^{\mu\rho}\mathcal{P}^{\nu\sigma}\left[\partial_\rho u_\sigma + \partial_\sigma u_\rho - \frac{2}{d}\eta_{\rho\sigma} \partial_\alpha u^\alpha \right] - \zeta \mathcal{P}^{\mu\nu} \partial_\alpha u^\alpha   \label{eq:Tmunuend}
\end{align}\end{subequations}
All coefficients in these equations are understood to be arbitrary up to the non-negativity of $\sigma_{\textsc{q}}$, $\eta$ and $\zeta$, and thermodynamic constraints, given in Section \ref{sec:thermo} for $n$, $\epsilon$ and $P$.
Together with the constraint $u^\mu u_\mu = -1$,   this forms a closed set of classical differential equations.   

\subsection{Hydrodynamic Modes}\label{sec:modes}
To gain some intuition into the hydrodynamic equations, let us solve the hydrodynamic equations within a linear response regime.  Namely, suppose that we are very close to equilibrium,  with $\mu = \mu_0$ and $T=T_0$ constants (with associated pressure $P_0$, density $n_0$, etc.) and velocity $u^\mu_0 = (1,\mathbf{0})$.   We now perturb \begin{subequations}\begin{align}
\mu(\mathbf{x},t) &= \mu_0 + \mdelta \mu(\mathbf{x},t), \\
T(\mathbf{x},t) &= T_0 + \mdelta T(\mathbf{x},t), \\
u^\mu(\mathbf{x},t) &= (1, \mdelta v^i(\mathbf{x},t)),
\end{align}\end{subequations}
and solve the hydrodynamic equations to linearized order in the $\mdelta$ variables. Note that the form of $u^\mu$ is restricted by (\ref{eq:umu2}).      For convenience, we will also write \begin{equation}
\mdelta n = \left(\frac{\partial n}{\partial \mu}\right)_{T,0} \mdelta \mu + \left(\frac{\partial n}{\partial T}\right)_{\mu,0} \mdelta T;  
\end{equation}
similar relations hold for other thermodynamic variables.  

At first order in the gradient expansion, and at first order in the $\mdelta$ variables, the linearized hydrodynamic equations become \begin{subequations}\label{eq:linres43}\begin{align}
0 &= \partial_t \mdelta n + \partial_i \left(n \mdelta v_i - \sigma_{\textsc{q}}\partial_i \left(\mdelta \mu - \frac{\mu_0}{T_0}\mdelta T\right)\right), \\
0 &=  \partial_t \mdelta \epsilon +  \partial_i \left((\epsilon_0+P_0)\mdelta v_i\right), \label{eq:linencons} \\
0 &=  \partial_t((\epsilon_0+P_0) \mdelta v_i)  + \partial_i \mdelta P - \partial_j \left( \eta_0  (\partial_j \mdelta v_i + \partial_i \mdelta v_j)\right) - \partial_i\left( \left(\zeta_0-\frac{2\eta_0}{d}\right) \partial_j \mdelta v_j\right).  \label{eq:linres43mom}
\end{align}\end{subequations}
Note that only two of the thermodynamic variables above are independent.  It is simplest to take $\mdelta n$ and $\mdelta P$ as the independent variables.   From (\ref{eq:epsdP}), we know that $\mdelta \epsilon = d\mdelta P$.

Let us begin by setting all the dissipative coefficients to vanish:  $\sigma_{\textsc{q}0} = \eta_0 = \zeta_0 = 0$.    Combining the last two equations of (\ref{eq:linres43}) leads to \begin{equation}
\partial_t^2 \mdelta P = \frac{v_{\mathrm{F}}^2}{d} \partial_i\partial_i \mdelta P.
\end{equation}
This equation describes sound waves which travel at a universal speed \begin{equation}
v_{\mathrm{s}} = \frac{v_{\mathrm{F}}}{\sqrt{d}}.  \label{eq:soundspeed}
\end{equation}
Keep in mind that $d=2$ for graphene.  Such waves are analogous to the ``cosmic sound" of an ultrarelativistic plasma in outer space  \cite{levitovsound}.   (\ref{eq:soundspeed}) follows generally from any theory with the thermodynamics described in Section \ref{sec:thermo}  \cite{kovtun, lucasplasma}.   The first equation of (\ref{eq:linres43}) describes no interesting dynamics.  In particular, in the absence of sound waves,  $\mdelta v_i = 0$ and hence $\partial_t \mdelta n = 0$.    Density fluctuations (independent from $\mdelta P$) are frozen in place.   Also frozen in place are divergenceless flows with $\partial_i \mdelta v_i = 0$.

Including the dissipative coefficients, one finds the following results after some algebra \cite{kovtun, lucasplasma}.   Looking for solutions to (\ref{eq:linres43}) of the form $\mathrm{e}^{\mathrm{i}kx-\mathrm{i}\omega t}$, we find three types of modes.   Firstly, there are sound waves of the previous paragraph have a dispersion relation \begin{equation}
\omega = v_{\mathrm{s}} k - \mathrm{i} \frac{(2-\frac{2}{d})\eta_0  + \zeta_0}{2(\epsilon_0+P_0)} k^2 + \mathrm{O}\left(k^3\right) \equiv v_{\mathrm{s}} k - \mathrm{i}\Gamma_{\mathrm{s}}k^2 + \mathrm{O}\left(k^3\right)  . \label{eq:gammas} 
\end{equation}
We have not written the $\mathrm{O}(k^3)$ terms because second order corrections to the gradient expansion will also contribute at this order,  just as the viscous effects contributed to the dispersion relation at $k^2$.   Indeed, recall that the hydrodynamics that we have developed should always be understood to be valid in the limit $k\rightarrow 0$.   Modes which were frozen in place -- transverse velocity fields and some charge fluctuations -- now diffuse: \begin{subequations}\begin{align}
\omega  &= -\mathrm{i} \frac{\eta_0}{\epsilon_0+P_0} k^2,  \;\;\;\;\; (\mdelta v_y) \label{eq:diffvy} \\
\omega &=  -\mathrm{i} \frac{(\epsilon_0 + P_0)\sigma_{\textsc{q}0}}{T(s\partial_\mu n - n\partial_\mu s)_0} k^2,  \;\;\;\;\; (\mdelta n) . \label{eq:diffcharge}
\end{align}\end{subequations}
We observe from (\ref{eq:diffvy}) that the viscosity is related to the diffusion constant for (transverse) momentum.  Indeed, the dynamical viscosity \begin{equation}
\nu \equiv \frac{\eta}{\epsilon+P},
\end{equation}
which has dimensions of $[\text{length}]^2 / [\text{time}]$ in any spatial dimension, is often the best way to compare how ``viscous" two fluids are, relative to one another.  We note that the precise definition of $\nu$ should always be taken as the diffusion constant for momentum in (\ref{eq:diffvy}), and depending on whether one is studying relativistic hydrodynamics or not, the relationship between $\nu$ and $\eta$ can change.

At the charge neutrality point $n_0 = \mu_0 = 0$.  The charge current $J_i$ is  entirely carried by the diffusive mode (\ref{eq:diffcharge}).   Because electrical transport is the simplest experiment to perform, this implies that it is rather subtle to detect the hydrodynamics of charge-neutral plasma in graphene, a point which we will return to in Section \ref{sec:DFtrans}.    Away from the charge neutrality point, the charge density can fluctuate in a sound wave.

\subsubsection{Nonlinear Hydrodynamics in Experiment?}
\label{sec:nonlinear}
The relativistic corrections to hydrodynamics can be safely treated in powers of the small parameter $v_{\mathrm{flow}} /v_{\mathrm{F}}$,   where $v_{\mathrm{flow}}$ is the value of the fluid velocity in a given setup.   In graphene,  typical flow velocities in simple experiments are of order $10^2$ m/s \cite{polini} ($v_{\mathrm{flow}} \sim 10^{-4}v_{\mathrm{F}}$),  although it is possible to reach $v_{\mathrm{flow}} > 0.1 v_{\mathrm{F}}$ \cite{kimflow, dorgan}, especially closer to the Dirac point.    Because we do not reach flows with $v\approx v_{\mathrm{F}}$,  it is generally a safe assumption to approximate the flow as non-relativistic, $v\ll v_{\mathrm{F}}$.  The relativistic dispersion relation of the electrons will leave its imprint in the form of the gradient expansion.

However, non-relativistic hydrodynamics is a nonlinear theory as well, due to convective terms in the hydrodynamic equations.     Of  particular interest are the nonlinear corrections to the momentum conservation equation $\partial_\mu T^{\mu i} = 0$, which (at quadratic order in velocity) read \begin{equation}
0 = \partial_t \left((\epsilon+P)v^i\right) + \partial_j \left((\epsilon+P)v^iv^j\right) + \partial^i P  - \partial_j \left(\eta \left(\partial^i v^j + \partial^j v^i - \mdelta_i^j \partial_k v^k\right)\right)  + \mathrm{O}\left(v^3\right).
\end{equation}
  One quantifies how important the quadratic nonlinear term is, relative to the linear viscous term, by computing \begin{equation}
\mathcal{R} = \left| \frac{(\epsilon+P) v_j \partial_j v_i}{\eta \partial_j \partial_j v_i}\right|.
\end{equation}
$\mathcal{R}$ is a generalization of the Reynolds number from conventional hydrodynamics \cite{landau}.   If $\mathcal{R} \ll 1$, then the nonlinear terms can be neglected;  if $\mathcal{R} \gg 1$, then the nonlinear terms are important.   It is instructive to estimate $\mathcal{R}$ as follows.   As we will show explicitly in Section \ref{sec:kinetic}, $\eta \sim v_{\mathrm{F}} \ell_{\mathrm{ee}} (\epsilon+P)$.  Letting $\ell_{\mathrm{flow}}$ denote the length scale of the flow (for example, the size of a sheet of graphene):  \begin{equation}
\mathcal{R} \sim \frac{v_{\mathrm{flow}}}{v_{\mathrm{F}}} \frac{\ell_{\mathrm{flow}}}{\ell_{\mathrm{ee}}}.
\end{equation}
The first fraction above is small, but the second could be relatively large.   Device sizes are typically not larger than 10 $\mmu$m;  using $\ell_{\mathrm{ee}} \gtrsim 100$ nm at realistic temperatures,\footnote{Our estimate  follows from the discussion below (\ref{eq:taueeintro}), using $\ell_{\mathrm{ee}} \sim v_{\mathrm{F}}\tau_{\mathrm{ee}}$; see  also  \cite{song, johannsen}.}    we estimate that $\ell_{\mathrm{flow}} \lesssim 100\ell_{\mathrm{ee}}$.   As noted in \cite{succiturb}, $\mathcal{R} \sim 10$ may be sufficient to observe hints of nonlinear hydrodynamics in graphene,  although we will see in Section \ref{sec:dis4turb} that disorder readily spoils this effect.

\subsection{The Fermi Liquid Limit}\label{sec:FLhydro}
Let us now discuss the limit $\mu \gg k_{\mathrm{B}}T$,  retaining the linearized approximation.    As we will see in Section \ref{sec:kinetic}, in this regime $\sigma_{\textsc{q}} \sim T^2 \eta/\mu^4$ is a small dissipative coefficient, and  \begin{equation}
P(\mu,T) = \frac{a\mu^{d+1}}{(\hbar v_{\mathrm{F}})^d} \left[1+ b\left(\frac{k_{\mathrm{B}}T}{\mu}\right)^2  +\cdots\right].
\end{equation}
Let us suppose that the background $\mu$ does not vary much from a constant value $\mu_0$.   In this limit, $\mdelta \mu$ and $\mdelta v_i$ dominate the hydrodynamic response of the fluid,  and the thermal response $\mdelta T$ is suppressed by a power of $T/\mu_0$.    To confirm this assertion, we must compare the charge and energy conservation equations, which read \begin{subequations}\begin{align}
\partial_t \left((\partial_\mu n)_0 \mdelta \mu \right) + \partial_i \left(n_0 \mdelta v_i\right) + \mathrm{O}(T^2, T^2\ell_{\mathrm{ee}}, T\mdelta T) &= 0, \\
\partial_t \left((\partial_\mu \epsilon)_0 \mdelta \mu \right) + \partial_i \left((\epsilon_0+P_0) \mdelta v_i\right) + \mathrm{O}(T^2, T^2\ell_{\mathrm{ee}}, T\mdelta T) &= 0.
\end{align}\end{subequations} 
In the $T\rightarrow 0$ limit we have $P\sim \mu^{d+1}$.  This means that $n = (d+1)P/\mu$ and $\partial_\mu n = dn/\mu$.   Together with (\ref{eq:epsdP}), we find 
\begin{subequations}\begin{align}
\partial_t \left(\frac{d(d+1)P_0}{\mu_0^2} \mdelta \mu \right) + \partial_i \left(\frac{(d+1)P_0}{\mu_0} \mdelta v_i\right) + \mathrm{O}(T^2, T\mdelta T) &= 0, \\
\partial_t \left(\frac{d(d+1)P_0}{\mu_0} \mdelta \mu \right) + \partial_i \left((d+1)P_0 \mdelta v_i\right) + \mathrm{O}(T^2, T\mdelta T) &= 0.
\end{align}\end{subequations} 
If the background $\mu_0$ is constant (independent of $\mathbf{x}$), then clearly these two equations are identical at leading order.   If the background $\mu$ varies a small amount, these two equations are inequivalent, and thermal effects cannot be neglected (see Section \ref{sec:hydrotrans}).   Nevertheless, away from the charge neutrality point, it may be a reasonable assumption in graphene that the chemical potential is approximately homogeneous.   And so we see that the hydrodynamic equations have reduced to  \begin{subequations}\label{eq:gal1}\begin{align}
d \partial_t \mdelta P + \partial_i ((d+1)P_0 \mdelta v_i) &= 0, \\
\partial_t ((d+1)P_0 \mdelta v_i) + \partial_i \mdelta P - \partial_j \left( \eta_0  (\partial_j \mdelta v_i + \partial_i \mdelta v_j)\right) - \partial_i\left( \left(\zeta_0-\frac{2\eta_0}{d}\right) \partial_j \mdelta v_j\right) &= 0.
\end{align}
\end{subequations}

In much of this paper, we will make the further assumption that fluid flows are static.  In practice, this means that we want the time scale of an experimental measurement to be slow compared to the decay times of hydrodynamic modes.   In  practice, this decay time is often set by momentum-relaxing scattering (see Section \ref{sec:dis4}) and can be of order 1 ps \cite{wang13}, which is quite fast.     If the background pressure and chemical potential are uniform, (\ref{eq:gal1}) further reduces to \begin{subequations}\begin{align}
\partial_i  \mdelta v_i &= 0, \\
\partial_i \mdelta P - \eta_0 \partial_j \partial_j \mdelta v_i &= 0.  \label{eq:simpleNS}
\end{align}\end{subequations} 
These equations are identical to the time-independent hydrodynamic equations that one finds for a Galilean invariant fluid \cite{landau}, upon neglecting thermal effects.   We will discuss particular solutions of these equations in Sections \ref{sec:FLhydroMR} and  \ref{sec:FLtrans}.

We emphasize that the inclusion of thermal effects differs between the Galilean and Lorentz invariant fluids.  In a Galilean invariant fluid, the charge current is proportional to the momentum density, while in a relativistic fluid the energy current is proportional to the momentum density.  Thus, in a Galilean invariant fluid, there is a diffusive mode associated to energy fluctuations, while in a Lorentz invariant fluid, the diffusive mode is associated with charge fluctuations, as we have seen in Section \ref{sec:modes}.

\subsection{Long-Range Coulomb Interactions}
One major oversight in our development thus far has been that we have neglected the long-range nature of the Coulomb interactions in graphene (and in many metals, more generally).   The standard way \cite{muller2, muller4} to account for such long range interactions is to simply couple the fluid dynamical equations to Maxwell's equations.\footnote{There has been some recent debate in the literature \cite{grozdanov1610, kovtun1703} over the validity of this procedure.   From the point of view of effective field theory, coupling Maxwell's equations to matter breaks the derivative expansion of hydrodynamics, and is therefore rather concerning.  In this review, we assume that the conventional approach is correct, which has been argued for from an effective field theory perspective in \cite{kovtun1703}.}   This implies that the electric field $E_i$ in (\ref{eq:Fmunu}) is given by  \begin{equation}
E_i = -\partial_i \mdelta \varphi,   \label{eq:Edphi}
\end{equation}
with $\varphi$ obeying Gauss' law in three spatial dimensions:  \begin{equation}
\partial_j \partial_j \mdelta \varphi(x,y,z) = -4\mpi \alpha \mdelta n(x,y) \mdelta(z).
\end{equation}
Neglecting charge puddles,  the immobile background ions imply that only fluctuations of the density contribute to the long-range Coulomb potential.   Note that the third dimension is important --  the physical space has three dimensions, even though the electrons in graphene are only mobile in two of them.  The coupling constant $\alpha$ is analogous to a dielectric constant.  We find 
\begin{equation}
\mdelta \varphi(x,y,0) = \int \mathrm{d}^2\mathbf{x}^\prime  \frac{\alpha}{\sqrt{(x-x^\prime)^2 + (y-y^\prime)^2}} \mdelta n(x^\prime, y^\prime).  \label{eq:1rpot}
\end{equation}
We now combine this equation with the hydrodynamic equations (\ref{eq:dJ0}) and (\ref{eq:dTFJ}) and constitutive relations (\ref{eq:genconst}) to obtain our linearized theory for the Coulomb-interacting Dirac fluid \cite{muller1, lucas3}:  \begin{subequations}\label{eq:EOMCou}\begin{align}
0 &= \partial_t \mdelta n + \partial_i \left(n_0 \mdelta v_i - \sigma_{\textsc{q}0}\partial_i \left(\mdelta \mu + \mdelta \varphi - \frac{\mu_0}{T_0}\mdelta T\right)\right), \\
0 &=  \partial_t \mdelta \epsilon +  \partial_i \left((\epsilon_0+P_0)\mdelta v_i\right), \label{eq:linencons} \\
0 &=  \partial_t((\epsilon_0+P_0) \mdelta v_i)  + n_0 \partial_i (\mdelta \mu + \mdelta \varphi) + s_0\partial_i \mdelta T - \partial_j \left( \eta_0  (\partial_j \mdelta v_i + \partial_i \mdelta v_j)\right) - \partial_i\left( \left(\zeta_0-\frac{2\eta_0}{d}\right) \partial_j \mdelta v_j\right).
\end{align}\end{subequations}
Due to (\ref{eq:1rpot}), these equations are nonlocal.  

We emphasize that $\mdelta \varphi$ enters (\ref{eq:EOMCou}) in a very special way. The spatial components of the charge and energy current are sensitive only to the gradient of the total electrochemical potential $\mdelta \mu + \mdelta \varphi$.   A static fluid is only sensitive to the `net' electric field.    Hence, we conclude that for time-independent flows,  long-range Coulomb interactions have \emph{no physical effect}, because the experimentalist can only measure the total electrochemical potential.   In contrast, the time-dependent thermodynamic response is only dependent on $\mdelta \mu$, not $\delta \varphi$.  As we will soon see,  this does lead to measurable consequences for dynamics.

A final subtlety is that the same parameter $\alpha$ governing the strength of Coulomb interactions also governs the thermodynamics of the electron fluid (see Section \ref{sec:kineticcoeff}).   Implicit in (\ref{eq:EOMCou}) is that the Coulomb interaction splits into a collective long range component, and a short-range component responsible for hydrodynamic and thermodynamic phenomena \cite{muller2, muller4}, and a careful check of this assumption is called for.   This ``splitting" does occur in a conventional Fermi liquid \cite{marston}.

\subsubsection{Plasmon-Like Corrections to Sound}
The simplest way to observe the consequences of Coulomb interactions is to study the dispersion relation of sound waves.   Neglecting dissipation, this can be done analytically \cite{levitovsound, lucasplasma}.  Using that $\mdelta \varphi(k) = 2\mpi \alpha |k|^{-1} \mdelta n(k)$ in Fourier space, we find that the dispersion relation $\omega^2 = v_{\mathrm{s}}k^2$ becomes modified to \begin{equation}
\omega^2 = v_{\mathrm{s}}^2 k^2 + \frac{2\mpi \alpha n^2}{\epsilon+P} |k|.  \label{eq:plasmon}
\end{equation}
As $k\rightarrow 0$, we therefore see that the dispersion relation of sound waves is severely altered.  In fact, the dispersion relation we have found is analogous to plasmons' dispersion relation in graphene.   The fact that plasmons disperse with the relation $\omega \sim \sqrt{k}$ is a consequence of the fact that the electrons are mobile in two dimensions \cite{hwangplasmon, sarmaplasmon}, while the Coulomb potential exists in three dimensions.\footnote{In a more conventional metal where the electrons are also mobile in three dimensions, one instead finds $\mdelta \varphi \sim \alpha k^{-2} \mdelta n $, and so $\omega^2 = \omega_{\mathrm{p}}^2 + v_{\mathrm{s}}^2k^2$, with $\omega_{\mathrm{p}}^2 \sim \alpha n^2/(\epsilon+P)$ the plasma frequency.}    \cite{soljacic2} is a recent review on plasmons in graphene;  they were observed experimentally in \cite{chen12, koppens12}.

We caution the reader that in the limit where $\omega \sim \sqrt{k}$,  the dispersion relation has an analogous form to the conventional plasmon,  but this mode is not the conventional plasmon of a two-component (electron-hole) plasma \cite{landaukinetic}.    At higher frequencies, we obtain ordinary sound from (\ref{eq:plasmon}).  See \cite{fogler3, fogler1} for more discussion on this point.

Dissipative corrections to the dispersion relation $\omega \sim \sqrt{k}$ are given by $\mdelta \omega \sim -\mathrm{i}\sigma_{\textsc{q}}|k|$ within hydrodynamics, instead of $\mdelta \omega \sim -\mathrm{i}\eta k^2$ as for the sound wave \cite{lucasplasma}. 

\subsubsection{Breaking Relativistic Invariance?}
\label{sec:userel}
Another possible issue is that because long-range Coulomb interactions break Lorentz invariance, the hydrodynamics of electrons in graphene will not be described by a Lorentz invariant hydrodynamics.   To check whether this is a problem, we may directly derive the energy current and the momentum density from the action (\ref{eq:SDF}) using Noether's Theorem.   Because only the kinetic terms contain derivatives, we see that the energy current and momentum density are identical to that of a free Dirac fermion.   This implies that $T^{ti}=T^{it}$ should be true within hydrodynamics, as it is an operator identity.   This equality alone is sufficient to recover the linearized hydrodynamic formalism of this section.

It may be the case that because the presence of nonlocal Coulomb interactions break Lorentz invariance, the thermodynamic and/or hydrodynamic properties of the Dirac fluid become more subtle.  Evidence for such an assertion can be found in  \cite{sodemann}.   In particular, this may correspond to interesting $\mathrm{O}(v^2)$ corrections to the hydrodynamic equations we have described so far.  Because the interactions do not break spatial isotropy, such effects cannot arise at linear order in velocities, which is the order to which we focus in this review.   Recent work \cite{sybesma} is beginning to develop hydrodynamics without boost invariance.

\subsection{Momentum Relaxation}\label{sec:dis4}
The second issue that we need to address is that the hydrodynamics we derived above assumed that momentum was an exactly conserved quantity.   Unfortunately, this is not true for the electrons in metals.  As we have seen in Section \ref{sec:graphene},  the scattering of electrons off of impurities and/or phonons cannot be neglected.    Continuing to work in the linearized approximation, the simplest thing to do is to modify the momentum conservation equation (\ref{eq:linres43mom}) to \begin{equation}
\partial_t((\epsilon+P) \mdelta v_i)  + \partial_i \mdelta P - \partial_j \left( \eta  (\partial_j \mdelta v_i + \partial_i \mdelta v_j)\right) - \partial_i\left( \left(\zeta-\frac{2\eta}{d}\right) \partial_j \mdelta v_j\right)  = -\frac{\epsilon+P}{\tau_{\mathrm{imp}}} \mdelta v_i.  \label{eq:46first}
\end{equation}
In this equation, we have dropped the 0 subscript on the background quantities, for simplicity, and will continue to do so for the rest of the paper to avoid clutter.
The parameter $\tau_{\mathrm{imp}}$ is a  relaxation time for the total momentum, as can be readily seen by integrating this equation over space.   It is often estimated to be the scattering rate between an electron/hole and an impurity or phonon, although we will see examples in Section \ref{sec:hydrotrans} where the impurity momentum relaxation time must be evaluated more carefully.

\subsubsection{Destruction of Sound Modes} 
Let us describe the consequences of momentum relaxation on the hydrodynamic modes described in Section \ref{sec:modes}.   Clearly, only the modes with $\mdelta v_i \ne 0$ will be affected.   These are the sound modes and diffusive shear modes.   The shear modes obtain a dispersion relation \begin{equation}
\omega = -\frac{\mathrm{i}}{\tau_{\mathrm{imp}}} - \mathrm{i} \frac{\eta}{\epsilon+P}k^2.
\end{equation}
On time scales long compared to $\tau_{\mathrm{imp}}$, this mode is `gapped' -- momentum is not long lived and will not play a role in the dynamics.     The sound modes become \cite{lucasplasma}  
\begin{equation}
\omega \approx \pm \sqrt{\frac{k^2}{d} -  \left(\frac{1}{2\tau_{\mathrm{imp}}} + \frac{\Gamma_{\mathrm{s}}k^2}{2}\right)^2} - \frac{\mathrm{i}}{2}\left(\frac{1}{\tau_{\mathrm{imp}}}+\Gamma_{\mathrm{s}}k^2\right).
\end{equation}
As $k\rightarrow 0$, these modes split into 
\begin{subequations}\begin{align}
\omega &\approx -\mathrm{i} v_{\mathrm{s}}^2\tau_{\mathrm{imp}} k^2, \\
\omega &\approx -\frac{\mathrm{i}}{\tau_{\mathrm{imp}}}.
\end{align}\end{subequations}
The interpretation of this effect is as follows.  The latter mode is gapped, and associated with the finite lifetime of momentum.   The former mode is diffusive, and describes the diffusion of energy.  At long wavelengths, energy is no longer transported by sound waves, but by diffusion.

The physical importance of this is as follows.   In conventional hydrodynamics, momentum is a long lived quantity.   On time scales large compared to $\tau_{\mathrm{imp}}$, and on distances long compared to $v_{\mathrm{F}}\tau_{\mathrm{imp}}$, we see that the dynamics of charge and energy will reduce to a set of diffusion equations.   This is what occurs in a conventional dirty metal.   Hence, in order to see hydrodynamics of electrons, we must find samples where we can observe flows on time and length scales short compared to these scales.

\subsubsection{Reduction to Ohmic Flow in the Fermi Liquid} \label{sec:FLhydroMR}
It is also instructive to study the Fermi liquid limit $T/\mu \rightarrow 0$, as described in Section \ref{sec:FLhydro}.    (\ref{eq:simpleNS}) generalizes to \begin{equation}
\partial_i \mdelta P - \eta \partial_j \partial_j \mdelta v_i  = -\frac{\epsilon+P}{\tau_{\mathrm{imp}}} \mdelta v_i,   \label{eq:gal460}
\end{equation}
while incompressibility continues to imply $\partial_i \mdelta v_i = 0$.    Relating $\mdelta P$ to $\mdelta \mu$ as in (\ref{eq:gal1}), we obtain \begin{equation}
\frac{\partial_i \mdelta \mu}{\mu} - \nu \partial_j \partial_j \mdelta v_i = -\frac{\mdelta v_i}{\tau_{\mathrm{imp}}}.  \label{eq:gal46}
\end{equation}
where we have defined the kinematic viscosity \begin{equation}
\nu \equiv \frac{\eta}{\epsilon+P}.
\end{equation}

A common trick to solve these equations is as follows \cite{lucas1612}.  Define the stream function $\psi$, so that 
\begin{equation}
\mdelta v_x = \partial_y \psi, \;\;\; \mdelta v_y = -\partial_x\psi.
\end{equation}
For a two dimensional incompressible flow,  we see that this ansatz automatically satisfies $\partial_i \mdelta v_i = 0$.   We can further take the curl of (\ref{eq:gal46}) to find \begin{equation}
\partial_i \partial_i \left(\partial_j \partial_j - \frac{1}{\lambda^2}\right)\psi = 0,  \label{eq:stream4}
\end{equation} 
where we have defined the ``Gurzhi length" or ``momentum relaxation length" 
\begin{equation}
\lambda \equiv \sqrt{\nu\tau_{\mathrm{imp}}}.
\end{equation}
When $\lambda$ is finite,  we can express solutions to this differential equation in the form \begin{equation}
\psi = \psi_0 + \psi_\lambda, \;\;\;\;  \partial_i \partial_i \psi_0 = 0, \;\;\;\; \partial_i \partial_i \psi_\lambda = \frac{\psi_\lambda}{\lambda^2}.
\end{equation}
Now suppose that $\lambda$ is small compared to the geometric scales in our problem.   The typical $\psi_\lambda$ is exponentially decaying away from the boundaries on the length scale $\lambda$:  $\psi(x) \sim \mathrm{e}^{-x/\lambda}$.   Hence, in the interior of the sample,  all velocity comes from $\psi_0$.   Furthermore, from (\ref{eq:gal46}), we see that on long distances compared to $\lambda$, the electric current $J_i \approx n\mdelta v_i$ is given by \begin{equation}
J_i \approx -\frac{ n \tau_{\mathrm{imp}}}{\mu} \partial_i \mdelta \mu  = - \sigma_{\mathrm{dc}} \partial_i \mdelta \mu
\end{equation}
where \begin{equation}
\sigma_{\mathrm{dc}} \equiv \frac{n^2 \tau_{\mathrm{imp}}}{\epsilon+P}  \label{eq:sigmadc46}
\end{equation}
is a constant which is, in fact, the (conventional, Ohmic) conductivity of an infinitely large sample, up to a factor of $e^2$ which we will mostly neglect.   We will return to this quantity, in some detail, in Sections \ref{sec:FLtrans} and \ref{sec:DFtrans}.   For now, we simply emphasize that on long distances, there is \emph{no physical distinction} between the equations of motion governing the flow of a hydrodynamic momentum-relaxing electron fluid,  and the motion of electrons in a ``conventional" Ohmic metal with an isotropic conductivity tensor.   Once again, we see how momentum relaxation ruins the interesting physics associated with hydrodynamic electron flow.

\subsubsection{Destruction of Turbulence}\label{sec:dis4turb}
As a final consequence of momentum relaxation, let us discuss the consequences of a finite $\tau_{\mathrm{imp}}$ on the development of turbulent flows.   Turbulence is one of the most dramatic phenomena in classical fluid dynamics:  the chaotic and `self-organizing' nonlinear dynamics of fluid vortices.  Turbulence has a rather peculiar character in two spatial dimensions \cite{boffetta}.  Let us give a qualitative description of the phenomenon.\footnote{Although this discussion focuses on the turbulence of Galilean invariant fluids, it appears as though relativistic (uncharged) fluids have qualitatively similar phenomena \cite{lehner}.}   Regions of positive vorticity $\Omega = \partial_x v_y - \partial_y v_x$ will merge together, as will regions of negative vorticity.  More quantitatively, \begin{equation}
\left\langle (\mathbf{v}(\mathbf{x}) - \mathbf{v}(\mathbf{0}))^2\right\rangle \sim (\epsilon x)^{2/3} \label{eq:2dturb}
\end{equation}
where we may think of \begin{equation}
\epsilon \sim \frac{v_{\mathrm{stir}}^3}{\ell_{\mathrm{stir}}} \label{eq:vstir}
\end{equation}
as a number characterizing the properties of a small-scale stirring of the fluid.  The average in (\ref{eq:2dturb}) is over statistical realizations of turbulence; we leave a precise defiinition to \cite{boffetta}.  

What happens if we now include a momentum relaxation time $\tau_{\mathrm{imp}}$?\footnote{In the fluid dynamics literature, the momentum relaxation time $\tau_{\mathrm{imp}}$ is commonly referred to as a `friction' term.  It is sometimes used to regulate simulations of two dimensional turbulence, and could mimic, for example, the drag on atmospheric flows due to the Earth's surface.}   We can form a second dimensionless number \begin{equation}
\mathcal{R}_\tau \equiv \frac{v_{\mathrm{flow}}\tau_{\mathrm{imp}}}{\ell_{\mathrm{flow}}} \label{eq:Rtau}
\end{equation}
which is the ratio of the nonlinear convective term relative to the momentum relaxing term in the nonlinear generalization of (\ref{eq:46first}).    From (\ref{eq:2dturb}) we estimate that $v_{\mathrm{flow}} \sim (\epsilon \ell_{\mathrm{flow}})^{1/3}$.   As $\ell_{\mathrm{flow}} \rightarrow \infty$,   $\mathcal{R}_\tau \rightarrow 0$, while $\mathcal{R} \rightarrow \infty$.  As we have seen throughout this subsection, the effects of momentum relaxation become most important at long distances.   Furthermore, by comparing $\mathcal{R}$ to $\mathcal{R}_\tau$, we conclude that the effects of momentum relaxation become more important than the effect of viscosity whenever $\ell_{\mathrm{flow}}^2 \gtrsim v_{\mathrm{F}} \ell_{\mathrm{ee}}\tau_{\mathrm{imp}} = \lambda^2$.   And when $\mathcal{R}_\tau \sim 1$, the effects of momentum relaxation become more important than the effects of momentum convection via the nonlinear terms in the Navier-Stokes equations.  Combining (\ref{eq:2dturb}), (\ref{eq:vstir}) and (\ref{eq:Rtau}), we observe that this occurs when \begin{equation}
\ell_{\mathrm{flow}}  \sim \sqrt{\frac{v_{\mathrm{stir}}^3\tau_{\mathrm{imp}}^3}{\ell_{\mathrm{stir}}}}.
\end{equation}
Obviously, to be in the hydrodynamic regime, we require $\ell_{\mathrm{stir}} \gtrsim \ell_{\mathrm{ee}}$.   We can now understand the difficulty for observing turbulent flows of electrons in metals.   We must find a metal where $\tau_{\mathrm{imp}}$ is large enough:  \begin{equation}
\frac{\tau_{\mathrm{imp}}}{\tau_{\mathrm{ee}}} \gtrsim \frac{v_{\mathrm{F}}}{v_{\mathrm{stir}}} \times \left(\frac{\ell_{\mathrm{flow}}}{\ell_{\mathrm{ee}}}\right)^{2/3}.
\end{equation}
Even if $\ell_{\mathrm{flow}}$ is not too much larger than $\ell_{\mathrm{ee}}$, given the discussion in Section \ref{sec:nonlinear},  we would conservatively require $\tau_{\mathrm{imp}} \gtrsim 10 \tau_{\mathrm{ee}}$ to see even a hint of turbulence.   This is very difficult to achieve experimentally.   We do not expect that electronic hydrodynamics will be in the nonlinear regime in the near term, if ever.

\section{Kinetic Theory}
\label{sec:kinetic}

In this section, we will present a microscopic `derivation' of the hydrodynamics of electrons in graphene, based upon kinetic theory.   The kinetic theory of electrons in graphene was recently reviewed in \cite{schmalian2}.   Kinetic theory is a framework for understanding the dynamics of weakly interacting quantum systems, as we will shortly review, and so while it can be useful for understanding Fermi liquid physics, one might question the legitimacy of such an approach for a strongly interacting quantum system such as the Dirac fluid.     Our view is that it is worth knowing the main results obtained using kinetic theory  -- even if some assumptions may break down at charge neutrality.  Kinetic theory gives us a controlled treatment of the ballistic-to-hydrodynamic crossover and will allow us to address questions such as the validity of  relativistic hydrodynamics (Section \ref{sec:userel}).
\subsection{The Boltzmann Equation} 
\label{sec:boltzeq}
Let us begin with a physically intuitive picture of the kinetic equations.   What follows can be derived more rigorously from quantum many-body theory \cite{kamenev}, and we will pause where appropriate and comment on the effects of quantum mechanics.  Indeed, what follows is often called ``quantum kinetic theory",  although we feel this is a misnomer.   The equations below are classical, while the coefficients of the classical equations can be microscopically computed in the quantum theory. 

The basic idea of the kinetic equations is that if there were no interactions -- namely, the many-body Hamiltonian in graphene was simply given by (\ref{eq:hopping}) -- then the number of fermions in every single-particle state would be conserved (no fermion can scatter into any other state).    One would like to construct a ``hydrodynamics" for  these conserved quantities.  However, these conserved quantities are spatially extended, and so we must be slightly careful.  The key observation is that if we are only interested in long wavelength physics on scales $\gg \mathrm{\Delta} x$,  and willing to only discriminate between fermions whose momenta are at least  $\mathrm{\Delta} p$ different, then whenever \begin{equation}
\mathrm{\Delta}x \mathrm{\Delta }p \gg \hbar,
\end{equation}
then we can assert that the number of fermions at every momentum is individually locally conserved.   This is simply the fact that quantum mechanics, and the wave like nature of the quasiparticles, only becomes important on length scales where Heisenberg's uncertainty principle cannot be ignored.   Writing $f_A(\mathbf{x},\mathbf{p})$ as the number density of fermions of flavor $A$ (spin/valley in graphene) and momentum $\mathbf{p}$, we can then write down conservation laws for $f_A$.   We further assert that $0\le f_A(\mathbf{x},\mathbf{p}) \le 1$ because the particles are fermions, and the Pauli exclusion principle forbids two of them from being in the same state.  The key observation is that we can write such conservation laws down explicitly, and not phenomenologically.   For simplicity, suppose that the single-particle Hamiltonian takes the form of \begin{equation}
H_{1A}(\mathbf{x},\mathbf{p}) = \epsilon(\mathbf{p}) + V_{\mathrm{imp}A}(\mathbf{x}).
\end{equation}
The time evolution of $f_A$ is given by the Liouville equation of classical mechanics:
\begin{equation}
\partial_t f_A  +  \mathbf{v}_A\cdot \frac{\partial f_A}{\partial \mathbf{x}} +  \mathbf{F}_A\cdot \frac{\partial f_A}{\partial \mathbf{p}}  = 0 \label{eq:liouville}
\end{equation}
where $\mathbf{v}$ is the group velocity of quasiparticles with momentum $\mathbf{p}$:
 \begin{equation}
\mathbf{v}_A = \frac{\partial \epsilon_{A}}{\partial \mathbf{p}}   \label{eq:H11}
\end{equation}
with $\epsilon_{A}$ the single-particle Hamiltonian for particles of flavor $A$.  The external force is given by \begin{equation}
\mathbf{F}_A = -\frac{\partial V_{\mathrm{imp}A}}{\partial \mathbf{x}}.  \label{eq:H12}
\end{equation}

With two important subtleties, this equation can also be derived more carefully from the quantum theory \cite{kamenev}.   First, the above derivation is only valid when the lifetime of quasiparticles $\tau_{\mathrm{ee}}$, which we will define below, obeys \begin{equation}
\tau_{\mathrm{ee}} \gg \frac{\hbar}{k_{\mathrm{B}}T}.  \label{eq:heisenbergtime}
\end{equation}
Such an assumption is sensible in a Fermi liquid, but less so in the Dirac fluid of graphene.   Second, we have overlooked the possibility that particles of different flavors $A$ may convert back and forth.  When such processes cannot be neglected (this is most commonly the case for spin degrees of freedom) one must generalize the distribution function to a matrix  in flavor indices $f_{AB}$. Although it is rarely done, in principle one can also compute the subleading contributions in $\hbar$ to (\ref{eq:liouville}).  Interestingly, they  will take the form of ``gradient" corrections to (\ref{eq:liouville}), involving higher derivatives of $\mathbf{x}$ and $\mathbf{p}$.

We now introduce the effects of interactions, assuming that the distribution function may be written as $f_A$.   The essential idea of kinetic theory is that if interactions occur very rarely (in a sense we will shortly make explicit), then interactions can be introduced into (\ref{eq:liouville}) perturbatively.   This is analogous to our treatment of hydrodynamics with weak momentum relaxation in Section \ref{sec:dis4}.   Even though the $f_A(\mathbf{p})$ will not all remain conserved, we can perturbatively correct the right hand side of (\ref{eq:liouville}):   \begin{equation}
\partial_t f_A  +  \mathbf{v}_A\cdot \frac{\partial f_A}{\partial \mathbf{x}} +  \mathbf{F}_A\cdot \frac{\partial f_A}{\partial \mathbf{p}}  = \mathcal{C}[f_A].  \label{eq:boltzmann5}
\end{equation}
This is called the Boltzmann equation, and $\mathcal{C}[f]$ is called the collision integral.  If all collisions between fermions are spatially local 2-body scattering events, such as Coulomb interactions, then 
\begin{align}
\mathcal{C}[f_A] &\equiv \int \frac{\mathrm{d}^2\mathbf{p}^\prime}{(2\mpi \hbar)^2}  \frac{\mathrm{d}^2\mathbf{q}}{(2\mpi \hbar)^2}  \frac{\mathrm{d}^2\mathbf{q}^\prime}{(2\mpi \hbar)^2}   \mdelta\left(\mathbf{p} + \mathbf{p}^\prime - \mathbf{q}-\mathbf{q}^\prime\right) \mdelta \left(\epsilon_{A}(\mathbf{p}) + \epsilon_{B}(\mathbf{p}^\prime) - \epsilon_{C}(\mathbf{q}) - \epsilon_{D}(\mathbf{q}^\prime)\right)   \times \notag \\
& |\mathcal{M}_{ABCD}(\mathbf{p},\mathbf{p}^\prime, \mathbf{q}, \mathbf{q}^\prime)|^2  \left[(1-f_A(\mathbf{p}))(1-f_B(\mathbf{p}^\prime))f_C(,\mathbf{q})f_D(\mathbf{q}^\prime) -f_A(\mathbf{p})f_B(\mathbf{p}^\prime) (1-f_C(\mathbf{q}))(1-f_D(\mathbf{q}^\prime))  \right]  \label{eq:collisionintegral}
\end{align}
To save space, we have suppressed the explicit $\mathbf{x}$-dependence of all factors of $f$  above.  Note that $\mathcal{C}$ carries arguments $A$, $\mathbf{x}$ and $\mathbf{p}$ which have been suppressed.   This equation looks more intimidating than it actually is.   What we are doing is counting how frequently 2 particles in flavor/momentum $A\mathbf{p}$ and $B\mathbf{p}^\prime$ scatter into $C\mathbf{q}$ and $D\mathbf{q}^\prime$.     The rate with which this occurs is given by $|\mathcal{M}_{ABCD}|^2 f_A f_B(1-f_C)(1-f_D)$  (we have suppressed momenta for simplicity):   $|\mathcal{M}_{ABCD}|^2$ is roughly proportional to the probability that such a scattering event would occur in the absence of all other particles, and it can be computed using Feynman diagrams in the microscopic quantum theory \cite{kamenev}.  We further assume  that  $f_A f_B$ is the probability that two fermions are in states $A$ and $B$,  and $(1-f_C)(1-f_D)$ is the probability that $C$ and $D$ are empty (the Pauli exclusion principle forbids two fermions from being in the same state).    We multiply all of the  resulting probabilities together to obtain the number of scattering events that occur.   The assumption that we can multiply such probabilities together, because the states of the incoming/outgoing particles are uncorrelated, is called molecular chaos.   Finally, noting that a scattering event of this type would destroy an $A$ and $B$, while creating a $C$ and $D$,  leads us to (\ref{eq:collisionintegral}):  the first term  arises from collisions where an $A$ is created, and the second term from collisions where $A$ is destroyed.    $\mathcal{C}$ straightforwardly generalizes to other kinds of collisions as well \cite{kamenev}.

We can self-consistently argue that this kinetic formalism is consistent with (\ref{eq:heisenbergtime}) by estimating the lifetime of a quasiparticle as \begin{align}
\frac{1}{\tau_{\mathrm{ee}A}(\mathbf{x},\mathbf{p})} &\equiv \int \frac{\mathrm{d}^2\mathbf{p}^\prime}{(2\mpi \hbar)^2}  \frac{\mathrm{d}^2\mathbf{q}}{(2\mpi \hbar)^2}  \frac{\mathrm{d}^2\mathbf{q}^\prime}{(2\mpi \hbar)^2}  \mdelta \left(\epsilon_{A}(\mathbf{p}) + \epsilon_{B}(\mathbf{p}^\prime) - \epsilon_{C}(\mathbf{q}) - \epsilon_{D}(\mathbf{q}^\prime)\right)   \times \notag \\
&\; \mdelta\left(\mathbf{p} + \mathbf{p}^\prime - \mathbf{q}-\mathbf{q}^\prime\right) |\mathcal{M}_{ABCD}(\mathbf{p},\mathbf{p}^\prime, \mathbf{q}, \mathbf{q}^\prime)|^2 f_A(\mathbf{p})  f_B(\mathbf{p}^\prime) (1-f_C(\mathbf{q}))(1-f_D(\mathbf{q}^\prime))  \label{eq:collisiontime}
\end{align}
(\ref{eq:heisenbergtime}) must then hold for any choice of $A$, $\mathbf{x}$ or $\mathbf{p}$ for kinetic theory to be valid.

The difficulty of the kinetic approach is that firstly, the factor of $|\mathcal{M}|^2$ in (\ref{eq:collisionintegral}) is generally very complicated.   As is the case in graphene, one may also need to make certain self-consistent approximations to further avoid spurious divergences in $\mathcal{M}$ and $\mathcal{C}$.   We will mostly not worry about such effects in this review.   Instead, what we observe is that the Boltzmann equation is a huge, highly nonlinear integro-differential equation.  It cannot possibly be solved in much generality:  even numerics pose a real challenge.    Still, there are two useful features one can prove in great generality about the kinetic approach.   Firstly, so long as one considers a stable phase of matter, one can prove an ``H-theorem", analogous to the second law of thermodynamics, that (\ref{eq:boltzmann5}) is a dissipative equation that tends towards thermal equilibrium.   Secondly, one can often find nonlinear solutions to (\ref{eq:boltzmann5})  of the form \begin{equation}
f_A(\mathbf{x},\mathbf{p}) = n_{\mathrm{F}}\left(\frac{\epsilon_{A}(\mathbf{p}) + V_{\mathrm{imp}A}(\mathbf{x}) - q_A\mu}{k_{\mathrm{B}}T}\right)  \label{eq:fermi1}
\end{equation}
where $q_A$ is the electric charge of a particle of type $A$ and $n_{\mathrm{F}}$ is the Fermi function \begin{equation}
n_{\mathrm{F}}(x) = \frac{1}{1+\mathrm{e}^{x}}.  \label{eq:fermi2}
\end{equation}
Indeed, using (\ref{eq:fermi1}) and (\ref{eq:fermi2}), one finds that the object  in square brackets in (\ref{eq:collisionintegral}) is proportional to \begin{equation*}
\exp\left[-\frac{\epsilon_{C}+\epsilon_{D} - q_C\mu - q_D\mu}{k_{\mathrm{B}}T}\right] - \exp\left[-\frac{\epsilon_{A}+\epsilon_{B} - q_A\mu-q_B\mu}{k_{\mathrm{B}}T}\right].
\end{equation*}
This vanishes due to the conservation of energy in two-body collisions found in (\ref{eq:collisionintegral}).  The left hand side of (\ref{eq:boltzmann5}) can also be shown to vanish on this solution using (\ref{eq:H11}) and (\ref{eq:H12}).  In fact, whenever $H_1$ does not depend on $\mathbf{x}$ at all,  momentum  is a good conserved quantity, and one can find (at least) a three-parameter family of equilibria, parameterized by $T$, $\mu$ and $\mathbf{u}$:  \begin{equation}
f_A(\mathbf{p}) = n_{\mathrm{F}}\left(\frac{\epsilon_{A}(\mathbf{p}) - \mathbf{u}\cdot \mathbf{p} - q_A\mu}{k_{\mathrm{B}}T}\right).
\end{equation}
The free parameters in this equation are exactly the temperature, chemical potential, and velocity that we introduced in hydrodynamics in Section \ref{sec:hydro}.
 In many metals, including graphene, only accounting for electron-electron collisions can lead to even more (approximate) conservation laws, as we will see.   Unlike the hydrodynamic approach, which required us to know a priori the conservation laws, the kinetic approach allows us to compute them.

\subsubsection{Hydrodynamic Limit of a Kinetic Theory}
\label{sec:hydrofromkt}
If quasiparticles are long-lived and the kinetic expansion is valid, one can understand hydrodynamics directly from kinetic theory.\footnote{Unfortunately, there are many incorrect statements in the literature.  We cannot stress strongly enough that hydrodynamic transport phenomena are computable in a correct and complete solution of the Boltzmann equation \cite{hartnoll1705}.}   Let us now quickly sketch how this is done, in general circumstances.   The approach follows textbook treatments of the derivation of hydrodynamics for a weakly interacting classical gas  \cite{kardar}.    In what follows we assume that $V_{\mathrm{imp}A}=0$ -- namely, there is no breaking of translational symmetry.

Suppose that we have identified the full nonlinear family of time-independent solutions to (\ref{eq:boltzmann5}), and that they take the form \begin{equation}
f_A(\mathbf{p}) = n_{\mathrm{F}}\left(-\lambda^IX^I_A(\mathbf{p})\right),
\end{equation}
where $X^I_A(\mathbf{p})$ label amount of conserved quantity $I$ carried by a particle of flavor $A$ and momentum $\mathbf{p}$, and $\lambda^I$ are corresponding free parameters; we have employed an Einstein summation convention on $I$.  For example, if the conserved quantities are energy, momentum and charge,  then we have \begin{equation}
X^I_A = (\epsilon_A,  \mathbf{p}, q_A)\;\;\; \text{and} \;\;\;  \lambda^I = \left(-\frac{1}{k_{\mathrm{B}}T}, \frac{\mathbf{u}}{k_{\mathrm{B}}T}, \frac{\mu}{k_{\mathrm{B}}T}\right).
\end{equation}
The equations of motion of zeroth order hydrodynamics are found by plugging in the ansatz \begin{equation}
f_A^0(\mathbf{x},\mathbf{p}) = n_{\mathrm{F}}\left(-\lambda^I(\mathbf{x})X^I_A(\mathbf{p})\right)
\end{equation}
into (\ref{eq:boltzmann5}), and one finds \begin{equation}
\partial_t \rho^I(\mathbf{x}) + \nabla \cdot \mathbf{J}^I(\mathbf{x}) = 0,  \label{eq:511cont}
\end{equation}
where \begin{subequations}\label{eq:kineticcurrents}\begin{align}
\rho^I &\equiv \sum_A \int \frac{\mathrm{d}^d\mathbf{p}}{(2\mpi \hbar)^d} X^I_A(\mathbf{p}) f_A^0(\mathbf{x},\mathbf{p}), \\
\mathbf{J}^I &\equiv \sum_A \int \frac{\mathrm{d}^d\mathbf{p}}{(2\mpi \hbar)^d} X^I_A(\mathbf{p}) \mathbf{v}_A(\mathbf{p}) f_A^0(\mathbf{x},\mathbf{p}), \label{eq:JI0}
\end{align}\end{subequations}
are the charge and current densities associated with each conserved quantity.   These equations can be understood intuitively as follows.   If we wait for times $t \gg \tau_{\mathrm{ee}}$,  then we qualitatively expect that the collision integral has relaxed away all non-equilibrium perturbations.

In Section \ref{sec:hydro},  we often expressed the hydrodynamic equations in terms of the variables $n(\mu,T)$, $u^\mu$, etc.   For example, if $\mathbf{u}=\mathbf{0}$, then from (\ref{eq:kineticcurrents}) we find that \begin{equation}
\epsilon(\mu,T) = \sum_A \int \frac{\mathrm{d}^d\mathbf{p}}{(2\mpi\hbar)^d}  \frac{\epsilon(\mathbf{p})}{1+\mathrm{e}^{(\epsilon(\mathbf{p})-\mu q_A)/k_{\mathrm{B}}T}}.  \label{eq:nmuTkinetic}
\end{equation}
The sum $A$ runs over electrons and  holes with $q_A = \pm 1$.   In this expression, we have assumed the dispersion relation (\ref{eq:disprel}) for all species of particles, as is appropriate for graphene.   At finite $\mathbf{u}$, it is easier to replace $\mathbf{u}$ with $u^\mu$, similarly to (\ref{eq:umu}).   The relativistic generalization of (\ref{eq:nmuTkinetic}) then becomes (in units with $
\hbar=v_{\mathrm{F}}=1$) \begin{equation}
T^{\mu\nu} = \sum_A \int \frac{\mathrm{d}^d\mathbf{p}}{(2\mpi\hbar)^d}   \frac{p^\mu p^\nu}{|\mathbf{p}|}  \frac{1}{1+\mathrm{e}^{(p^\mu u_\mu-\mu q_A)/k_{\mathrm{B}}T}},\;\;\; \text{where } \;\; p^\mu = (|\mathbf{p}|, \mathbf{p}).
\end{equation}
Performing this integral, one finds the form (\ref{eq:T0}) along with the identity (\ref{eq:epsdP}).  

As we saw in Section \ref{sec:hydro}, the hydrodynamic equations can be slightly pathological at ideal order, and to find a set of dissipative equations which truly settle to thermal equilibrium we must account for some dissipation.  This can be accounted for by perturbatively solving (\ref{eq:boltzmann5}) in the ``small parameter"  $\tau_{\mathrm{ee}}$.   Because $\tau_{\mathrm{ee}}$ is defined implicitly via (\ref{eq:collisiontime}), the easiest way to do this is to write \begin{equation}
f_A(\mathbf{x},\mathbf{p}) = f_A^0(\mathbf{x},\mathbf{p}) + f_A^1(\mathbf{x},\mathbf{p}).
\end{equation}
with $f_A^1$ characterizing the small correction to the distribution function arising from the fact that $\mathcal{C}[f]$ should not exactly vanish at all times.   We enforce \begin{equation}
\sum_A \int \frac{\mathrm{d}^d\mathbf{p}}{(2\mpi \hbar)^d} X^I_A(\mathbf{p}) f_A^1(\mathbf{x},\mathbf{p}) = 0  \label{eq:kineticlandau}
\end{equation}
because any local fluctuation of a conserved quantity should be absorbed into the local value of $\lambda^I(\mathbf{x})$.   This is the analogue of the ``Landau frame" of hydrodynamics.
We then approximate that at first order in $\mathrm{\Delta} t$:   
\begin{equation}
\mathcal{C}[f^1_A] \approx \partial_t f_A^0 + \mathbf{v}_A \cdot \nabla f_A^0.  \label{eq:bgkapprox0}
\end{equation}
It is instructive to make a relaxation time approximation \cite{bgk} \begin{equation}
\mathcal{C}[f^1_A] \approx -\frac{f^1_A}{\tau_{\mathrm{ee}}}.  \label{eq:bgkapprox}
\end{equation}
Combining (\ref{eq:kineticlandau}), (\ref{eq:bgkapprox0}) and (\ref{eq:bgkapprox}), we find that the equations of motion are still the continuity equations (\ref{eq:511cont}) but with a dissipative contribution to the current $\mathbf{J}^I$: \begin{equation}
\widehat{\mathbf{J}}^I = -\tau_{\mathrm{ee}}\sum_A \int \frac{\mathrm{d}^d\mathbf{p}}{(2\mpi \hbar)^d} X^I_A(\mathbf{p})  \mathbf{v}_A(\mathbf{p}) \left(\partial_t f^0_A + \mathbf{v}_A(\mathbf{p})\cdot \nabla f^0_A\right).  \label{eq:JI1}
\end{equation}
To leading order in $\tau_{\mathrm{ee}}$ one can then use the zeroth order equations of hydrodynamics to simplify the integrals on the right hand side.    Relative to (\ref{eq:JI0}), (\ref{eq:JI1}) contains an extra derivative.  Thus $\widehat{\mathbf{J}}^I$ contains viscous dissipation, among other first order corrections to hydrodynamics.  In the limit where $\tau_{\mathrm{ee}} \rightarrow 0$, we indeed recover zeroth order hydrodynamics.

More carefully, one could linearize (\ref{eq:bgkapprox0}), replacing $\tau^{-1}_{\mathrm{ee}}$ with a matrix in both $A$ and $\mathbf{p}$ indices.   A generalization of the remaining steps gives a more accurate determination of the first order corrections to the hydrodynamic equations.

\subsection{Collisions in Graphene}\label{sec:collision}
We now turn to the application of kinetic theory to graphene \cite{muller2, muller4, foster, mirlin11, svinstov, narozhny}.    As in any kinetic theory, some of the subtlety arises from the explicit form of the collision integral.  In graphene, the most important electron-electron interactions are long-range Coulomb interactions, as we discussed in Section \ref{sec:phases}.   So we can compute the collision integral in graphene using (\ref{eq:collisionintegral}) with \cite{muller2} \begin{equation}
\left|\mathcal{M}_{ABCD}(\mathbf{k}_1,\mathbf{k}_2, \mathbf{q},\omega)\right|^2 = \left(\frac{2\mpi \alpha}{\epsilon(\mathbf{q},\omega)|\mathbf{q}|}\right)^2  \times \mathcal{Y}_{ABCD}(\mathbf{k}_1,\mathbf{k}_2,\mathbf{q},\omega) 
\end{equation}
with $\epsilon(\mathbf{q},\omega)$ a frequency-dependent effective permittivity \cite{hwangplasmon}, which we approximate as \cite{muller2}
 \begin{equation}
\epsilon(\mathbf{q},\omega) \approx \left(1+\frac{c_1 \alpha \mu}{q}\right)\left(1 + \frac{c_2\alpha}{\sqrt{1-(\omega/v_{\mathrm{F}}q)^2}}\right).  \label{eq:effperm}
\end{equation}
The factor of $\mathcal{Y}$ is typically just a O(1) constant but is rather complicated: see \cite{muller2}.

However, in graphene one of the most important effects is ``geometric".   This paragraph follows the discussion in \cite{muller4}; see also \cite{vafek2008}.   As discussed above, the collision integral contains $\mdelta$-functions for energy and momentum conservation.   We now consider the possibility that the two incoming and two outgoing momenta are nearly collinear:  e.g., $\mathbf{k}_1 = k_{1\parallel} \hat{\mathbf{x}} + k_{1\perp}\hat{\mathbf{y}}$ with $k_{1\parallel} \gg k_{1\perp}$.   Without loss of generality we set $k_{2\perp}=0$.   If the momentum exchange during the collision is $\mathbf{q}$, and incoming quasiparticles have momenta $\mathbf{k}_{1,2}$, then the energy $\mdelta$-function is \begin{align}
\mdelta(|\mathbf{k}_1| + |\mathbf{k}_2| -&|\mathbf{k}_1+\mathbf{q}| - |\mathbf{k}_2-\mathbf{q}|) \approx \mdelta \left( \frac{k_{1\perp}^2}{2|k_{1\parallel}|} - \frac{(k_{1\perp}+q_\perp)^2}{2|k_{1\parallel} + q_\parallel|} - \frac{q_\perp^2}{2|k_2-q|}  \right) \notag \\
&=   \frac{1}{k_{1\perp}} \times \left[f_1 (k_{1\parallel}, k_{2\parallel},q_\parallel) \mdelta(q_\perp - q_{\perp1}^*) + f_2 (k_{1\parallel}, k_{2\parallel},q_\parallel) \mdelta(q_\perp - q_{\perp2}^*)\right]
\end{align}
To get the second step, we have used $\mdelta$-function identities.  The functions $f_{1,2}$ and $q_{\perp1,2}^*$ are unimportant, and follow from solving a quadratic equation.   The key point is that the collision integral (\ref{eq:collisionintegral}) will involve integrals over $\mathbf{k}_2$ and $\mathbf{q}$.   The $k_{1\perp}$ integral will have a logarithmic divergence, which is sensitive to the fact that there are $d=2$ spatial dimensions.  This is called the (foward) collinear scattering singularity.     The integral is not truly divergent so long as the scattering amplitude $\mathcal{M}$ vanishes for collinear scattering.   The form of (\ref{eq:effperm}) ensures this does happen.  Because screening is proportional to $\alpha$, we conclude that as $\alpha \rightarrow 0$ collinear scattering will lead to rapid thermalization at every angle.   For example, at charge neutrality, we find \cite{trushin} \begin{equation}
\frac{1}{\tau_{\mathrm{collinear}}} \sim \frac{\alpha^2 k_{\mathrm{B}}T}{\hbar \log \alpha^{-1}}.
\end{equation}   
The consequences of this collinear scattering for graphene will be discussed in Section \ref{sec:kineticcoeff}.  Experimental signatures of rapid collinear scattering were observed in \cite{brida}.   A toy model of the kinetic theory of graphene accounting for this rapid collinear scattering can be found in \cite{svintsov1710}.


\subsection{The Imbalance Mode}
\label{sec:imbalance}
\begin{figure}
\centering
\includegraphics[width=3in]{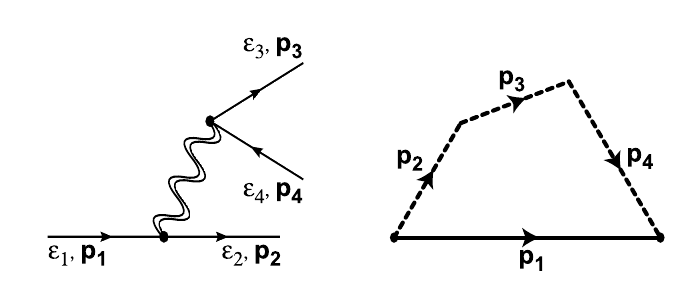}
\caption{A proposed $\mathrm{e}\rightarrow \mathrm{e}+\mathrm{e}+\mathrm{h}$ scattering event.   It is only possible to end up with on-shell quasiparticles if $|\mathbf{p}_1|= |\mathbf{p}_2| + |\mathbf{p}_3| + |\mathbf{p}_4| $ and $\mathbf{p}_1 = \mathbf{p}_2 + \mathbf{p}_3 + \mathbf{p}_4 $ with graphene's relativistic dispersion relation.   Scattering events must then be collinear. Figure taken from \cite{foster} with permission.}
\label{fig:imbalance}
\end{figure}

Beyond the divergences in the collinear limit, the relativistic dispersion of graphene leads to another important effect:  the separate (approximate) conservation of electrons and holes \cite{foster, svinstov, narozhny}.   In particular, at strong coupling, a simple expectation is that an electron will sometimes ``spontaneously convert" into $n$ electrons and $n-1$ holes; sometimes the reverse process could occur.   There is no fundamental symmetry that prevents this from happening.   However, in the weak coupling limit, this is very unlikely to happen for kinematic reasons.   With a linear dispersion relation, conservation laws demand that both $\mathbf{p}_1 = \mathbf{p}_2+\mathbf{p}_3+\mathbf{p}_4$ and $|\mathbf{p}_1| = |\mathbf{p}_2| + |\mathbf{p}_3| + |\mathbf{p}_4|$.   This is only possible if all four momenta are collinear: see Figure \ref{fig:imbalance}.   
Because the decay rate for a single electron would involve an integral over all of phase space, the contribution of the collinear decays is vanishingly small.   We conclude that on times of order $\alpha^2$, the number of electrons $n_{\mathrm{e}}$ and holes $n_{\mathrm{h}}$ are separately conserved.   It is more conventional to write the two conserved densities as \begin{subequations}\begin{align}
n &= n_{\mathrm{e}} - n_{\mathrm{h}}, \\
n_{\mathrm{imb}} &= n_{\mathrm{e}} + n_{\mathrm{h}},
\end{align}\end{subequations}
with $n_{\mathrm{imb}}$ the ``imbalance density".

The discovery of this imbalance mode at weak coupling is, in our view, the most important contribution of the kinetic theory of graphene hydrodynamics.   We will discuss whether the imbalance mode is really present in the Dirac fluid at the end of this section, but the predictions of a theory of imbalance hydrodynamics can always be compared directly to experiment.

Finally, one might ask whether the presence of two valleys in graphene, which we have so far neglected, provides \emph{extra} imbalance modes.   The answer is yes -- but with an important caveat.   Because the two valley fluids in graphene have identical properties, so long as we are only interested in disorder or experimental probes which are smooth on atomic scales,  the valley imbalance degree of freedom will decouple from any measurement.    For this reason, we have neglected the presence of multiple valleys.

\subsubsection{Hydrodynamics with an Imbalance Mode}
Following the logic of Section \ref{sec:hydro}, it is straightforward to construct the theory of hydrodynamics with an additional imbalance mode;  see also \cite{mirlin2015}.  Because both the electron fluid and hole fluid have a relativistic dispersion relation, and the instantaneous Coulomb interactions do not contribute to the energy current or momentum density, we conclude that $T^{\mu\nu}$ continues to be symmetric.   We have two conserved charges $n^a = (n,n_{\mathrm{imb}})$ -- $a$, $b$ indices will label the electric/imbalance charge indices.   Still, the key observation is that nothing in the derivations of Section \ref{sec:hydro} changes if we simply replace $\mu \mathrm{d}N$ in (\ref{eq:firstlawthermo}) with $\mu^a \mathrm{d}N^a$,   replace (\ref{eq:gibbsduhem}) with $\epsilon+P = Ts + \mu^an^a$, etc.    We conclude that the nonlinear hydrodynamics with imbalance modes is given by the equations \begin{subequations}\begin{align}
\partial_\mu J^{a\mu} &= 0, \label{eq:imbcons} \\
\partial_\nu T^{\mu\nu} &= F^{a\mu\nu}J_\nu
\end{align}\end{subequations}
For compactness, we have written $F^{a\mu\nu} = (F^{\mu\nu},0)$ -- only the electric part of the conserved charges may realistically be externally sourced in experiments.\footnote{Theorists should also not get confused by the notation $F^a_{\mu\nu}$ -- the presence of multiple conservation laws here does not imply any emergent non-Abelian gauge invariance.}   To first order in the derivative expansion, one finds $T^{\mu\nu}$ is given by (\ref{eq:Tmunuend}), while \begin{equation}
J^{a\mu} = n^a u^\mu - \sigma^{ab}_{\textsc{q}} \mathcal{P}^{\mu\nu} \left[\partial_\nu \mu^b - \frac{\mu^b}{T} \partial_\nu T - F^b_{\nu\rho}u^\rho\right],
\end{equation}
where $\sigma^{ab}_{\textsc{q}}$ is a positive-definite $2\times 2$ matrix.   Many authors use different conventions than us (see the recent review \cite{schmalian2} for an example), but one can show that their equations are equivalent after suitable relabelings.   

The analysis of these hydrodynamic equations  in linear response is a straightforward generalization of our analysis in Section \ref{sec:modes}.   One obtains sound modes with dispersion relation (\ref{eq:gammas}), together with diffusive modes $\omega = -\mathrm{i}Dk^2$ with diffusion constants \begin{equation}
D = \mathrm{eigenvalues}\left[  \frac{1}{2} \left(\frac{\partial}{\partial \mu^a} \left(\frac{n^d}{s}\right) \right)^{-1} \sigma_{\textsc{q}}^{db}\left(\mdelta^{bc} + \frac{\mu^b n^c}{Ts}\right) \right].
\end{equation}
Consistency of hydrodynamics requires that all of these eigenvalues are positive.

The possibility of imbalance modes in a far broader class of materials was discussed in \cite{hartnoll1704, hartnoll1705}, along with their experimental implications;  see also Section \ref{sec:drag1}.

\subsubsection{Decay of the Imbalance Mode}
\begin{figure}
\centering
\includegraphics[width=5in]{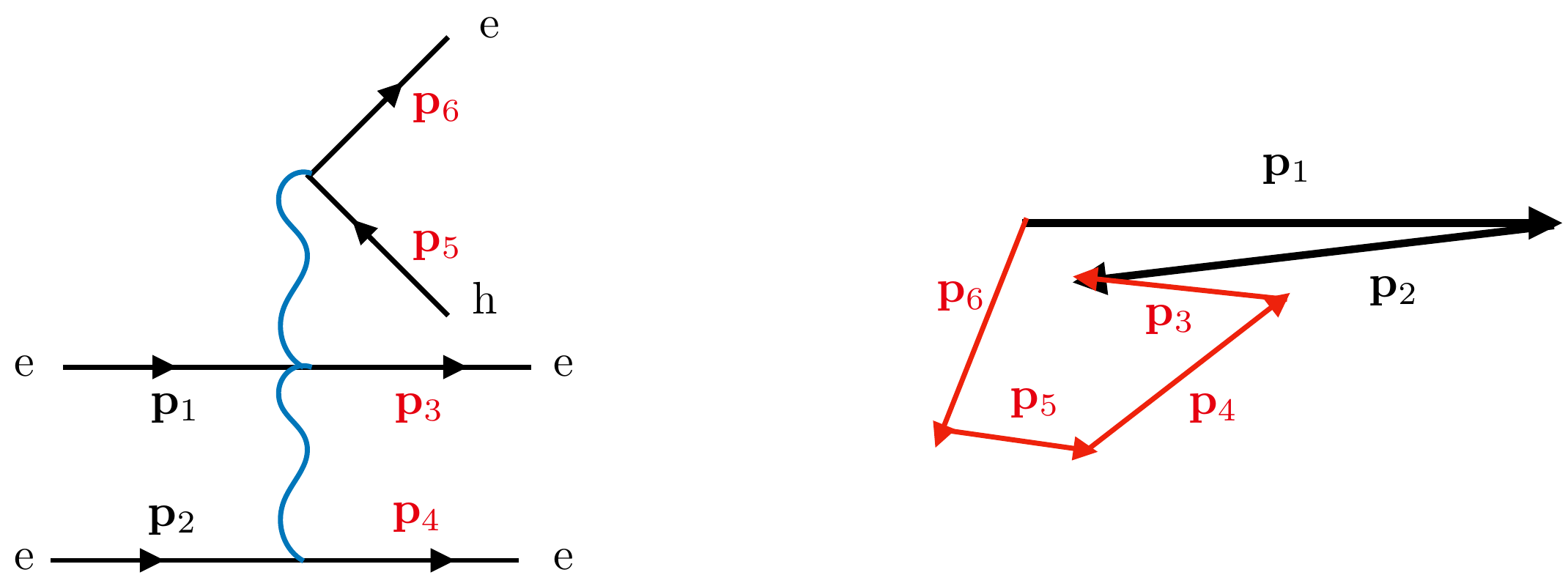}
\caption{The imbalance mode can decay through a higher order (3-body) scattering event (left).  Energy and momentum conservation laws no longer forbid such processes with graphene's relativistic dispersion relation, so long as the scattering event is of the form $\mathrm{e}+\mathrm{e} \rightarrow \mathrm{e}+\mathrm{e}+\mathrm{e}+\mathrm{h}$ and not $\mathrm{e}+\mathrm{e} \rightarrow \mathrm{e}+\mathrm{e}+\mathrm{h}+\mathrm{h}$ (right).}
\label{fig:imbalance2}
\end{figure}

The reason that we did not include this imbalance mode explicitly in the hydrodynamics of Section \ref{sec:hydro} is that this conservation law is not exact.   In particular, consider the higher order scattering process shown in Figure \ref{fig:imbalance2}.   In this case, we have 2 electrons scattering into 3 electrons and 1 hole -- in other words, the ``assisted decay" of an energetic electron into  electrons and holes.   
In order to compute the rate of  this scattering process, one must include higher order Feynman diagrams in the collision integral (\ref{eq:collisionintegral}), and an explicit evaluation becomes more and more cumbersome.   At the neutrality point, where the imbalance mode is likely to be most important, we can estimate the decay rate of such scattering events as \begin{equation}
\frac{1}{\tau_{\mathrm{imb}}} \sim \frac{\alpha^4 k_{\mathrm{B}}T}{\hbar},  \label{eq:tauimb}
\end{equation}
up to $\log \alpha^{-1}$ prefactors.   We see that while such processes are quite suppressed when $\alpha \ll 1$,  so long as $\alpha \sim 1$ the imbalance decay rate can be relatively fast.   Keeping in mind our previous estimate $\alpha \sim 0.4$, we conclude that the imbalance mode's lifetime is about 5 times longer than other non-conserved degrees of freedom.   It could lead to quantitative changes in experiment, but likely not qualitative changes.  Another recent discussion of $\tau_{\mathrm{imb}}$ can be found in \cite{svintsov1709}.

Another decay channel for the imbalance mode is disorder-assisted two-body scattering \cite{foster}.
Because we are typically interested in fluid dynamics where momentum is very long-lived, we will find it more useful to think about imbalance decay via higher order scattering events, as in (\ref{eq:tauimb}).

If the imbalance mode is a long-lived degree of freedom, then it may be useful to maintain it in the equations of motion.   This is analogous to keeping track of viscous effects in a momentum relaxing fluid -- on short enough length scales, viscous effects may have experimental consequences.   A simple model which accounts for the decay of the imbalance mode on a time scale $\tau_{\mathrm{imb}}$ is to replace (\ref{eq:imbcons}) with \begin{subequations}\begin{align}
\partial_\mu J^\mu &= 0, \\
\partial_\mu J^{\mathrm{i}\mu} &=  - \frac{n_{\mathrm{imb}} - n_{\mathrm{imb,eq}}(\mu, T)}{\tau_{\mathrm{imb}}}.
\end{align}\end{subequations}

\subsection{The Hydrodynamic Coefficients of Graphene}
\label{sec:kineticcoeff}
In this section, we collect some explicit results (without proof) for the thermodynamic and hydrodynamic coefficients of graphene, as computed in kinetic theory.   The numerical coefficients below may not be exact in the Dirac fluid regime.   Let us emphasize that relativistic dimensional analysis can be used to predict almost everything below up to the O(1) constant prefactors.   Below our purpose is to explicitly give such prefactors.

\subsubsection{Thermodynamics}

The thermodynamics of graphene in both the Fermi liquid and Dirac fluid, at least when $\alpha \ll 1$, is well approximated by the thermodynamics of a free Fermi gas of suitable Fermi velocity $v_{\mathrm{F}}$ \cite{vafek, schmalian}.   This can be understood by recalling that at weak coupling, the only scale (temperature) dependent parameter in the action (\ref{eq:SDF}) was the Fermi velocity $v_{\mathrm{F}}$, given by (\ref{eq:vFeff}).   Within kinetic theory, we simply use the standard Fermi gas formula for the pressure \begin{equation}
P = -k_{\mathrm{B}}T \sum_A \int\frac{\mathrm{d}^2\mathbf{p}}{(2\mpi\hbar)^2} \log \left(1+\mathrm{e}^{(\mu-\epsilon_A(\mathbf{p}))/k_{\mathrm{B}}T}\right) = \frac{2}{\mpi} \frac{(k_{\mathrm{B}}T)^3}{( \hbar v_{\mathrm{F}})^2} \left[\mathrm{Li}_3\left(-\mathrm{e}^{\mu/k_{\mathrm{B}}T}\right)+\mathrm{Li}_3\left(-\mathrm{e}^{-\mu/k_{\mathrm{B}}T}\right)\right]  \label{eq:54P}
\end{equation}
where $\mathrm{Li}$ denotes the polylogarithm function.  In (\ref{eq:54P}), we are implicitly using the $T$ and $\mu$ dependent $v_{\mathrm{F}}$ described in (\ref{eq:vFeff}) and (\ref{eq:vFlog}).  As we described in Section \ref{sec:thermo} this can be used to derive all other thermodynamic properties.   To leading logarithmic order we can neglect the $\mu$ and $T$ dependence of  $v_{\mathrm{F}}$ when computing thermodynamic derivatives.

In the Fermi liquid regime ($\mu \gg k_{\mathrm{B}}T$), we can Taylor expand (\ref{eq:54P}) to obtain \begin{equation}
P = \frac{1}{3\mpi} \frac{|\mu|^3}{( \hbar v_{\mathrm{F}})^2} \left(1 + \mpi^2 \left(\frac{k_{\mathrm{B}}T}{\mu}\right)^2 + \cdots \right).
\end{equation}
Using (\ref{eq:vFlog}), we may write the inverse compressibility as $T\rightarrow 0$ as \cite{sarma0703}
\begin{equation}
\frac{1}{\chi} = \left(\frac{\partial \mu}{\partial n}\right)_T \approx  \hbar \tilde v_{\mathrm{F,0}} \sqrt{\frac{\mpi}{4|n|}} \left(1+\frac{\alpha_0}{8} \log \frac{n_\Lambda}{|n|}\right).
\end{equation}
Similarly, we find the specific heat \begin{equation}
c \approx T\left(\frac{\partial s}{\partial T}\right)_n = \frac{4k_{\mathrm{B}}T}{3\mpi} \frac{|\mu|}{(\hbar \tilde v_{\mathrm{F,0}})^2}   \left(1+\frac{\alpha_0}{8} \log \frac{n_\Lambda}{|n|}\right)^{-2}.
\end{equation}

In the Dirac Fluid regime ($\mu \ll k_{\mathrm{B}}T$), we instead find \begin{equation}
P= \frac{(k_{\mathrm{B}}T)^3}{( \hbar v_{\mathrm{F}})^2} \left[\frac{3\mzeta(3)}{\mpi} + \frac{2\log 2}{\mpi}\left(\frac{\mu}{k_{\mathrm{B}}T}\right)^2 + \cdots\right],
\end{equation}
where $\mzeta(x)$ is the Riemann zeta function.  We find inverse compressibility \begin{equation}
\frac{1}{\chi}  = \frac{ (\hbar v_{\mathrm{F,0}})^2}{k_{\mathrm{B}}T}  \left(1+\frac{\alpha_0}{4} \log \frac{T_\Lambda}{T}\right)^{2}\approx 1.13 \frac{\mpi (\hbar v_{\mathrm{F}})^2}{(4\log 2)k_{\mathrm{B}}T}
\end{equation}
and specific heat \begin{equation}
c = \frac{k_{\mathrm{B}}^2T^2}{(\hbar v_{\mathrm{F}})^2} \frac{18\mzeta(3)}{\mpi} \approx 6.89 \frac{k_{\mathrm{B}}^2T}{(\hbar v_{\mathrm{F,0}})^2}\left(1+\frac{\alpha_0}{4} \log \frac{T_\Lambda}{T}\right)^{-2}.
\end{equation}

\subsubsection{Dissipative Coefficients}

In the remainder of the section, we study the dissipative hydrodynamic coefficients.  First, on general grounds we anticipate that these dissipative coefficients will scale as $\eta \sim \sigma_{\textsc{q}} \sim \alpha^{-2}$, up to logarithmic factors.   The reason for this is two-fold.   Firstly, we saw in Section \ref{sec:hydro} that hydrodynamics is a derivative expansion  in the small parameter $k\ell_{\mathrm{ee}}$, where $k$ is the wave number of spatial variations and $\ell_{\mathrm{ee}}$ is the mean free path.   Because dissipative coefficients like $\eta$ and $\sigma_{\textsc{q}}$ show up at first order in the gradient expansion, we obtain $\eta,\sigma_{\textsc{q}} \sim \ell_{\mathrm{ee}}$.  We can estimate $\ell_{\mathrm{ee}} \sim v_{\mathrm{F}} \tau_{\mathrm{ee}}$, where $\tau_{\mathrm{ee}}$ is the typical lifetime of quasiparticles.   In Section \ref{sec:collision}, we estimated that $\tau_{\mathrm{ee}} \sim \alpha^{-2}$ (up to possible logarithms).   Hence we find that at weak coupling ($\alpha \rightarrow 0$), viscosity becomes very large, while at strong coupling ($\alpha \sim 1$) viscosity is small.

Why is it viscosity -- a hydrodynamic, collective effect -- is largest when interactions are weak?   In Section \ref{sec:modes}, we showed that the diffusion constant of transverse momentum was proportional to shear viscosity.  For example, assuming that momentum is ``randomly" exchanged during collisions, we estimate the diffusion constant $D \sim v_{\mathrm{F}} \ell_{\mathrm{ee}}$ by dimensional analysis:  $v_{\mathrm{F}}$ is the velocity of the particles carrying away the momentum from a ``source", and $\ell_{\mathrm{ee}}$ is the typical distance they travel before being scattered.   The weaker interactions are, and the larger  $\ell_{\mathrm{ee}}$ becomes, the farther particles can travel before they get scattered.  This is why viscosity is so large for a weakly interacting quantum gas.    For contrast, some of the most viscous classical liquids that we know are very strongly interacting; their viscosity is high because they are closer to a crystallization or jamming/glassy transition \cite{teitel}.   

Let us now provide more quantitative results, beginning with the properties of graphene in the Fermi liquid limit ($\mu \gg k_{\mathrm{B}}T$).   As we noted in Section \ref{sec:FLhydro}, in the Fermi liquid we can neglect the distinction between charge and energy conservation (and imbalance) to leading order in $k_{\mathrm{B}}T/\mu$.   Therefore, the most important dissipative coefficients are the viscosities $\eta$ and $\zeta$.    Firstly, on general grounds we expect that $\zeta \approx 0$ as graphene is an approximately scale invariant quasirelativistic plasma \cite{polini1506}.   Secondly, due to the rapid enhancement of the phase space of scattering in a 2d Fermi liquid, one finds that \cite{wilkins, quinn, jungwirth, zala} \begin{equation}
\frac{1}{\tau_{\mathrm{ee}}} \sim \frac{T^2}{\log (k_{\mathrm{B}}T/\mu)}. \label{eq:taueelog}
\end{equation}
Based on the general arguments above, we might expect that $\eta \sim 1/\tau_{\mathrm{ee}}$.   However, this turns out to not be true.   The dominant collisions in a 2d Fermi liquid are either ``head on" collisions where two quasiparticles of nearly opposite momenta scatter into two other quasiparticles of nearly opposite momenta \cite{ledwith1, ledwith2}, or collinear scattering.  The logarithmic enhacement in (\ref{eq:taueelog}) can be traced to collinear scattering, but such collisions do not efficiently dissipate transverse momentum.   Thus, it turns out that in graphene the shear viscosity is \cite{polini1506}
\begin{equation}
\eta \approx  \frac{3}{64\mpi \alpha^2 \log \alpha^{-1}} \frac{\hbar \mu^2 |n|}{(k_{\mathrm{B}}T)^2}.   \label{eq:FLvisc}
\end{equation}
It may even be possible that $\eta$ is enhanced to scaling as $T^{-2} (\log(\mu/k_{\mathrm{B}}T))^2$ \cite{novikov} in certain 2d Fermi liquids, due to possible further suppression of head on collisions \cite{aleiner06}.    The $T^{-2}$ scaling of dissipative coefficients is a classic result of Fermi liquid theory \cite{abrikosov}.

Earlier, we (correctly) observed that $\sigma_{\textsc{q}}$ will be negligible in the Fermi liquid limit.  However, it may still useful to keep track of the first non-zero correction to $\sigma_{\textsc{q}}$ in $T/\mu$.  The reason is simple:  as discussed below (\ref{eq:histframe}), a textbook (Galilean-invariant) fluid \cite{landau} has a dissipative coefficient $\kappa_{\textsc{q}}$, related to the flow of heat, in the absence of momentum flow, in a temperature gradient.   A Galilean-invariant Fermi liquid has $\kappa_{\textsc{q}} \sim T\tau_{\mathrm{ee}} \sim 1/T$ \cite{abrikosov}.  Making the frame choice (\ref{eq:histframe}), we then expect from the form of (\ref{eq:genconst}) that $\sigma_{\textsc{q}} \sim T^0$.  Although not directly stated, the authors of \cite{muller2}  indeed find this $T$-dependence of $\sigma_{\textsc{q}}$ in an explicit calculation.

Next we turn to the Dirac fluid regime ($\mu \ll k_{\mathrm{B}}T$).   In this regime, one finds that \cite{muller2, muller4, mirlin11}  
\begin{equation} 
\sigma_{\textsc{q}} \approx \frac{0.12}{\alpha(T)^2} \frac{e^2}{\hbar}.
\end{equation}  
It is not easy to directly measure this result experimentally, unfortunately:  as we will see in later sections, the experimentally measured conductivity is also affected by (e.g.) disorder.
If we assume that the imbalance mode is also conserved, then we must compute the other coefficients of the matrix $\sigma_{\textsc{q}}^{ab}$.  At the neutrality point, one finds that off-diagonal components of this matrix vanish, and \begin{equation}
\sigma^{\mathrm{imb}}_{\textsc{q}} \approx \frac{5.3}{\alpha(T)^2} \frac{e^2}{\hbar}.
\end{equation}
One also finds a very small value for the viscosity at the neutrality point \cite{muller3}: \begin{equation}
\eta \approx 0.45 \frac{(k_{\mathrm{B}}T)^2}{\hbar v_{\mathrm{F}}^2 \alpha^2}. 
\end{equation}
Note that while $\alpha$ and $v_{\mathrm{F}}$ are temperature-dependent, $v_{\mathrm{F}}\alpha$ is temperature-independent.   When $\alpha \sim 1$, one finds that $\eta/s \sim \hbar/k_{\mathrm{B}}$, as is typical of a strongly coupled quantum fluid \cite{kss}.   Finally, as in the Fermi liquid, we find that $\zeta \approx 0$.

\section{Transport in the Fermi Liquid}
\label{sec:FLtrans}

In this section, we now turn to the experimentally observable consequences of viscous fluid flow.   We focus on the Fermi liquid regime where $\mu \gg k_{\mathrm{B}}T$.  We also assume a ``mean field" treatment of disorder, and do not consider an inhomogeneous charge puddle landscape.   The equations that we must solve, subject to appropriate boundary conditions, were derived in Section \ref{sec:hydro}: \begin{subequations}\label{eq:sec6EOM}\begin{align}
\partial_i \mdelta v_i &= 0, \\
n \partial_i \mdelta \mu - \eta  \partial_j \partial_j  \mdelta v_i &= -\frac{\epsilon+P}{\tau_{\mathrm{imp}}} \mdelta v_i.
\end{align}\end{subequations}
One crucial property of these equations is that they contain only a single true ``fit" parameter, the viscosity $\eta$.  The density $n$ is measured experimentally, and $(\epsilon+P)/\tau_{\mathrm{imp}}$ is also measured experimentally through the (temperature-dependent) dc conductivity, via (\ref{eq:sigmadc46}).    The hydrodynamic equations (\ref{eq:sec6EOM}) provide clear and direct  predictions for experiments, which we detail in this section.

To the extent that generic collisions should not have additional conservation laws besides charge and momentum (we saw in Section \ref{sec:FLhydro} that energy conservation is mostly irrelevant in a Fermi liquid), we caution that this universality is only true when the electronic mean free path $\ell_{\mathrm{ee}}$ is very small compared to all other length scales.  When $\ell_{\mathrm{ee}}$ is comparable to other length scales, the dynamics of correlated electrons can become much more exotic, obtaining sensitive dependence on the details of the Fermi surface \cite{hartnoll1705}.    Due to the simplicity of the Fermi surface, however, we expect that graphene is a very good candidate material for observing the simple viscous hydrodynamics described in Section \ref{sec:FLhydro}.

%

\subsection{Flow Through Narrow Channels}\label{sec:narrow}

\begin{figure}
\centering
\includegraphics[width=3in]{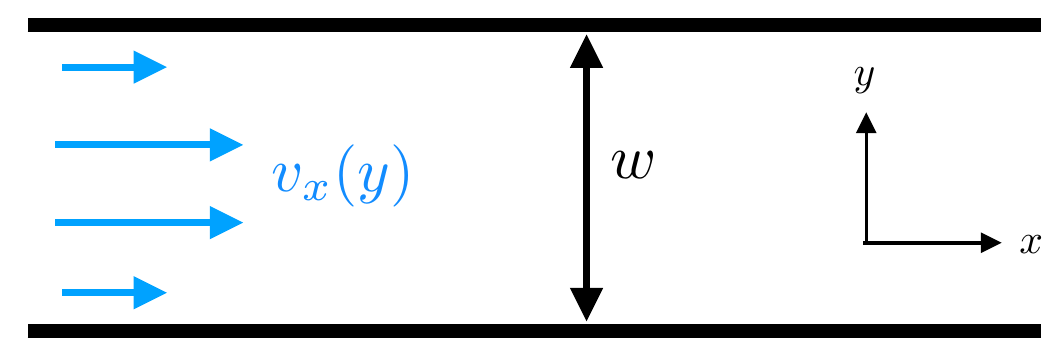}
\caption{Fluid flow through a narrow channel of width $w$.}
\label{fig:channelflow}
\end{figure}

The simplest signature of hydrodynamics in the Fermi liquid occurs in the flow of an electron fluid through a narrow channel, driven by an electric field:  see Figure \ref{fig:channelflow}.    We postulate that the flow is independent of position $x$ along the channel, which is sensible if the channel is long.  In the presence of an external electric field,  $\partial_i P = - nE_i$, and so the $x$-component of (\ref{eq:gal460}) reads \begin{equation}
- \partial_y^2 v_x = \frac{n}{\eta}E_x - \frac{1}{\lambda^2} v_x.
\end{equation}
Interestingly, a similar equation shows up in magnetohydrodynamics in the Hartmann flow geometry (see e.g. \cite{lock}).   This equation can be exactly solved for various boundary conditions.   Let us focus on two:  (\emph{i}) if the velocity is pinned to zero at the edges of the channel at $y = \pm w/2$ (``no slip"),  then by symmetry we conclude that the solution is given by \begin{equation}
v_x = \frac{n\tau}{\epsilon+P} E_x \left(1 - \frac{\cosh(\frac{y}{\lambda})}{\cosh(\frac{w}{2\lambda})}\right).
\end{equation} 
We will discuss the physics of this solution shortly.   (\emph{ii}) if the \emph{momentum flux} through the boundary is fixed to zero (``no stress"), then \begin{equation}
T_{xy} = -\eta \partial_y v_x\left(y=\pm \frac{w}{2}\right) = 0.
\end{equation}
We can see by inspection that, in this case, the solution to the equations of motion is \begin{equation}
v_x = \frac{n\tau}{\epsilon+P} E_x.
\end{equation}

An experimentally easy quantity to measure is the resistance per unit length $\mathcal{R}$ of this channel: \begin{equation}
\mathcal{R} = E_x \left(\int\limits_{-w/2}^{w/2} \mathrm{d}y\; nv_x(y)\right)^{-1}.
\end{equation} 
Using the results above we find \begin{equation}
\mathcal{R} = \left\lbrace \begin{array}{ll} \displaystyle \dfrac{1}{\sigma_{\mathrm{dc}} (w-2\lambda \tanh (\frac{w}{2\lambda}))} &\  \text{no slip} \\ \displaystyle \dfrac{1}{\sigma_{\mathrm{dc}} w} &\ \text{no stress} \end{array}\right.,  \label{eq:R61}
\end{equation}
with $\sigma_{\mathrm{dc}}$ defined in (\ref{eq:sigmadc46}).    Let us now discuss the physical consequences of hydrodynamics.   If there are no stress boundary conditions, then the flow down the channel is \emph{perfectly Ohmic}: the second row of (\ref{eq:R61}) is what one finds by solving Ohm's law in a channel of width $w$.   The fluid hardly feels the boundary at all.   However, with no slip boundary conditions, the fluid is pinned to the boundary at the edges of the channel.   Because the fluid is viscous, the stationary fluid at the edges of the channel ``pulls back" on the fluid that tries to flow down the center of the channel.   Hence, the resistance \emph{increases} as the effective width of the channel becomes smaller.   To be more quantitative, we have seen in Section \ref{sec:FLhydroMR} that the length scale $\lambda$ controls the onset of viscous effects in a momentum relaxing fluid.   When $w\gg \lambda$, one might expect that the flow is approximately Ohmic, but in an channel of effective width $w-2\lambda$ -- there is a region of size $\lambda$ on each end of the channel where viscous drag effectively forbids current from flowing.   The explicit computation (\ref{eq:R61}) confirms this.  When $w\lesssim \lambda$, then the viscous drag effects permeate the whole channel.   Then one finds \begin{equation}
\mathcal{R} = \frac{12\lambda^2}{\sigma_{\mathrm{dc}}w^3} = \frac{12 \eta}{ n^2 w^3}.   \label{eq:Rchannel}
\end{equation}
Now the resistance is much more sensitive to the width of the channel than before.   This limit is famous in  the fluid dynamics literature, where it goes by the name of Poiseuille flow \cite{landau}.  The velocity profile $v_x(y)$ is approximately parabolic: \begin{equation}
v_x(y) \approx \frac{nE_x}{2\eta} \left(\frac{w^2}{4}-y^2\right).
\end{equation}
Furthermore, the resistance is proportional to the viscosity alone, and is finite even when $\tau = \infty$.   Such effects were first noted by Gurzhi over 50 years ago \cite{gurzhi}, and this is sometimes called the Gurzhi effect.

The most dramatic experimental signature of the Gurzhi effect (\ref{eq:Rchannel}) is the fact that \begin{equation}
\frac{\partial \mathcal{R}}{\partial T} < 0,  \label{eq:dRdTneg}
\end{equation}
because (up to logarithms) $\eta \sim T^{-2}$ in a Fermi liquid (see (\ref{eq:FLvisc})).   This result violates a ``theorem" of the conventional theory of semiclassical transport \cite{ziman}, where adding more microscopic collisions \emph{always} increases the resistivity; the resolution of this paradox is discussed in some detail in \cite{hartnoll1705}.  Although the prediction (\ref{eq:dRdTneg}) was made for viscous electron flows (in Fermi liquids), it was not seen for a long time, and so historically there has been almost no interest in the hydrodynamic theory of transport in condensed matter physics.   However, (\ref{eq:Rchannel}) also predicts interesting width dependence of $\mathcal{R}$, which can also serve as an important experimental signature.

In order to see the effects of viscous flow in such channels, one needs to allow for momentum dissipation at the edges of the channel.   Are such boundary conditions generic in metals?   A simplistic answer would be yes -- atomically rough edges could act as microscopic impurities, allowing for electrons to lose their momentum at the edges of the channel.   We will discuss this question in more detail from a microscopic perspective in Section \ref{sec:narrow2}.   Experimentally, it is not definitively known what the correct boundary conditions are.   The answer is likely sensitive to the material at hand, and perhaps even to details of device fabrication.  There are reasons to believe that in graphene no stress boundary conditions are more appropriate:  in strong magnetic fields, quasiparticle orbits have been imaged which scatter almost perfectly off of the boundary \cite{dgg}.  This suggests that graphene has atomically smooth edges, which (under pristine conditions) has been observed experimentally \cite{zettledge}.

There is some experimental evidence of viscous electron flow through narrow channels, but we defer discussion to Section \ref{sec:narrow2}.

\subsection{Flow Through Constrictions}\label{sec:constrict}

\begin{figure}[h]
\centering
\includegraphics[width=3in]{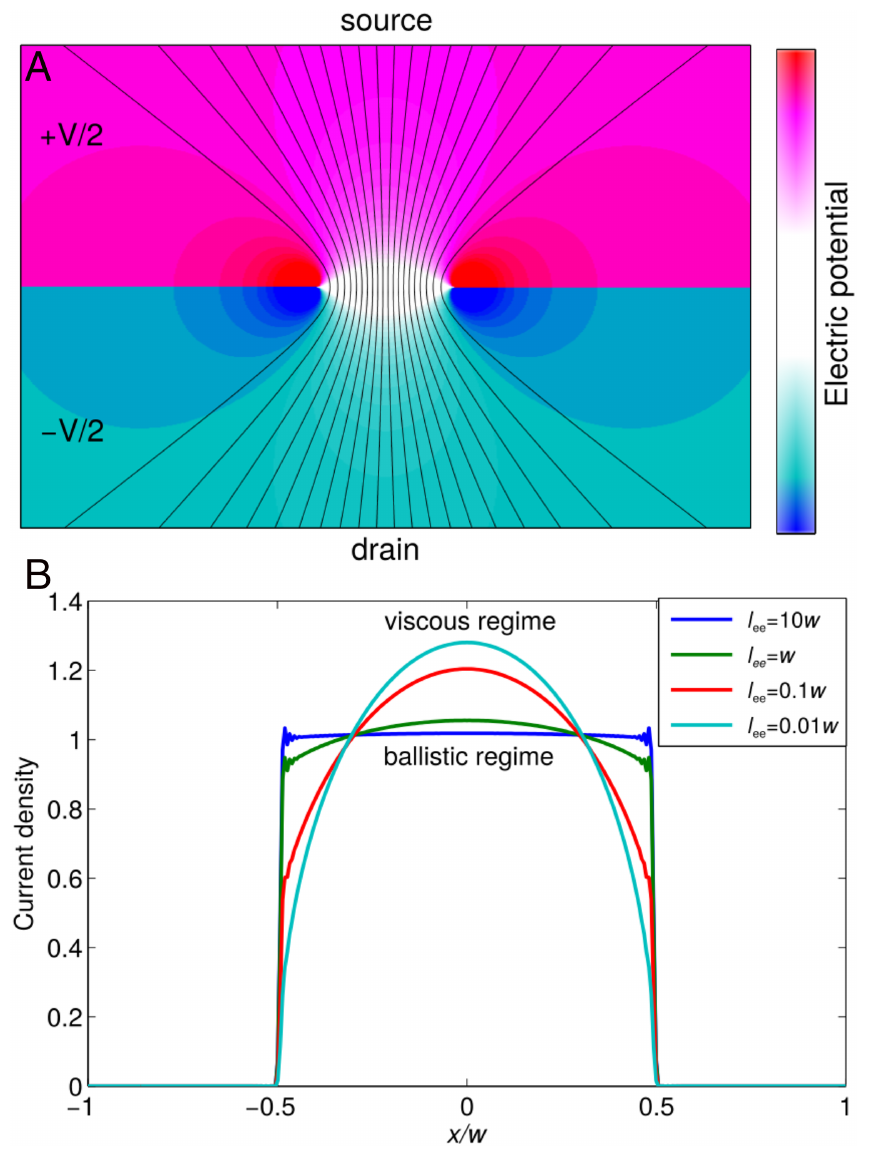}
\caption{Top panel: the chemical potential of the electron fluid as current flow from the top to the bottom through a narrow constriction of width $w$.   The curved black lines denote the streamlines along which elements of fluid will flow.   Bottom panel:  the spatial distribution of the electrons as they flow through the constriction; different  lines correspond to different ratio $\ell_{\mathrm{ee}}/w$. In the ballistic limit, the non-interacting electrons are equally likely to be found anywhere in the constriction, but in the viscous regime they are clustered at the center of the constriction.   The wiggles in the ballistic regime of the  lower panel are a consequence of the truncation of the kinetic equations described in Section \ref{sec:balhydro},  and should not detract from the physics.    Figure taken from \cite{levitov1607} with permission.}
\label{fig:guo}
\end{figure}

We now turn to a slightly different set-up: flows through constrictions or narrow openings into a broader region of fluid.   The simplest example of such a flow is depicted in the top panel of Figure \ref{fig:guo};  this set-up was considered in \cite{levitov1607}.   Fluid flows from a region of high chemical potential to a region of lower chemical potential through a constriction of width $w$.   We assume that the current is blocked from flowing away from the constriction by an infinitely thin barrier (of course, this is a mathematical simplification).  Neglect momentum-relaxing electronic collisions, we simply need to solve (\ref{eq:stream4}) in the limit $\lambda=\infty$, subject to suitable boundary conditions.  The solution of the problem is rather technical and relies on techniques from complex analysis, and we will not describe it here.   The bottom panel of Figure \ref{fig:guo} shows the current distribution through the constriction:  in the hydrodynamic regime of flow, it is given by $J(x) \propto \sqrt{(w/2)^2 - x^2}$, where $x$ is the distance from the center of the constriction.

The simplest thing to measure, however, is again the total electrical resistance $\mathcal{R}$, defined as the ratio of the voltage difference (far from the constriction) between the top and bottom half planes, divided by the current flowing through the constriction.   One finds \cite{levitov1607} \begin{equation}
R = \frac{32\eta}{\mpi e^2n^2 w^2}.  \label{eq:Rvisccon}
\end{equation}
This formula is somewhat similar to (\ref{eq:Rchannel}), up to the difference in the power of $w$.  The smaller power of $w$ appearing here is a consequence of the fact that the constriction is infinitely thin, while the channel is infinitely thick.   Many of the same signatures of viscous flow that can, in principle, be seen in flows through narrow channels are the same signatures of viscous flows through constrictions:  most notably, the temperature dependence $R\sim T^{-2}$.   However, as we will see in Section \ref{sec:narrow2}, the flow through a constriction has proven a more useful setting to look for signatures of viscous flows experimentally.

An alternative way to measure the impact of viscosity on transport may be to study the flow around a circular obstacle \cite{spivak02, spivak06, lucas1612, levitov1612}.  Here, one measures the resistance associated with the obstacle is proportional to viscosity, similar to (\ref{eq:Rvisccon}).

\subsubsection{Negative Nonlocal Resistance}
\begin{figure}
 \centering
 \includegraphics[width=4.5in]{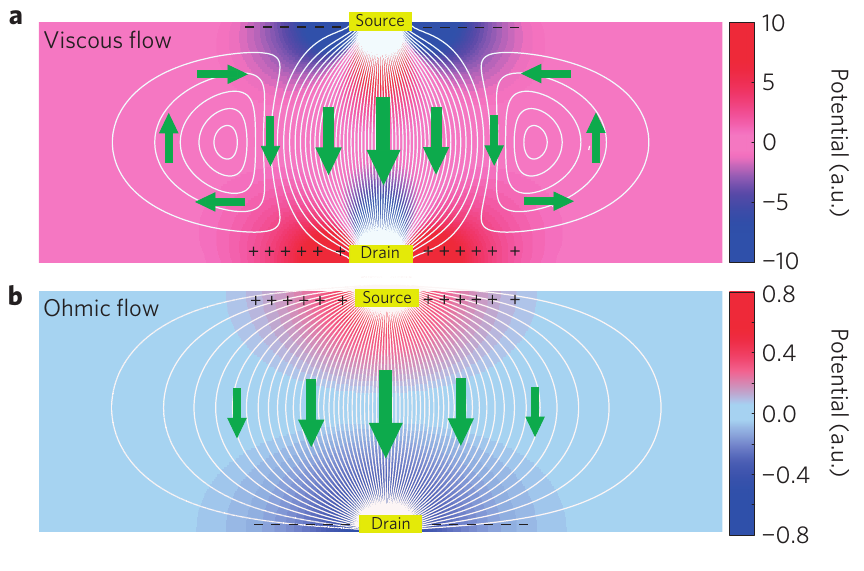}
\caption{The voltage or chemical potential of a Fermi liquid in the viscous (top) or Ohmic (bottom) regime.   In the Ohmic regime, the potential is a monotonically decreasing function from source to drain, while in the viscous regime the voltage drop is strongest just to the sides of the source.   This negative nonlocal resistance is not possible in an Ohmic limit.  Figures taken from \cite{levitovhydro} with permission.}
\label{fig:nonlocalresistance}
\end{figure}

A slight variation on the flow through a single constriction is the flow of current between a narrow source and a narrow drain \cite{polini, levitovhydro}.   The precise nature of such a flow depends on the specific geometry studied.   The simplest case is depicted in Figure \ref{fig:nonlocalresistance}:  a source and drain of current are located on opposite sides of an infinitely long slab of width $W$.    Solving the momentum-relaxing Navier-Stokes equation (\ref{eq:gal460}), one finds two qualitatively different behaviors depending on whether the momentum relaxation length $\lambda$ is large or small compared to $W$.   If $\lambda \ll W$,   then the flow of current is essentially Ohmic, and the electrochemical potential (and thus voltage measured) will decrease monotonically along flow lines from source to drain.   However, if $\lambda \gg W$,  momentum relaxation is negligible and one typically finds sign-changing voltage profiles, as shown in Figure \ref{fig:nonlocalresistance}.   The observation of such negative nonlocal resistance is a key signature that Ohmic transport theory is not applicable.   Another important, possibly experimentally accessible difference between Ohmic/viscous transport arises from studying local Joule heating \cite{levitovhydro}.

Perhaps a more direct signature of viscous current flow is the ``backflow" of current, or the formation of vortices.  The arrows in Figure \ref{fig:nonlocalresistance} depict the formation of vortices in such a setup.   A natural question is whether vortices always form whenever nonlocal resistance is obtained.   Unfortunately, the answer is no \cite{levitovjuly, torre}.   Negative nonlocal resistance is often observed in viscous regimes near the source or drain of current, and is not necessarily sensitive to the locations of other boundaries or sources \cite{torre}.  In contrast, the existence of vortices is found to be much more sensitive to global boundary conditions.  Intuitively, vortices form when the fluid flow can interfere with itself as it flows around the geometry.    Some geometries turn out to only exhibit backflow with a large enough viscosity $\eta$, whereas others have backflow for any $\eta \ne 0$ \cite{torre}.

Another subtlety with predicting vortex flow from nonlocal resistance measurements is that multiple current distributions, obeying different boundary conditions, can lead to the same potential distributions \cite{levitovjuly}.  A measurement of nonlocal negative resistance is not sufficient to predict the current flow.  This is a consequence of the fact that (\ref{eq:stream4}) is a fourth order differential equation, and not second order as in the Ohmic case.   Different current distributions that lead to the same potential distribution can be distinguished by a magnetic field \cite{levitovjuly}.  

  \begin{figure}
 \centering
 \includegraphics[width=2.5in]{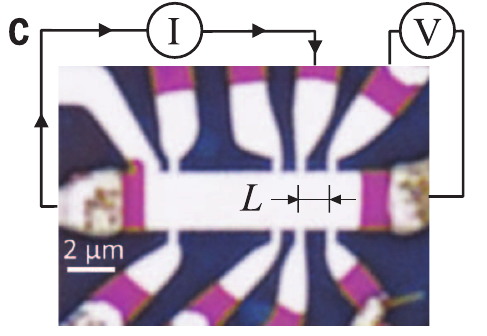}
 \hspace{0.5in}
  \includegraphics[width=2.8in]{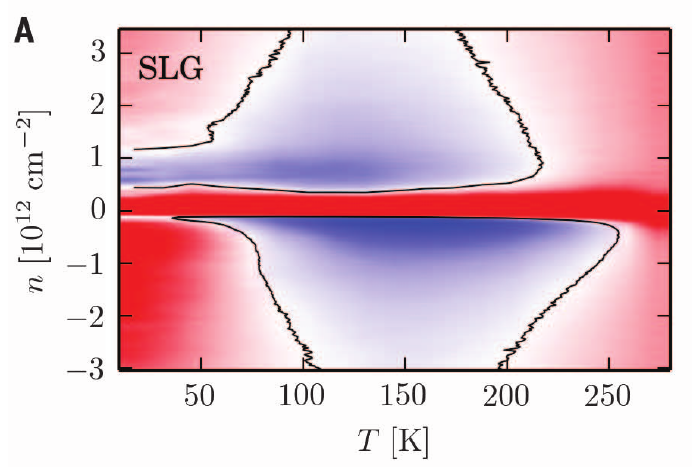}
\caption{Left: experimental set-up of current source/drain and voltage probes for the nonlocal resistance measurements.  Right: measured nonlocal resistance as a function of temperature $T$ and density $n$.  Red denotes a positive, Ohmic resistance measurement;  blue denotes a negative, viscous resistance.  Only in the Fermi liquid, for temperatures where electron-electron scattering length is smaller than both the electron-impurity and electron-phonon scattering length, is nonlocal resistance observed.  Figures taken from \cite{bandurin} with permission.}
\label{fig:bandurin}
\end{figure}

This negative nonlocal resistance has been observed experimentally in graphene \cite{bandurin}, using the geometry  depicted in Figure \ref{fig:bandurin}.   For the most part, we observe that a critical temperature is required before the onset of nonlocal resistance.  This is consistent with the intuition that hydrodynamic, viscous effects are necessary to see this negative voltage, although we note that ballistic effects can also give negative voltages \cite{torre}.   At higher temperatures, electron-phonon scattering becomes non-negligible and transport becomes phonon-dominated and conventional.
Finally, observe that there is no nonlocal resistance in the Dirac fluid ($n\approx 0$).  We will explain why this is so at the start of Section \ref{sec:DFtrans}.    

Using their experimental data, the authors of \cite{bandurin} estimated the dynamical viscosity of the graphene Fermi liquid (for moderate doping) to be \begin{equation}
\nu \sim 0.1\; \frac{\mathrm{m^2}}{\mathrm{s}}.
\end{equation}
This result is consistent with theoretical predictions discussed in Section \ref{sec:kineticcoeff}.   For comparison, the dynamical viscosity of water at room temperature is $\nu \sim 10^{-6} \mathrm{m}^2/\mathrm{s}$.    $\nu$ is so large for electrons in  graphene due to both the weak electron-electron interactions of the Fermi liquid, and the very high $v_{\mathrm{F}}$.

 \subsection{Viscometry}  \label{sec:viscometry}
 So far, we have discussed experiments that see clear hints of hydrodynamic flow, but for practical reasons they turn out to be rather non-ideal measurements of the electronic viscosity directly.  The essential challenge is that (as we will see in great detail later) transport measurements are quite sensitive to momentum relaxation, and it can be challenging to disentangle this effect.
 
 So let us now briefly discuss a few alternative ideas for how to measure the electronic viscosity of a metal.  One proposal which has received a lot of attention is the Dyakonov-Shur instability \cite{DS}, which occurs in  a fluid with a uniform background velocity flow, subject to a pair of exotic boundary conditions at the edges of the device (density $n$ is fixed at one end of the flow, and current $J_x$ is fixed at the other).   Such an instability, which would be experimentally observed via the detection of spontaneous ac current in electronics, is a dramatic and surprising feature of fluid motion.  It arises from the amplification of sound waves as they reflect back and forth between the two walls, where very different boundary conditions are imposed \cite{DS}.   Taking viscous dissipation into account, one estimates that within the hydrodynamic limit, this effect is only visible when the background velocity $v_0$ is large enough: \cite{DS} \begin{equation}
 v_0 \ge \frac{\mpi^2 \nu}{8L},  \label{eq:DS}
 \end{equation}
 with $L$ the length of the device, and $\nu$ the dynamical viscosity.  As we have noted previously, it is possible to make $v_0$ rather large in many metals, and so detection of this instability ought to be possible.  (\ref{eq:DS}) gives a simple and unambiguous measure for the dynamical viscosity of the electron fluid, and so could be a very precise viscometer, in theory.  In practice, the usefulness of (\ref{eq:DS}) will be limited by both electron-impurity (momentum-relaxing) scattering.  There is also debate in the literature as to whether this instability could reappear in the ballistic limit \cite{DSkin, giliberti}, which may further complicate matters.   A discussion  of this instability in graphene can be found in \cite{svintsov13}.
 
  \begin{figure}
 \centering
 \includegraphics[width=2.5in]{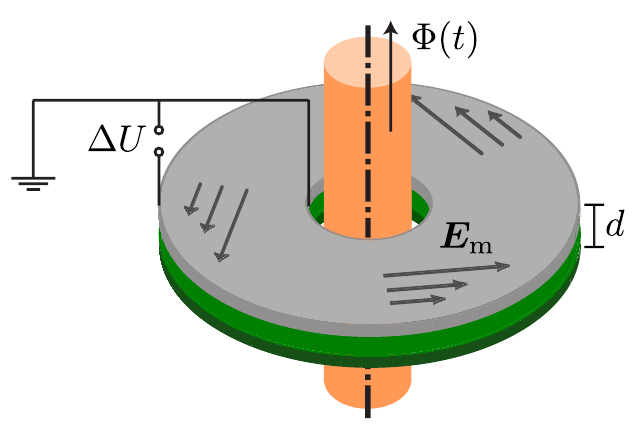}
 \hspace{0.5in}
  \includegraphics[width=2.5in]{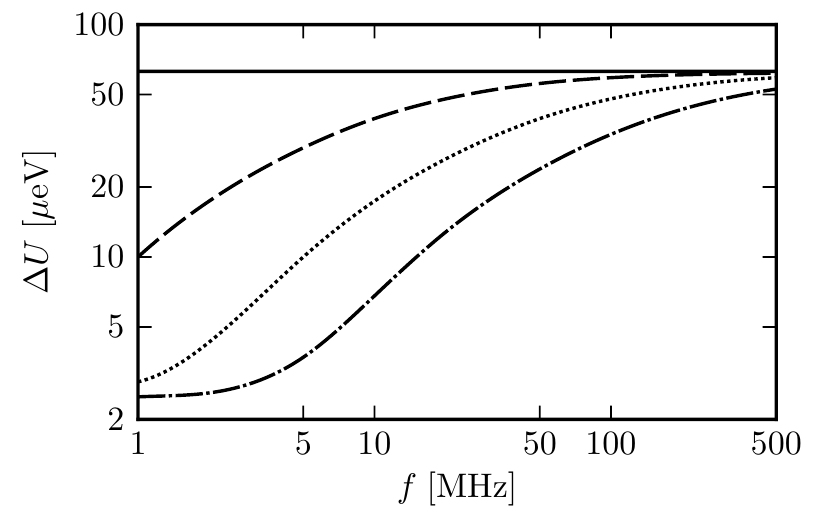}
\caption{Left: the Corbino disk geometry, with time-dependent magnetic flux $\Phi(t)$, and potential measurements across the disk.  Right:  the measured electric potential drop as a function of frequency.   The particular scales on the axes will vary from material to material but the shape of the curves is universal in the hydrodynamic limit.   In particular, the drop in the voltage at low frequency allows us to estimate the viscosity.  Figures taken from \cite{tomadin} with permission.}
\label{fig:corbino}
\end{figure}

One interesting observation about the Dyakonov-Shur instability is that it relies on the nonlinear structure of hydrodynamics.  Another proposal for measuring $\nu$ that also relies on nonlinear terms in the Navier-Stokes equations is to study the frequency dependent response of an electron fluid in the Corbino disk geometry: see Figure \ref{fig:corbino}.   The inspiration for this measurement is two-fold.   First, we note that angular fluid flows cause to radial pressure (and hence voltage) drops:  schematically, \begin{equation}
\mathrm{\Delta} P \propto \int \mathrm{d}r \; v_\theta(r)^2.  \label{eq:corbino1}
\end{equation}
Secondly, we observe that $v_\theta(r)$ will exponentially decay in a viscous flow oscillating with frequency $\omega$ on length scales larger than \begin{equation}
\xi = \sqrt{\frac{\nu}{\omega}}.
\end{equation}
Then, the simple observation is that the pressure drop $\mathrm{\Delta}P$, computed in (\ref{eq:corbino1}) will get much smaller once $\xi$ is small compared to the relative radii of the Corbino disk geometry.   See Figure \ref{fig:corbino} for the quantitative result found by solving the Navier-Stokes equations.   In order to drive the fluid in such an angular flow, one threads a time-dependent magnetic field through the center of the disk.   Maxwell's equations then apply a time-dependent angular electric field, even outside of the solenoid where the magnetic field is threaded.   This electrical forcing is useful as it allows us to only drive the electronic degrees of freedom.  Unfortunately, it is not easy to fabricate graphene in the annular shape required to realize such an experiment, and so such a viscometer has not been created yet.   A similar proposal in a different geometry was recently given in \cite{surowka}.

One final way to measure the viscosity of an electron fluid in two dimensions is to study magnetotransport.  We will discuss this in more detail in Section \ref{sec:magnetic}, but let us emphasize for now that such a viscometer is likely not going to give a quantitative measurement of  $\eta$.  However, it may be sufficient to obtain the qualitative magnitude and temperature dependence of $\eta$.   Along these lines, we note that a very simple model of viscous electronic flow is able to explain certain non-trivial features of magnetotransport in GaAs \cite{alekseev}.

 \subsection{The Ballistic-to-Hydrodynamic Crossover}
 \label{sec:balhydro}
So far, we have seen some simple signatures of viscous electron flow in a Fermi liquid.   However, as we observed in Section \ref{sec:kinetic},  to be cleanly in the hydrodynamic regime requires $\ell_{\mathrm{ee}}$ to be very small compared to all length scales in the problem.   This is a challenge for graphene, where device sizes are generally $\lesssim 10\; \mmu \mathrm{m}$, while $\ell_{\mathrm{ee}} \gtrsim 0.5\; \mmu\mathrm{m}$ is not an unreasonable expectation for the Fermi liquid regime.   As such, it is useful to have a solvable model that interpolates between hydrodynamics, and a theory of a non-interacting Fermi gas.    The discussion in Section \ref{sec:kinetic} makes clear that kinetic theory provides one such approach.   In this section, we will describe a particular toy model of kinetic theory, appropriate for graphene.  Although this model has some history \cite{molenkamp}, it has been studied comprehensively much more recently  \cite{levitov1607, lucas1612, levitov1612}.   

As we have already seen in our discussion of the Fermi liquid limit of hydrodynamics in Section \ref{sec:FLhydro}, in a homogeneous fluid the effects of energy conservation provide only $\mathrm{O}((T/\mu)^2$ corrections to the low temperature dynamics.   Hence we propose (without proof) that in kinetic theory, the full distribution function $f(\mathbf{x},\mathbf{p})$ may be approximated as (at low temperature) 
\begin{equation}
f(\mathbf{x},\mathbf{p}) \approx \mathrm{\Theta}(\mu - v_{\mathrm{F}}p - \Phi(\mathbf{x},\theta) )  \label{eq:fTheta}
\end{equation}
Here $\mathrm{\Theta}(x)$ is the Heaviside step function, $p=|\mathbf{p}|$ and $\theta \equiv \arctan(p_y/p_x)$.   For simplicity, we have kept track of the distribution function $f$ only in the conduction band.    What (\ref{eq:fTheta}) asserts is that, to good approximation, the distribution function of the low temperature Fermi liquid can be approximated by the spatially inhomogeneous ``sloshing" of a sharp Fermi surface.     One can then approximate the solution to the full Boltzmann equation by
 \begin{equation}
\partial_t \Phi + v_{\mathrm{F}} \left(\cos\theta \partial_x \Phi + \sin\theta \partial_y \Phi\right) = \mathcal{C}[\Phi].
\end{equation}
Rather than performing a complicated microscopic calculation of $\mathcal{C}[\Phi]$, we will simply guess the answer using a ``relaxation time approximation" \cite{bgk},\footnote{We caution the reader that the relaxation time approximations commonly employed in condensed matter physics, as in conventional textbooks \cite{ashcroft}, often do not carefully account for conservation laws.} which relaxes $\Phi$ towards thermal equilibrium at a uniform rate $\tau^{-1}_{\mathrm{ee}}$.   To be more explicit, it is instructive to write \begin{equation}
\Phi(\mathbf{x},\theta,t) = \sum_{n=-\infty}^\infty a_n(\mathbf{x},t) \mathrm{e}^{\mathrm{i}n\theta}.
\end{equation}
Then one finds the equations \begin{equation}
\partial_t a_n + \frac{v_{\mathrm{F}}}{2}\left((\partial_x + \mathrm{i}\partial_y) a_{n+1} + (\partial_x - \mathrm{i}\partial_y) a_{n-1}\right) = \left\lbrace \begin{array}{ll} 0 &\ n=0 \\  \displaystyle -\dfrac{a_n}{\tau_{\mathrm{imp}}}  &\   n=\pm1 \\ \displaystyle -\dfrac{a_n}{\tau_{\mathrm{ee}}} &\  \text{otherwise} \end{array}\right..  \label{eq:anboltzmann}
\end{equation}
These equations can (in some circumstances) be efficiently solved using fancy techniques \cite{levitov1607, lucas1612, levitov1612}.    We have also taken the liberty to account for momentum relaxation in these equations.   Of course, this model does not yet specify \emph{what} any of these relaxation times are.   This can be actually a cumbersome question to address, even within kinetic theory, as one must explicitly evaluate the linearized collision integral.  Along these lines, recent work \cite{ledwith1, ledwith2} has presented a particularly efficient scheme for evaluating such integrals.  They also find that the relaxation time for even and odd harmonics $a_n$ could be parametrically different, leading to novel physics on intermediate length scales.  We refer the readers to this pair of papers for more details;  in what follows, we use the simpler model of (\ref{eq:anboltzmann}).

Following the procedure of Section \ref{sec:hydrofromkt}, the qualitative physics of (\ref{eq:anboltzmann}) can be understood relatively easily.    For simplicity, we set $\tau_{\mathrm{imp}}^{-1} = 0$ for the moment.   On time scales $t\ll \tau_{\mathrm{ee}}$,   the right hand side is very small, and one approximately finds the equation $\partial_t \Phi + v_i(\theta) \partial_i \Phi = 0$.    This simply states that particles propagate ballistically at a fixed velocity.    On time scales $t\gg \tau_{\mathrm{ee}}$,  one can show that $a_{\pm 2} \approx - \frac{1}{2}v_{\mathrm{F}}\tau_{\mathrm{ee}} (\partial_x \mp \mathrm{i}\partial_y) a_{\pm 1}$, and that (\ref{eq:anboltzmann}) approximately closes to a set of equations for $a_0$, $a_1$ and $a_{-1}$.   Furthermore, the fluctuations in the electronic number density are given by \begin{equation}
\mathrm{\Delta} n = \int \frac{\mathrm{d}^2\mathbf{p}}{(2\mpi\hbar)^2} \mdelta(\mu - v_{\mathrm{F}}p)  \Phi = \frac{p_{\mathrm{F}}}{2\mpi \hbar^2 v_{\mathrm{F}}} a_0,
\end{equation}
and a similar calculation shows that $a_1 + a_{-1} \propto v_x$, while $(a_1 - a_{-1})/\mathrm{i} \propto v_y$.    A straightforward calculation then reveals that the resulting equations for $a_{-1}$, $a_0$ and $a_1$ are exactly (\ref{eq:gal1}), the equations of hydrodynamics in the low temperature limit.   The explicit expression for the speed of sound and  dynamical viscosity are \begin{equation}
v_{\mathrm{s}} = \frac{v_{\mathrm{F}}}{\sqrt{2}}, \;\;\; \nu = \frac{v_{\mathrm{F}}^2\tau_{\mathrm{ee}}}{4},
\end{equation}
and are consistent with our previous discussions.   A pedagogical treatment of these points, and more computational details, may be found in \cite{levitov1607, lucas1612}.  

Let us emphasize once more, however, that the use of this model is that it is relatively numerically tractable to study the entire crossover between the ballistic and hydrodynamic regimes.   The assumption of a circular Fermi surface is also particularly well-suited to graphene,  although it may not be well-suited to other materials \cite{hartnoll1705}.

\subsubsection{Flow through Narrow Channels and Constrictions, Revisited}\label{sec:narrow2}
With this toy model for the ballistic-to-hydrodynamic crossover at hand, let us return to the question of flow through narrow channels and constrictions.   We are now ready to address more  experimental observations of hydrodynamic electron flow.

We begin by discussing the flow of electrons through a narrow channel, as discussed in Section \ref{sec:narrow}.   Recall that the boundary conditions must be chosen such that momentum can relax at the boundary -- for example, the conventional no-slip boundary conditions of hydrodynamics.   The equations (\ref{eq:anboltzmann}) can be solved numerically in this setting, and here we simply focus on the qualitative features.   The resistance of the channel is (up to dimensional prefactors) is \begin{equation}
R \sim \frac{1}{w} \times \min\left(\frac{1}{\tau_{\mathrm{imp}}}, \; \frac{v_{\mathrm{F}}}{w}, \; \frac{v_{\mathrm{F}}^2\tau_{\mathrm{ee}}}{w^2}\right).  \label{eq:Rwchannel}
\end{equation}
This equation can be understood as follows.  The resistance $R\sim 1/(w\tau_{\mathrm{mom}})$ where $\tau_{\mathrm{mom}}$ is the time scale it takes for momentum to relax, including relaxation due to the presence of a boundary.   If $\tau_{\mathrm{imp}}^{-1}\rightarrow 0$, then there is a simple competition between ballistic and hydrodynamic effects.   If $v_{\mathrm{F}} \tau_{\mathrm{ee}} \gg w$,  then quasiparticles will approximately bounce back and forth between the edges of the channel, and at each bounce will lose much of their forward momentum.   The time between bounces is given by $w/v_{\mathrm{F}}$ -- the time it takes to travel between the two sides of the channel.      When $v_{\mathrm{F}}\tau_{\mathrm{ee}} \ll w$, then collisions occur frequently and the theory is described by hydrodynamics.   Now the time scale required to relax momentum is set by the diffusion of transverse momentum.  This mode was described in (\ref{eq:diffvy}).  Because dissipation is now diffusive, the time scale is $w^2/D_{\mathrm{mom}}$, where $D_{\mathrm{mom}} \sim v_{\mathrm{F}}^2\tau_{\mathrm{ee}}$ is the diffusion constant for transverse momentum.     Accounting for a finite $\tau_{\mathrm{imp}}$, we see that if the channel width $w$ is too large, then bulk momentum-relaxing scattering dominates $R$.   This is, in principle, an easy effect to account for experimentally because the bulk resistivity can be measured in a multitude of other geometries independently.

\begin{figure}
\centering
\includegraphics[width=5in]{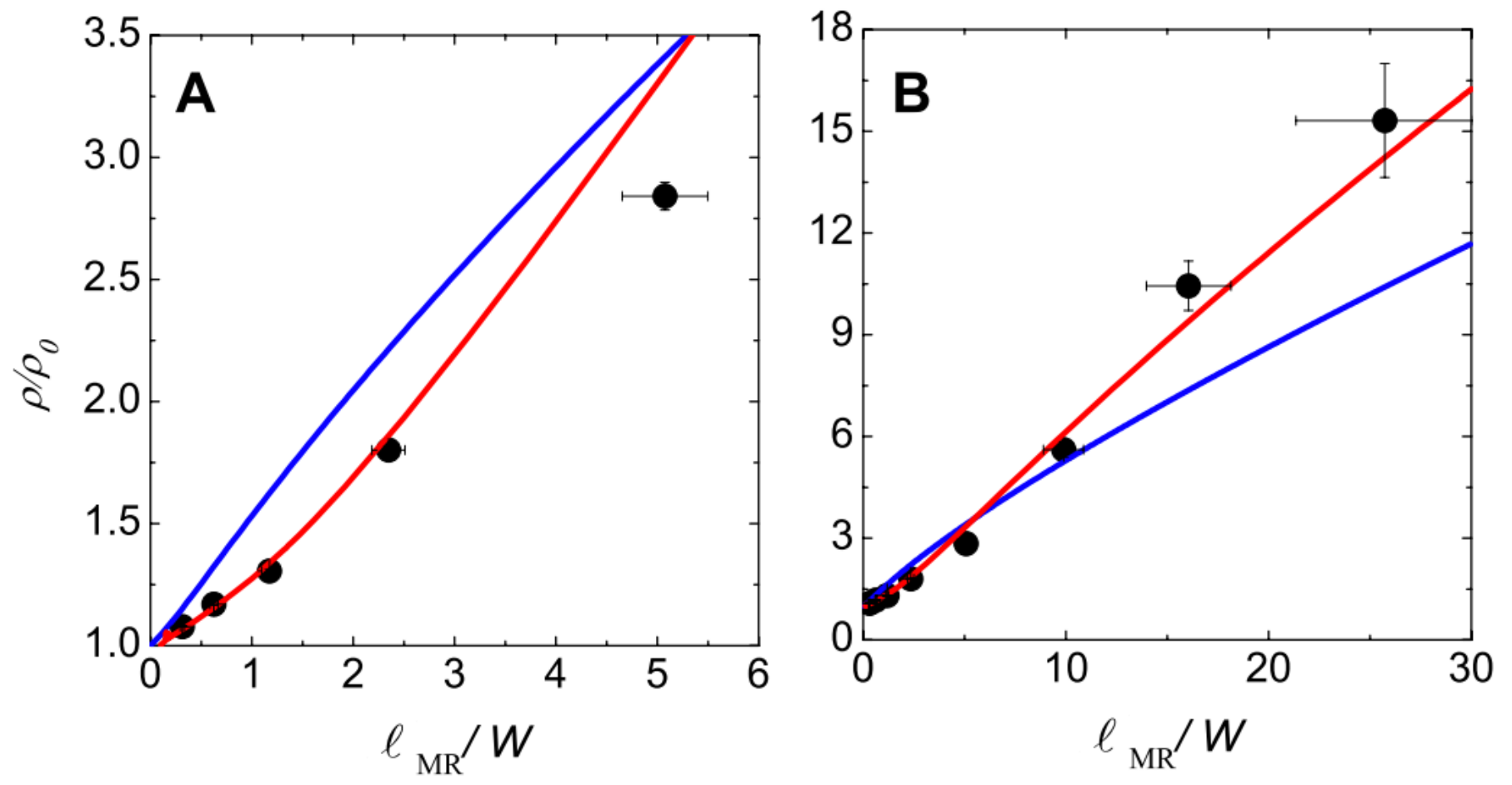}
\caption{A normalized resistance $wR$ as a function of inverse channel with $1/w$.  When $1/w$ is larger, we observe that $wR$ grows faster than $1/w$.  This is consistent with a hydrodynamic regime with $\tau_{\mathrm{ee}} =10 \tau_{\mathrm{imp}}$ (red curve), but not $\tau_{\mathrm{ee}} = \tau_{\mathrm{imp}}$ (blue curve).  Reprinted from \cite{mackenzie} with permission.}
\label{fig:moll}
\end{figure}

The main experimental evidence for (\ref{eq:Rwchannel}) comes from looking for the various powers of $R\propto w^{-n}$ in (\ref{eq:Rwchannel}):   $n=1$ signifies conventional Ohmic flow,  $n=2$ signifies relatively conventional ballistic flow, and $n=3$ signifies hydrodynamic flow.      As shown in Figure \ref{fig:moll},  experiments have seen evidence for $R(w)$ decaying faster than $1/w^2$ in GaAs \cite{molenkamp}, $\mathrm{PdCoO}_2$ \cite{mackenzie} and $\mathrm{WP}_2$ \cite{felser};  evidence in $\mathrm{WP}_2$ is particularly striking.    While the $w$-dependence of $R(w)$ is evidence for some kind of `hydrodynamic' effect,  on the other hand, the expected non-monotonic temperature dependence of $R(T)$ for a fixed channel width $w$ is not seen cleanly in the above experiments.   Indeed, it is unclear whether the toy model above is appropriate for either $\mathrm{PdCoO}_2$ or $\mathrm{WP}_2$.   Both have more complicated band structures than graphene, and the effective hydrodynamics at the ballistic crossover could be more complicated \cite{hartnoll1705}.  

\begin{figure}
\centering
\includegraphics[width=2.2in]{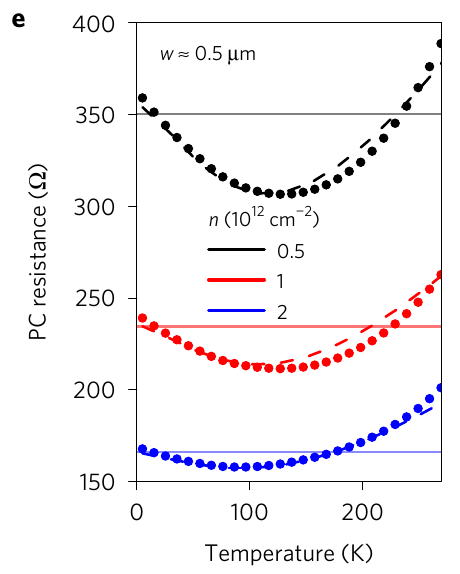}
\caption{The non-monotonic temperature dependence of the resistance $R(T)$ of a $w=500$ nm constriction  is a clear signature of viscous effects.  Reprinted from \cite{levitov1703} with permission.}
\label{fig:consexp}
\end{figure}

Next, let us return to the flow of electrons through a narrow constriction of width $w$, as discussed in Section \ref{sec:constrict}.   Numerical evidence \cite{levitov1607} suggests that the generalization of (\ref{eq:Rvisccon}) to account for ballistic effects is simply \begin{equation}
\frac{1}{R} \approx \frac{1}{R_0} + \frac{\mpi e^2n^2 w^2}{32\eta}.
\end{equation}
Here $R_0$ is the resistance in the non-interacting electron gas, entirely due to ballistic flow through the constriction.   Viscous effects enhance transport beyond the ballistic limit.    This effect has been observed directly in the flow of electrons through constrictions cut into samples of graphene \cite{levitov1703}: see Figure \ref{fig:consexp}.   The dramatic non-monotonic temperature dependence is the most clear indication observed yet in experiment of the onset of viscous transport regime of a Fermi liquid.

\section{Transport in the Dirac Fluid}
\label{sec:DFtrans}
The phenomena we have observed in Section \ref{sec:FLtrans} are specific to the doped Fermi liquid regime, at least in graphene.    As we noted in Section \ref{sec:modes}, the dynamics of charge becomes decoupled from energy and momentum in a charge-neutral relativistic fluid.   This regime of charge neutrality is experimentally accessible in graphene.   The linear response phenomena described in Section \ref{sec:FLtrans} all have analogues at charge neutrality, but energy density and temperature play the role of charge density and chemical potential:  see, for example, \cite{jerome1}.   This makes experiments to detect such flows challenging.   Indeed, at charge neutrality, the linearized equations governing charge transport are (neglecting charge puddles) simply 
\begin{equation}
\partial_i \left(\sigma_{\textsc{q}} \left(E_i - \partial_i \mu \right)\right) = 0.
\end{equation}
This is, of course, the equation governing Ohmic charge diffusion in a medium with conductivity $\sigma_{\textsc{q}}$.   Electrical transport in exceptionally pure Dirac fluid looks identical to transport in a conventional metal.

The interesting hydrodynamic response at charge neutrality is in the energy-momentum sector.   In this section, we will discuss the signatures of the Dirac fluid in thermoelectric transport.   These are more complicated experiments to perform than measurements of electrical resistance, and so we will also review the subtleties required to observe these phenomena experimentally.

\subsection{``Mean-Field" Hydrodynamic Model of Thermoelectric Transport}
\label{sec:MFdisorder}

Let us begin with a review of a ``mean field" model of thermoelectric transport within hydrodynamics, following \cite{hkms}.    By thermoelectric transport, we mean the following:  consider a system perturbed by a background, time-dependent electric field and/or temperature gradient, which may be time-dependent.   As a response to this perturbation, charge currents $J_i$ and heat currents $Q_i$ will begin to flow in the system.    If the electric fields and temperature gradients are small, then we can write 
\begin{equation}
\left(\begin{array}{c}  J^i(t) \\ Q^i(t) \end{array}\right) = \int \mathrm{d}t^\prime \; \left(\begin{array}{cc}  \sigma^{ij}(t-t^\prime) &\ \alpha^{ij}(t-t^\prime) \\ T\bar\alpha^{ij}(t-t^\prime) &\ \bar\kappa^{ij}(t-t^\prime) \end{array}\right) \left(\begin{array}{c}  E_j(t^\prime) \\ -\partial_j T(t^\prime) \end{array}\right)  + \cdots
\end{equation}
Due to time-translation invariance, the matrix of coefficients above only depends on the difference, $t-t^\prime$, and so we will often find it useful to compute the Fourier transform the above expression:  \begin{equation}
\left(\begin{array}{c}  J^i(\omega) \\ Q^i(\omega) \end{array}\right) = \left(\begin{array}{cc}  \sigma^{ij}(\omega) &\ \alpha^{ij}(\omega) \\ T\bar\alpha^{ij}(\omega) &\ \bar\kappa^{ij}(\omega) \end{array}\right) \left(\begin{array}{c}  E_j(\omega) \\ -\partial_j T(\omega) \end{array}\right)  + \cdots  \label{eq:thermoelectric}
\end{equation}
We note without proof the following useful facts:  (\emph{i}) that the heat current in graphene can be defined as \begin{equation}
Q^i = T^{ti} - \mu J^i,  \label{eq:7heat}
\end{equation}
and is in fact equivalent to $Ts^i$, where $s^i$ is the spatial part of the entropy current defined in (\ref{eq:entropy4});  (\emph{ii}) the coefficients $\sigma^{ij}$ are complex-valued and must be analytic in the upper-half complex plane.

 As we introduced in Section \ref{sec:dis4}, a simple way to account for momentum relaxation is to simply add a term to the hydrodynamic equations that spoils momentum conservation equation, as in (\ref{eq:46first}).   One of our goals in Section \ref{sec:hydrotrans} will be to rigorously assess (in a certain limit) the validity of this approximation.   Nevertheless, we proceed for the moment assuming the validity of (\ref{eq:46first}), and use these equations to evaluate $J^i$ and $Q^i$ in the presence of background electric fields and temperature gradients.

The calculation is actually very straightforward.   Because (\ref{eq:46first}) do not break spatial homogeneity, we may look for an ansatz where the velocity $v^i$ is a constant.  The resulting equation can be rearranged to the following simple form: \begin{equation}
\frac{\epsilon+P}{v_{\mathrm{F}}^2}\left(\frac{1}{\tau_{\mathrm{imp}}} - \mathrm{i}\omega\right) v_i =  nE_i + s\partial_i T  \label{eq:71mom}
\end{equation}
Using our formal expression for the charge current from Section \ref{sec:422}, and simplifying to the linear response limit where $E^i$ and $v^i$ are small (recall that $E^i$ is analogous to $-\partial^i\mu$):
 \begin{equation}
J^i = \sigma_{\textsc{q}}\left(E^i- \frac{\mu}{T}\partial_i T\right) + n v^i.  \label{eq:71charge}
\end{equation}
Using (\ref{eq:7heat}) along with the two equations above, simple algebra leads us to \begin{subequations}\label{eq:7drude}\begin{align}
\sigma^{ij} &= \mdelta^{ij} \left[\sigma_{\textsc{q}} + \frac{n^2 v_{\mathrm{F}}^2 \tau_{\mathrm{imp}}}{(\epsilon+P)(1-\mathrm{i}\omega \tau_{\mathrm{imp}})}\right], \\
\alpha^{ij}  = \bar\alpha^{ij} &= \mdelta^{ij} \left[-\frac{\mu}{T}\sigma_{\textsc{q}} + \frac{n s v_{\mathrm{F}}^2 \tau_{\mathrm{imp}}}{(\epsilon+P)(1-\mathrm{i}\omega \tau_{\mathrm{imp}})}\right], \\
\bar\kappa^{ij} &= \mdelta^{ij} \left[\frac{\mu^2}{T}\sigma_{\textsc{q}} + \frac{Ts^2 v_{\mathrm{F}}^2 \tau_{\mathrm{imp}}}{(\epsilon+P)(1-\mathrm{i}\omega \tau_{\mathrm{imp}})}\right]. 
\end{align}\end{subequations}

In the limit $\tau_{\mathrm{imp}} \rightarrow \infty$, the terms proportional to $\sigma_{\textsc{q}}$ can be neglected in (\ref{eq:7drude}).   We then find a conventional Drude peak in the conductivity.   Indeed, in a non-relativistic theory, one would replace $\epsilon+P \approx nmv_{\mathrm{F}}^2$, and $\rho = -ne$, with $n$ the number density of quasiparticles. The conventional argument for the Drude peak assumes that there are long-lived quasiparticles, and in many cases relies on a crude ``relaxation time" approximation \cite{ashcroft} that is \emph{not equivalent}, in any sense, to the derivation above.   It is clear from our derivation of the Drude peak that the time scale $\tau_{\mathrm{imp}}$ in the conductivity is related to \emph{momentum relaxation}.  This intuition has been made quite precise in \cite{hofman, lucasMM}; see \cite{lucasrmp} for a review.    Interestingly, while it is quite common to interpret the dc conductivity according to the Drude formula in ordinary metals,  most ordinary metals do not exhibit a sharp Drude peak in their ac conductivity.  This is due to the many competing scattering pathways, such as interband transitions, that complicate dirty samples of conventional metals.  Often, sharp Drude peaks are observed experimentally only in very pure systems, or in correlated electron materials \cite{scheffler}.   While in much of the literature, scattering rates are estimated from the value of the dc conductivity, we emphasize that the extraction of a meaningful scattering time from the dc conductivity can be difficult, as we will see below.

\begin{figure}
\centering
\includegraphics[width=4.3in]{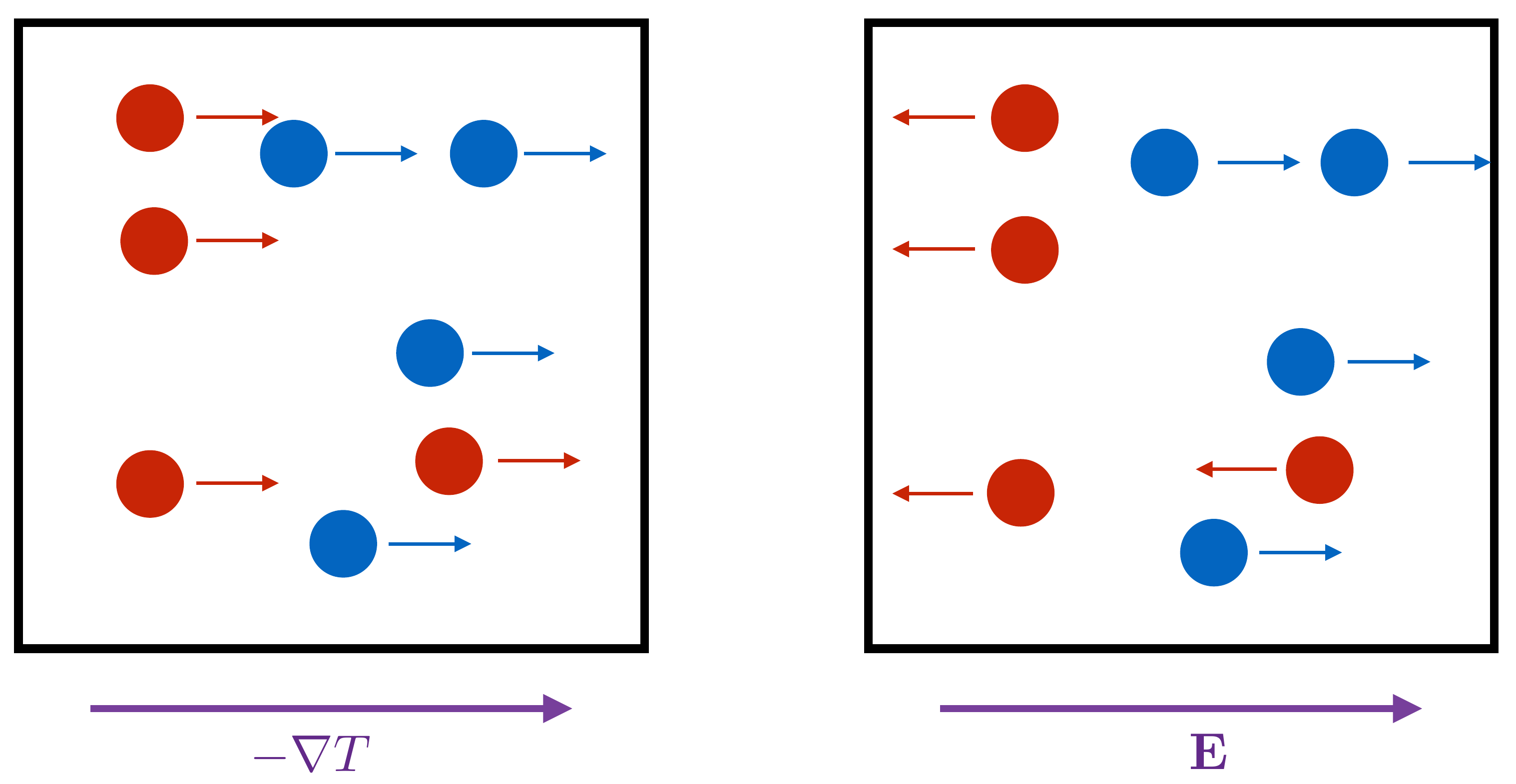}
\caption{The flow of electrons (red) vs. holes (blue) in the charge neutral Dirac fluid.   Both electrons and holes move in the same direction in a temperature gradient, but move in opposite directions in an electric field.}
\label{fig:chargevsheatflow}
\end{figure}

Nevertheless, let us assume the validitiy of (\ref{eq:7drude}) beyond the strict $\tau_{\mathrm{imp}} \rightarrow \infty$ limit.  Each of the thermoelectric conductivities consists of a \emph{sum} of two different kinds of terms -- a Drude peak which is sensitive to momentum relaxation, and a diffusive contribution which is not.   This sum behavior violates Mattheisen's rule, which argues that the resistivity is a sum of all of the possible scattering mechanisms.\footnote{Mattheisen's rule is commonly used in condensed matter physics.  However, it is not true, even in any perturbative limit, and this is but one of many counter-examples.  There are even more dramatic examples that can arise in kinetic theory \cite{hartnoll1705}.}   The logic behind this sum is as follows.  There are two kinds of mechanisms that can contribute to the conductivities in graphene (or any other relativistic fluid).  Firstly, a charged/thermal fluid will collectively flow unimpeded until it scatters off of obstacles, and this is responsible for the Drude behavior.   But there is a second, \emph{parallel} way for charge current to flow -- particles and holes can move in opposite directions.  This flow carries no momentum or energy, and so should not be sensitive to momentum relaxation.   This flow, schematically depicted in Figure \ref{fig:chargevsheatflow}, is responsible for non-vanishing $\sigma_{\textsc{q}}$, and can contribute to the conductivities.

It is common in experiments to not measure $\bar\kappa^{ij}$, but to measure \begin{equation}
\kappa^{ij} = \bar\kappa^{ij} - T\bar\alpha^{ik}\sigma^{-1}_{kl} \alpha^{lj}.
\end{equation}
This can be understood as the coefficient of proportionality between a heat current and temperature gradient, subject to the boundary conditions that no electric current flows:
\begin{equation}
Q^i = \left.-\kappa^{ij}\partial_j T\right|_{J^i =0}.
\end{equation}
Using (\ref{eq:7drude}), we find that \begin{equation}
\kappa^{ij}(\omega=0) = \mdelta^{ij} \frac{v_{\mathrm{F}}^2(\epsilon+P)\tau_{\mathrm{imp}}}{T} \frac{\sigma_{\textsc{q}}}{\sigma(n)}.
\end{equation}
Interestingly, the $\tau_{\mathrm{imp}}$ from the denominator in $\sigma(n)$ cancels the overall prefactor $\tau_{\mathrm{imp}}$ whenever $ n \ne 0$.   Thus,  this measured $\kappa$ is finite while $\sigma$ is infinite in a clean theory.   

This leads to a dramatic violation of the Wiedemann-Franz law, which states that in a conventional metal \cite{ashcroft} \begin{equation}
\frac{\kappa}{T\sigma} \equiv \mathcal{L}
\end{equation}
is given by \begin{equation}
\mathcal{L} = \mathcal{L}_{\mathrm{WF}} = \frac{\mpi^2 k_{\mathrm{B}}^2}{3e^2}.  \label{eq:LWF}
\end{equation}
The coefficient $\mathcal{L}$ is often called the Lorenz number, and it is  1 whenever elastic impurity or phonon scattering dominates transport.   However, in the hydrodynamic regime we find that \begin{equation}
\mathcal{L} = \frac{\mathcal{L}_0}{(1+(n/n_0)^2)^2}, \;\;\;\; \mathcal{L}_0 = \frac{v_{\mathrm{F}}^2 (\epsilon+P)\tau_{\mathrm{imp}}}{T^2\sigma_{\textsc{q}}}, \;\;\;\;\; n_0^2 = \frac{(\epsilon+P)\sigma_{\textsc{q}}}{e^2v_{\mathrm{F}}^2\tau_{\mathrm{imp}}}.  \label{eq:Lhydro}
\end{equation}
At $n=0$, we see that $\mathcal{L} \rightarrow \mathcal{L}_0$,  but for large enough $n$ we observe that $\mathcal{L}\rightarrow 0$.   Hence, a relativistic plasma like the Dirac fluid with long-lived conserved momentum violates the Wiedemann-Franz law both from above and below.  The violation from above is due to the fact that the charge neutral plasma has an intrinsically finite $\sigma$, but a diverging $\kappa$ which is kept finite only by disorder.   The violation from below, for densities $n\ne 0$, is due to the locking of charge and heat currents:  the most efficient way to transport both charge and heat is to create local momentum density.   We will discuss the experimental observation of (\ref{eq:Lhydro}) in Section \ref{sec:exptherm}.   We also note that $\mathcal{L} \ll \mathcal{L}_{\mathrm{WF}}$ in a very clean Fermi liquid where the momentum relaxation rate is much longer than the electron-electron scattering rate \cite{vignale}.  For further discussion of the Wiedemann-Franz law in correlated electron systems with Fermi surfaces, see \cite{mahajan}.

\subsection{Hydrodynamic Theory of Transport through Charge Puddles}\label{sec:hydrotrans}
\begin{figure}
\centering
\includegraphics[width=3in]{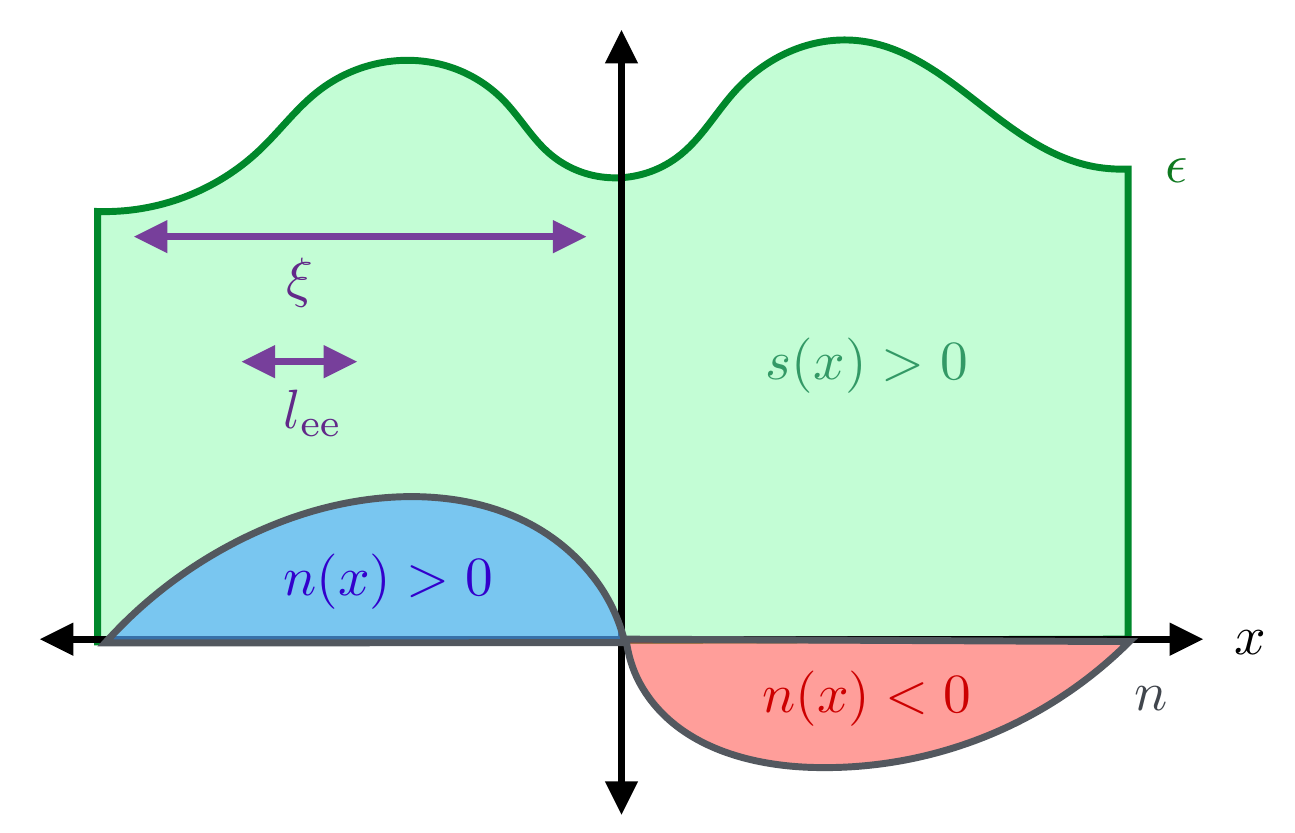}
\caption{When the mean free path for electronic collisions is short compared to the size of charge puddles, then transport is described by the hydrodynamic equations.   Momentum relaxation arises from the spatial variations in the thermodynamic coefficients such as entropy density (green) and charge density (blue/red).  In graphene, the charge density may locally switch signs.   Figure adapted from \cite{lucas3} with permission.}
\label{fig:chargepuddlehydro}
\end{figure}

In this section, we will finally relax the assumption that momentum relaxes in a ``homogeneous" way.  Indeed, as we saw in Section \ref{sec:puddle},  experimental graphene is not homogeneous, but is described by a landscape of ``charge puddles" where the local value of the chemical potential $\mu(\mathbf{x})$ varies from one point to the next: see Figure \ref{fig:chargepuddlehydro}.    This inhomogeneity is sufficient to relax momentum, and render the transport coefficients finite.  In this section, we describe the transport coefficients in such an inhomogeneous medium, when the inhomogeneity is very long wavelength.  In particular, if $\xi$ is the typical size of a charge puddle, and $\xi \gg \ell_{\mathrm{ee}}$, then following \cite{andreev, lucas} we can compute the conductivity of the medium by solving the hydrodynamic equations.   

Indeed, all non-hydrodynamic modes will relax on length scales short compared to the inhomogeneity, and will play no role in transport.  This follows from the fact that transport is sensitive to momentum relaxation, and not directly to electron-electron momentum-conserving scattering, as we saw previously;  it can be observed explicitly within the full kinetic theory of transport  \cite{hartnoll1705}.  We now folllow \cite{lucas3} and describe the solution to the transport problem within relativistic hydrodynamics.

Our starting point is to generalize (\ref{eq:EOMCou}), in the stationary limit, to study the response of an inhomogeneous fluid, in a background electric field or temperature gradient.    For simplicity we assume that the disorder couples to the chemical potential, as in \cite{lucas3};  the case with inhomogeneous strain disorder is described in \cite{scopelliti, jerome2}.   We also focus only on time-independent solutions.  The procedure is straightforward, and we find \begin{subequations}\label{eq:linres72}\begin{align}
-E_i \partial_i\sigma_{\textsc{q}} &= \partial_i \left(n\mdelta v_i - \sigma_{\textsc{q}} \partial_i \left(\mdelta \mu - \frac{\mu_0}{T}\mdelta T\right)\right), \\
-\zeta_i \partial_i(\mu_0 \sigma_{\textsc{q}}) &= \partial_i \left(s\mdelta v_i + \frac{\mu_0 \sigma_{\textsc{q}}}{T} \partial_i \left(\mdelta \mu - \frac{\mu_0}{T}\mdelta T\right)\right), \\
nE_i + Ts \zeta_i &=   n\partial_i \mdelta \mu + s\partial_i \mdelta T - \partial_j \left( \eta  (\partial_j \mdelta v_i + \partial_i \mdelta v_j)\right) - \partial_i\left( \left(\zeta-\frac{2\eta}{d}\right) \partial_j \mdelta v_j\right).
\end{align}\end{subequations}
Here $\mdelta \mu$ denotes the deviation of the chemical potential from the inhomogeneous background $\mu_0(\mathbf{x})$.   The coefficients such as $n$ and $s$ in the above equations are functions of $n(\mathbf{x}) = n(\mu_0(\mathbf{x}),T)$, for example.    These equations are a set of elliptic partial differential equations and are straightforward to solve numerically, as in \cite{lucas3}.

Suppose that the local fluctuations in the chemical potential are small:  \begin{equation}
\mu_0(\mathbf{x}) = \bar\mu_0 + u \hat\mu(\mathbf{x}),
\end{equation}
with $\hat\mu$ an O(1) function and $u\ll T, \bar\mu_0$ a perturbatively small parameter governing the strength of chemical potential.   In this case, one can perturbatively solve (\ref{eq:linres72}) order by order in $u$.  The dc thermoelectric conductivities are given by (\ref{eq:7drude}) with $\omega=0$ and (see \cite{lucas3} for a precise formula)\footnote{This formula can also be derived \cite{dsz} using the memory matrix formalism \cite{lucasMM, lucasrmp}.} \begin{equation}
\frac{1}{\tau_{\mathrm{imp}}} \approx \frac{v_{\mathrm{F}}^2}{2} u^2 \left(\frac{\partial n}{\partial \mu}\right)^2 \left[\frac{e^2}{\sigma_{\textsc{q}}(\epsilon+P)} + \frac{\eta+\zeta}{\xi^2} \frac{4\eta \mu^2}{(\epsilon+P)^3}\right].  \label{eq:hydrotautime}
\end{equation}
Note that in the above formula, all thermodynamic coefficients are evaluated in the homogeneous background with uniform chemical potential $\bar\mu_0$;  $\xi$ is the typical size of a charge puddle.   A few comments are in order.  Most importantly, we see that the momentum relaxation time $\tau_{\mathrm{imp}}$ which we have -- so far -- treated as a simple constant is in fact quite non-trivial.  It will generally depend on $\mu$ and $T$ in a complicated way.   In the smooth disorder limit $\xi \rightarrow \infty$,  the conductivity becomes limited by $\sigma_{\textsc{q}}$.  An analogous effect was first observed in Galilean-invariant fluids in \cite{andreev}, and was found in general settings in \cite{hartnoll1704, hartnoll1705}.   However, for fluids which are not deep in the hydrodynamic limit (in particular, $\xi \sim \ell_{\mathrm{ee}}$),  the two terms above may be comparable and even the temperature dependence of the transport coefficients becomes quite sensitive to microscopic details of the sample.

A non-perturbative technique for analyzing (\ref{eq:linres72}) is to bound the transport coefficients using a variational principle \cite{lucas, hartnoll1704}.\footnote{Note that this variational principle is distinct from a separate variational principle which states that the probability of finding a particle in a given state, in thermodynamic equilibrium,  is given by the distribution that maximizes entropy given the density of the conserved quantities:  energy, charge  and momentum \cite{camiola, barletti}.  The variational principle (\ref{eq:variational}) states that \emph{dissipative} processes minimize entropy production, not that \emph{thermodynamic} ensembles maximize entropy.}   For simplicity, let us describe the bound on the electrical conductivity.  Intuitively, looks for a flow of charge and heat current which minimizes entropy production;  rigorously, one finds \begin{equation}
\frac{1}{\sigma_{xx}} \le \frac{V_2}{(\int \mathrm{d}^2\mathbf{x} \; J_x)^2} \int \mathrm{d}^2\mathbf{x} \left[\frac{1}{\sigma_{\textsc{q}}} \left(\frac{TsJ_i - nQ_i}{\epsilon+P}\right)^2 + \eta_{ijkl} \partial_i \left(\frac{Q_j + \mu J_j}{\epsilon+P}\right) \partial_k \left( \frac{Q_l + \mu J_l}{\epsilon+P}  \right)\right] \label{eq:variational}
\end{equation}
where $V_2$ is the total volume of the two spatial dimensional region of interest, $J_i$ and $Q_i$ are \emph{arbitrary} trial charge and heat currents, up to the constraints $\partial_i J_i = \partial_i Q_i = 0$,  and \begin{equation}
\eta_{ijkl} = (\zeta-\eta) \mdelta_{ij}\mdelta_{kl} + \eta (\mdelta_{ik}\mdelta_{jl} + \mdelta_{il} \mdelta_{jk}).
\end{equation}
These conductivity bounds can be useful for obtaining qualitative estimates of the transport coefficients in the strong disorder limit,  but one must keep in mind that there is no guarantee that a bound obtained from (\ref{eq:variational}) is sharp.   We note that one can derive (\ref{eq:hydrotautime}) by plugging in the ansatz that $J_i$ and $Q_i$ are $\mathbf{x}$-independent into (\ref{eq:variational}).  Thus, (\ref{eq:hydrotautime}) is a lower bound on the true conductivity, which becomes perturbatively exact at weak disorder.

\subsection{Experimental Measurement of Thermal Conductivity}
\label{sec:exptherm}
The breakdown of Wiedemann-Franz (WF) law (\ref{eq:LWF}) provides strong evidence for  electronic hydrodynamics  in graphene \cite{crossno}.    Indeed, note that $\mathcal{L}_{\mathrm{WF}}$ depends only on fundamental constants, and not on material-specific parameters such as carrier density and effective mass of the charge carriers.   Experimentally, the WF law holds robustly in many conductors, and this has confirmed the validity of our standard picture of transport for ordinary metals \cite{ashcroft}.   The breakdown of the WF law in graphene is especially significant, in our view, because specific predictions for the nature of this breakdown were made based on hydrodynamics many years prior \cite{hkms}, and were subsequently observed \cite{crossno}.  

\begin{figure}
\centering
\includegraphics{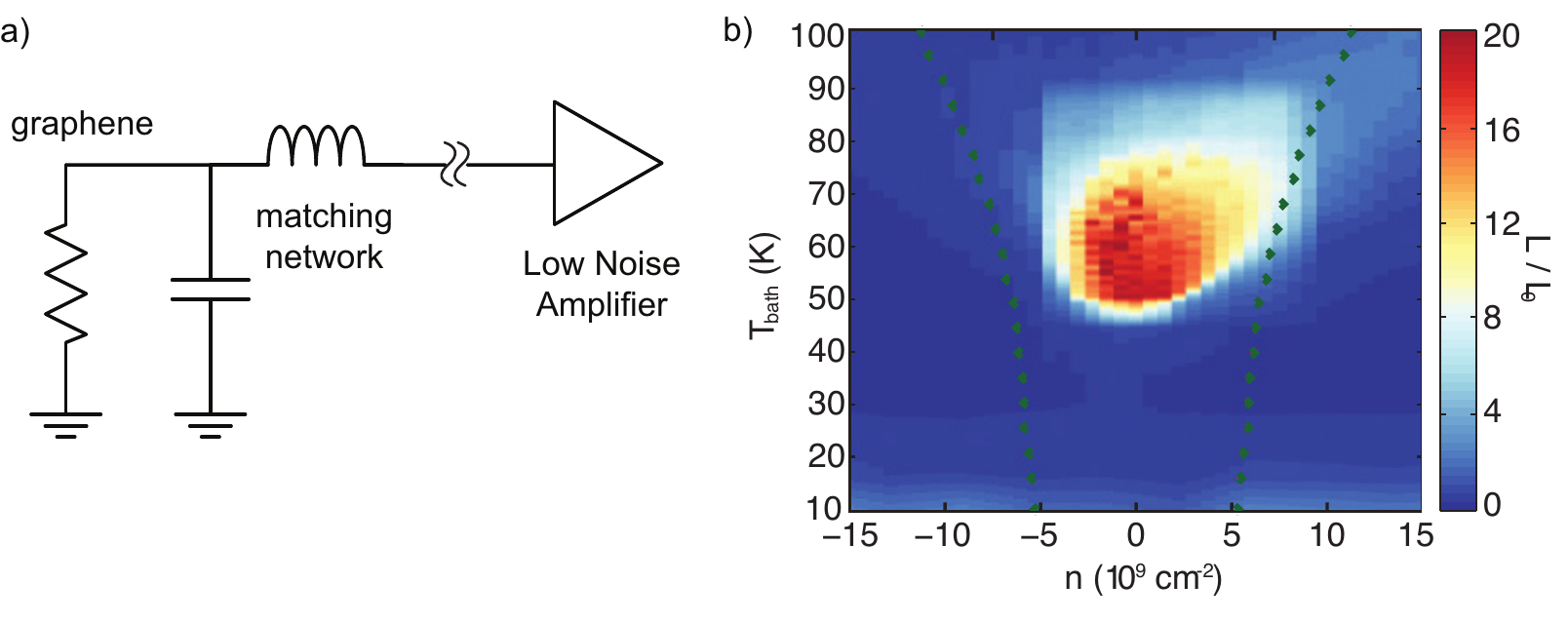}
\caption{(a) The core components of the experimental setup to measure the electronic thermal conductivity in graphene. (b) Measured Lorenz number as function of temperature and carrier density in graphene.  The dramatic peak in the Lorenz number measured at charge neutrality and intermediate temperatures was a direct prediction of hydrodynamics (\ref{eq:Lhydro}).   Figures taken from \cite{crossno} with permission.}
\label{fig:WFexpt}
\end{figure}

Testing the WF law is not always easy.  Implicit in the standard theory is the assumption that the dominant contribution to $\kappa$ is electronic.   In many materials, including graphene, electrons contribute $\lesssim 1\%$ of the total $\kappa$ as measured at room temperature: the dominant contribution arises from  phonons.  

We will now describe a technique that allows for the direct measurement of only the electronic contributions to $\kappa$.   The thermal conductivity can be measured based on Fourier's law, by taking the ratio of the input heating power to the temperature change of the electrons.\footnote{Technically one has to be careful about the boundary conditions because we are dealing with thermoelectric transport.  One can confirm that under experimental boundary conditions the techniques described below measure $\kappa$, the open-circuit thermal conductivity \cite{crossno}.}   To overcome limitations on the sensitivity of resistive thermometry in graphene \cite{yigen}, a Johnson noise radiometer was developed \cite{Fong:2012ut, Fong:2013hl}. Figure \ref{fig:WFexpt}a shows the schematic diagram of the core of the experimental setup. It is analogous to the microwave radiometer used to measure the temperature of the cosmic microwave background \cite{dicke}.  While in the astrophysical setting one measures temperature by collecting blackbody radiation from space with an antenna,  in graphene, \cite{Fong:2012ut, Fong:2013hl} employ a reactive impedance matching network with a center frequency in the microwave range.  The Johnson noise across the resistor, which is directly proportional to the temperature \cite{johnsonnoise}, is then measured. The coupling of the Johnson noise and the noise from the low noise amplifier determines the sensitivity of the temperature measurement,
which is often $\sim 1$ mK.
The first application of this Johnson noise thermometry was to understand electron-phonon coupling in graphene \cite{Fong:2012ut, Fong:2013hl, crossno2}, and to identify the regimes where such effects are small.

Figure \ref{fig:WFexpt}b plots the measured Lorenz number $\mathcal{L}$ in graphene as function of temperature and carrier density \cite{crossno}. The data is normalized to $\mathcal{L}_{\mathrm{WF}}$:  in much of the plot, $\mathcal{L}/\mathcal{L}_{\mathrm{WF}} \approx 1$ means the WF law is satisfied.   This law is always satisfied away from the charge neutrality point, as was mostly expected -- graphene is often a conventional Fermi liquid away from the neutrality point.   The quantitative agreement also confirms the accuracy of the Johnson noise thermometer.   
At low temperatures below 50 K, the WF law also holds well at the charge neutrality point.  We expect that the reason for this is that the local fluctuations in the chemical potential due to charge puddles are of order 50 K (in suitable units) \cite{crossno, lucas3}.   Raising the temperature, however, the Lorenz number is enhanced by a factor of more than 20 at temperature around 60 K at neutrality.  This is the signature of the hydrodynamic Dirac fluid;  the physics behind this effect was discussed below (\ref{eq:Lhydro}).
Above 100 K, experimental data shows that heat transfer from electrons to phonons is comparable to the electronic diffusion \cite{Fong:2012ut, Fong:2013hl, crossno2}.  This electron-phonon scattering degrades the Dirac fluid and so the decrease of $\mathcal{L}$ at $T\sim 100$ K is not surprising.   

\begin{figure}
\centering
\includegraphics{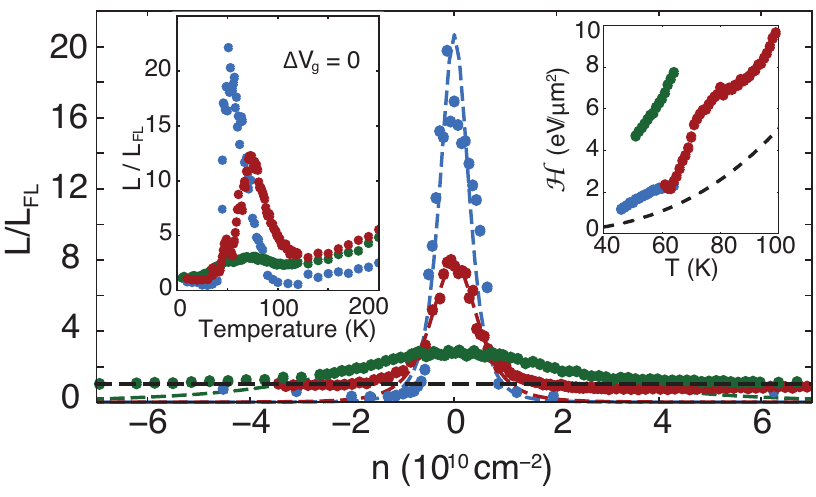}
\caption{The measured Lorenz number from three samples at $T = $60 K as function of charge density $n$. The data fits well to (\ref{eq:Lhydro}) (dashed lines). Data in blue, red, and green are from samples with increasing amounts of charge puddles. All samples return to the Fermi liquid value (black dashed line) at high density. Insets show (left) the measured Lorenz number as function temperatures and (right) the fitted enthalpy density $\epsilon+P$ as a function of temperature,  compared to the theoretical value in clean graphene (black dashed line).  Figure taken from \cite{crossno} with permission.}
\label{fig:WFFit}
\end{figure}

Figure \ref{fig:WFFit} shows a more quantitative comparison of the measured thermal conductivity to (\ref{eq:Lhydro}).   For the cleanest sample, the extracted momentum relaxation lengths are on the order of 1 $\mmu$m. This is comparable to the size of the charge puddles that can be probed directly using scanning probe microscopes \cite{xue, crommie}.   This suggests that these charge puddles will ultimately be the main obstacle to observing the Dirac fluid in futuer studies.   Away from neutrality, $\mathcal{L}$ drops sharply from its enhanced value to $\sim\mathcal{L}_{\mathrm{WF}}/4$, before rising back up to $\mathcal{L}_{\mathrm{WF}}$ at a higher density.  This suppression at finite carrier density $n$ is consistent with (\ref{eq:Lhydro}) and also the theory of \cite{foster2, vignale}.  

\subsection{Experimental Measurement of Thermoelectric Conductivity}
Another window into the hydrodynamic regime arises in thermoelectric transport -- in particular, by studying the cross-coefficient $\alpha_{ij}$ in (\ref{eq:thermoelectric}).  In this subsection, we will assume that $\alpha_{ij} = \alpha \mdelta_{ij}$ is isotropic.   In experiment, one often measures not $\alpha$ but the Seebeck coefficient \begin{equation}
\mathcal{S} = -\frac{\mathrm{\Delta} V}{\mathrm{\Delta} T} = \frac{\alpha}{\sigma}.
\end{equation}
In the hydrodynamic limit, we predict  \begin{equation}
\mathcal{S} = \frac{s}{en},  \label{eq:Shydro}
\end{equation} 
using (\ref{eq:7drude}).  This formula is very different from the Fermi liquid result (Mott relation) \cite{ashcroft} 
\begin{equation}
\mathcal{S} = -\frac{\mpi^2 k_{\mathrm{B}}^2T}{3e\sigma} \frac{\mathrm{d}\sigma}{\mathrm{d}\mu},  \label{eq:mott}
\end{equation}
and so again provides a specific, experimentally testable prediction of hydrodynamics.  In particular, (\ref{eq:Shydro}) diverges as $n\rightarrow 0$, and is expressable in terms of thermodynamic coefficients.  Such a simple formula is a consequence of the ``mean field" treatment of disorder in Section \ref{sec:MFdisorder}, but we expect it to be qualitatively reasonable.

\begin{figure}[t]
\centering
\includegraphics{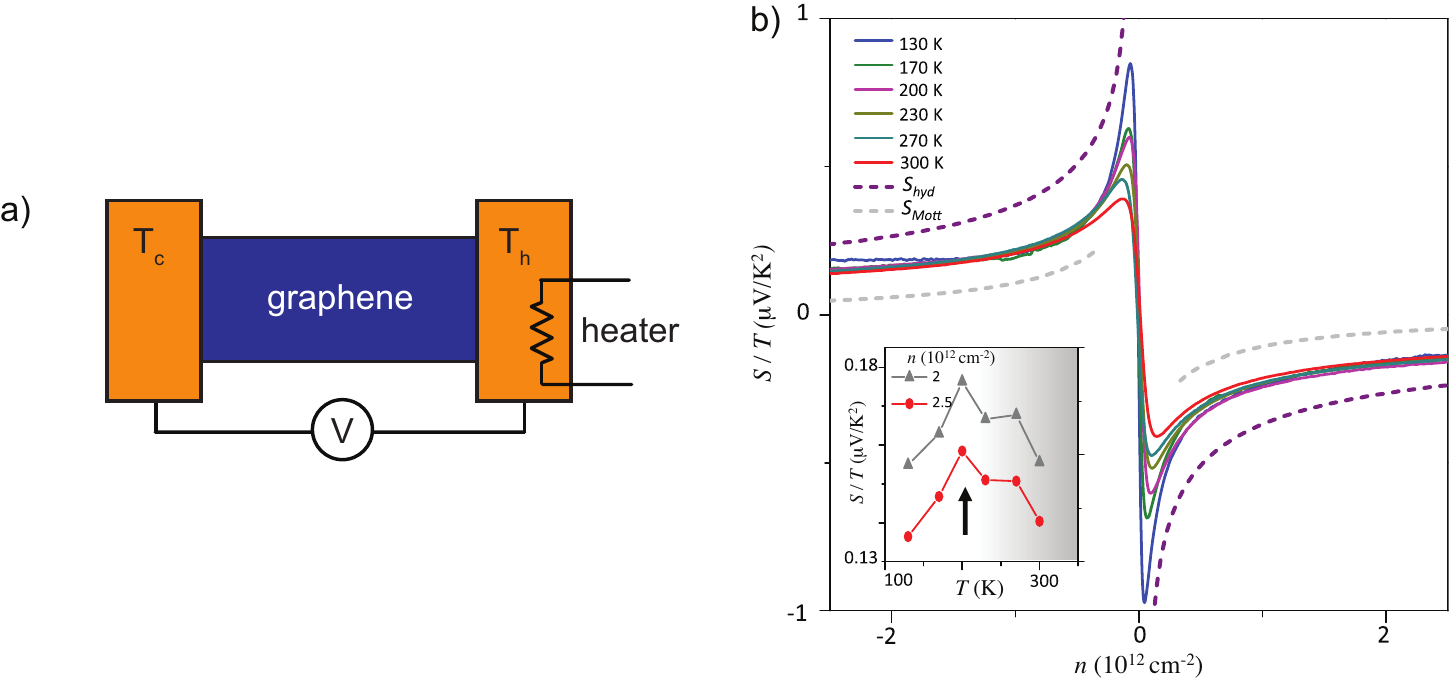}
\caption{(a) Schematic diagram of the thermoelectric power measurement in graphene.  Induced voltage is measured with a temperature gradient produced by a heater and controlled by resistive thermometers on the two ends of the sample. (b) Measured Seebeck coefficient-to-temperature ratio (solid lines) as function of carrier density in comparison to the hydrodynamic and Mott relation (\ref{eq:mott}).   Figure taken from \cite{ghahari} with permission.}
\label{fig:MottExpt}
\end{figure}

Figure \ref{fig:MottExpt}a shows the schematic diagram of the experimental setup used to measure $\mathcal{S}$ in \cite{ghahari}. The induced voltage is measured after a heater, applied to one end of the sample, generates a steady temperature gradient.   Figure \ref{fig:MottExpt}b shows the experimentally observed enhacement of $\mathcal{S}$ near the neutrality point is larger than predicted by (\ref{eq:mott}).   The enhancement diminishes as the system crosses over to the Fermi liquid regime at large $n$, or with increasing disorder or phonon scattering.  Indeed, early experiments with impurity-scattering-limited graphene on $\mathrm{SiO}_2$ substrates show excellent agreement with (\ref{eq:mott}) \cite{seebeck1, seebeck2, seebeck3}.  However, using the higher quality BN substrated samples, noticable deviations from (\ref{eq:mott}) may be observed \cite{ghahari}.

Using kinetic theory, \cite{foster2} attributes the discrepancy of $\mathcal{S}$ from both (\ref{eq:Shydro}) and (\ref{eq:mott}) as a consequence of both the crossover between ballistic and hydrodynamic regimes of transport, together with the scattering off of optical phonons at higher temperatures.   The temperature range where the optical phonon scattering occurs is consistent to the heat transfer measurement from graphene electrons to phonons \cite{crossno2}.

\subsection{Beyond Relativistic Hydrodynamics}
In this subsection, we describe some of the complications that can arise for the models of hydrodynamic transport in (weakly) interacting electron fluids.
\subsubsection{Imbalance Modes}
As we noted in Section \ref{sec:imbalance}, there can be a long lived imbalance mode in graphene.   \cite{foster} noted that this can lead to finite size corrections to the  theory of transport derived in Section \ref{sec:MFdisorder}.   The precise form of these corrections depends on details of the boundary conditions, but we can qualitatively understand the effects of an imbalance mode as follows.   Let us assume a homogeneous sample with momentum relaxation time $\tau_{\mathrm{imp}}$ and imbalance relaxation time $\tau_{\mathrm{imb}}$, with boundaries located at $x=0,L$.  The equations one must solve then  take the schematic form of \begin{subequations}\begin{align}
J^\prime &= 0, \\
J_{\mathrm{i}}^{\prime} &= -\frac{n_{\mathrm{imb}}-n_{\mathrm{imb,eq}}}{\tau_{\mathrm{imb}}}, \\
\left[Ts v(x) -\frac{\mu^a}{T} \sigma_{\textsc{q}}^{ab} \left(E^b  - \mu^{b\prime} - \frac{\mu^b}{T}T^\prime \right)\right]^\prime &= 0, \\
\rho^a (E^a  - \mu^{a\prime}) - s T^\prime &= \frac{(\epsilon+P)v}{\tau_{\mathrm{imp}}}
\end{align}\end{subequations}
where primes denote $x$-derivatives and \begin{equation}
J^a(x) = n^a v(x) + \sigma_{\textsc{q}}^{ab} \left(E^b  - \mu^{b\prime} - \frac{\mu^b}{T}T^\prime \right).
\end{equation}
For simplicity, we have neglected viscous effects above.   The exact solution to the equations above is sensitive to the boundary conditions on $n$ and $n_{\mathrm{imb}}$.  However, typical boundary conditions will all share the following features.   Firstly, in the limit $\tau_{\mathrm{imb}} \rightarrow \infty$, these equations are solved by constant $v$ as well as constant $\mu^{a\prime}$ and $T^\prime$ -- gradients are homogeneous throughout the sample.   However, because the imbalance mode decays, the non-equilibrium imbalance gradient will be concentrated near the edges of the sample.   The length scale over which the imbalance mode will decay obeys \begin{equation}
\ell_{\mathrm{imb}} \le \sqrt{\sigma^{\mathrm{imb}}_{\textsc{q}} \tau_{\mathrm{imb}} \frac{\partial \mu_{\mathrm{imb}}}{\partial n_{\mathrm{imb}}}}.
\end{equation}
We have employed slightly different notation than \cite{foster}.   What one then finds is that the electrical and thermal resistivities of the slab of length $L$ can be written in the following schematic form:  \begin{equation}
R = \mathcal{R} L + R_{\mathrm{contact}} \tanh \frac{L}{2\ell_{\mathrm{imb}}},
\end{equation}
where $\mathcal{R}$ is the resistance per unit length of the infinite sample (where imbalance modes do not play any role),  and $R_{\mathrm{contact}}$ is a finite contribution arising from imbalance modes that are excited near the contacts.

These imbalance modes limit the extent to which the large violations of the Wiedemann-Franz law described in Section \ref{sec:exptherm}, can be observed.   Indeed, the fact that such a large violation of the Wiedemann-Franz law was observed experimentally in \cite{crossno} suggests that $\ell_{\mathrm{imb}} < 1 \; \mmu \mathrm{m}$.

Possible nonlinear hydrodynamic signatures of the imbalance mode in graphene are discussed in \cite{mirlin2015}.  Observable signatures of an imbalance mode appear related to the presence of an extra diffusive hydrodynamic mode.   An imbalance mode can also lead to changes to the theory of transport through charge puddles \cite{hartnoll1704, hartnoll1705}, although we expect such effects to be less significant in the Dirac fluid.

A final perspective on imbalance modes can be found in \cite{sangjin}.   The authors consider an exotic model for the Dirac fluid coming from the AdS/CFT correspondence.   The model considered has the flavor of a ``momentum relaxation time" model, but with a different assumption on the thermodynamics of the imbalance mode than \cite{foster}, which allows this mode to couple to charge transport beyond the edges of the sample.   It remains an open question whether such a model is appropriate for the Dirac fluid realized in experiment.

\subsubsection{Bipolar Diffusion}

Another possible complication of the hydrodynamic description, is the possibility that the electron and hole fluids essentially decouple.   In the hydrodynamic langauge, this corresponds to the assumption that the charge/energy of the electrons/holes are \emph{separately conserved}.   The assumption that the energies of the two fluids are separately conserved is difficult to microscopically justify away from a non-interacting limit, but from a phenomenological, hydrodynamic perspective, we can take this as a postulate.   This decoupling of electron and hole fluids leads to a phenomenon called bipolar diffusion  \cite{goldsmid}, which also leads to an  enhancement of $\mathcal{L}$ relative to the WF law.

It is simple to derive the effect.  The conventional open-circuit thermal conductivity $\kappa$ was defined under the assumption that no charge current flows.   However, if there are electron and hole fluids, then we find \begin{align}
\kappa &= \bar\kappa_{\mathrm{e}} + \bar\kappa_{\mathrm{h}} - \frac{T(\alpha_{\mathrm{e}}+\alpha_{\mathrm{h}})^2}{\sigma_{\mathrm{e}}+\sigma_{\mathrm{h}}} = \left(\bar\kappa_{\mathrm{e}} - \frac{T\alpha^2_{\mathrm{e}}}{\sigma_{\mathrm{e}}}\right) + \left(\bar\kappa_{\mathrm{h}} - \frac{T\alpha^2_{\mathrm{h}}}{\sigma_{\mathrm{h}}}\right) + \frac{T\sigma_{\mathrm{e}}\sigma_{\mathrm{h}}}{\sigma_{\mathrm{e}}+\sigma_{\mathrm{h}}}\left(\frac{\alpha_{\mathrm{e}}}{\sigma_{\mathrm{e}}} - \frac{\alpha_{\mathrm{h}}}{\sigma_{\mathrm{h}}} \right)^2 \notag \\
&= \kappa_{\mathrm{e}}+\kappa_{\mathrm{h}}+ \frac{T\sigma_{\mathrm{e}}\sigma_{\mathrm{h}}}{\sigma_{\mathrm{e}}+\sigma_{\mathrm{h}}}\left(\frac{\alpha_{\mathrm{e}}}{\sigma_{\mathrm{e}}} - \frac{\alpha_{\mathrm{h}}}{\sigma_{\mathrm{h}}} \right)^2
\end{align}
If $\sigma=\sigma_{\mathrm{e}}+\sigma_{\mathrm{h}}$, and the electron/hole fluids separately obey the WF law,  the combined electron-hole fluid need not obey the WF law due to the presence of the third ``bipolar diffusion" contribution to $\kappa$.   In particular, the presence of the bipolar diffusion term suggests that $\mathcal{L}\ge 1$.

\begin{figure}
\centering
\includegraphics{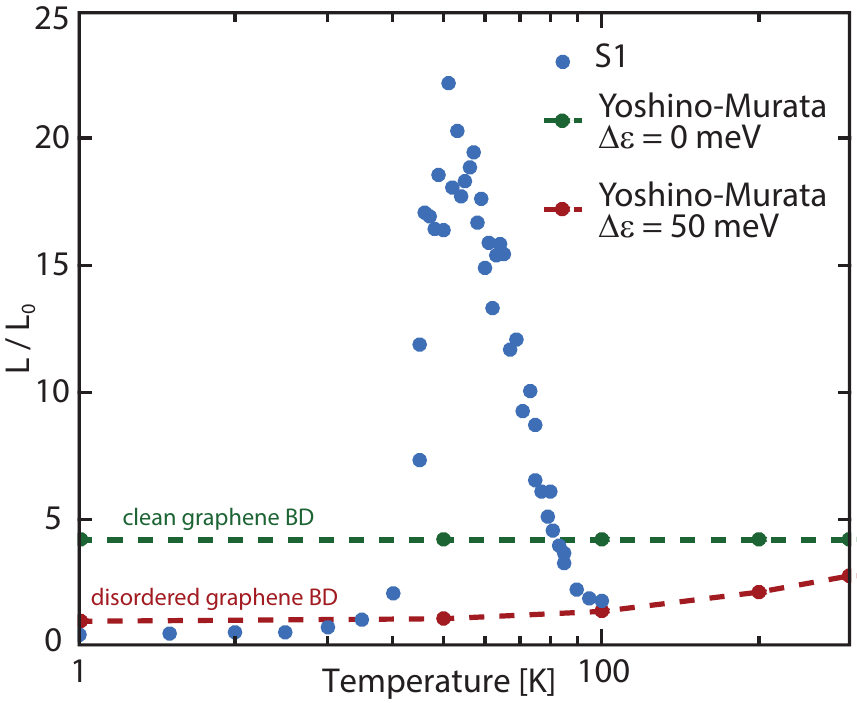}
\caption{The theory of bipolar diffusion in graphene \cite{yoshino} does not describe the experimentally measured $\kappa$.  This data is instead consistent with a hydrodynamic transport theory: see Figure \ref{fig:WFFit}.  Figure taken from \cite{crossno} with permission.}
\label{fig:LorenzComparison}
\end{figure}

The bipolar diffusion effect is \emph{not} adequate to explain the experimentally observed phenomena in graphene.  Let us summarize the reasons why \cite{crossno} (see Figure \ref{fig:LorenzComparison}):  (\emph{i}) the Lorenz number associated with  bipolar diffusion in disorder-free graphene is about 4.2$\mathcal{L}_{\mathrm{WF}}$ at the Dirac point \cite{yoshino}, and significantly smaller than what is seen experimentally;  (\emph{ii}) $\mathcal{L}$ is weakly dependent on temperature $T$ in a theory of bipolar diffusion, but also a monotonic function of $T$, contradictory to experimental data;  (\emph{iii}) the theory of bipolar diffusion predicts disorder amplitudes an order of magnitude larger than experimentally observed, given the approximate observation of the WF law for $T<50$ K.   Instead, thermal transport in graphene is described by a Dirac fluid, possibly with an electron-hole imbalance mode.

\subsection{Magnetic Fields}\label{sec:magnetic}
In this final subsection, we briefly discuss hydrodynamic magnetotransport phenomena.   These are most interesting near the Dirac point in graphene.

We first begin by extending the model of Section \ref{sec:MFdisorder} to include the effects of a magnetic field, following \cite{hkms}.   A magnetic field is introduced by adding a uniform background magnetic field in the external electromagnetic tensor $F^{\mu\nu}$ introduced in Section \ref{sec:extF}:  $F^{xy} = -F^{yx}=B$.   In the presence of a background magnetic field, (\ref{eq:71mom}) generalizes to include the magnetic Lorentz force: \begin{equation}
\frac{\epsilon+P}{v_{\mathrm{F}}^2}\left(\frac{1}{\tau_{\mathrm{imp}}} - \mathrm{i}\omega\right) v_i =  nE_i + s\partial_i T   + B\epsilon_{ij}J_j,
\end{equation}
where $J_j$ is the charge current and $\epsilon_{ij}$ is the Levi-Civita tensor:  $\epsilon_{xx}=\epsilon_{yy} =0$, $\epsilon_{xy}=-\epsilon_{yx}=1$.   The charge current is modified from (\ref{eq:71charge}) to \begin{equation}
J^i = \sigma_{\textsc{q}} \left(E^i + B\epsilon^{ij}v_j - \frac{\mu}{T}\partial_i T\right) + nv^i.
\end{equation} 
To see this equation, note that $F^{i\nu}u_\nu = F^{it}u_t + F^{ij}u_j = E^i + B\epsilon^{ij}v_j $.  It is straightforward to combine these two equations to compute the thermoelectric conductivity matrix, within the relaxation time approximation.  For simplicity we write down the components of $\sigma_{ij}$ alone: \begin{subequations}\label{eq:hallsigmaQ}\begin{align}
\sigma_{xx} = \sigma_{yy} &=   \dfrac{\displaystyle \frac{\epsilon+P}{v_{\mathrm{F}}^2} \left(\frac{1}{\tau_{\mathrm{imp}}} - \mathrm{i}\omega\right) +  n^2 +  B^2\sigma_{\textsc{q}}^2}{\displaystyle n^2B^2 + \left[\frac{\epsilon+P}{v_{\mathrm{F}}^2}\left(\frac{1}{\tau_{\mathrm{imp}}} - \mathrm{i}\omega\right) + B^2\sigma_{\textsc{q}}\right]^2 }  \frac{\epsilon+P}{v_{\mathrm{F}}^2} \left(\frac{1}{\tau_{\mathrm{imp}}} - \mathrm{i}\omega\right), \\
\sigma_{xy} = -\sigma_{yx} &= \dfrac{\displaystyle 2\frac{\epsilon+P}{v_{\mathrm{F}}^2} \left(\frac{1}{\tau_{\mathrm{imp}}} - \mathrm{i}\omega\right) + n^2 +  B^2\sigma_{\textsc{q}}^2}{\displaystyle n^2B^2 + \left[\frac{\epsilon+P}{v_{\mathrm{F}}^2}\left(\frac{1}{\tau_{\mathrm{imp}}} - \mathrm{i}\omega\right) + B^2\sigma_{\textsc{q}}\right]^2 }  Bn.
\end{align}\end{subequations}

When $\sigma_{\textsc{q}} \rightarrow 0$,  these equations can be derived rigorously \cite{lucasMM, lucasrmp} in the limit of perturbatively weak disorder.\footnote{In a certain ``cartoon" limit, these equations can also be derived at finite $\sigma_{\textsc{q}}$ \cite{lucasMM, lucasrmp}.  However, the derivation of these equations as $\sigma_{\textsc{q}}\rightarrow 0$ is generic and valid for general quantum systems, not just graphene.}   If we assume their validity for finite $\sigma_{\textsc{q}}$,  then we predict a number of novel new phenomena.  In particular, note that there are poles (divergences) in all components of $\sigma_{ij}$ whenever \begin{equation}
\hbar \omega = \pm \frac{en B v_{\mathrm{F}}^2}{\epsilon+P}  - \frac{\mathrm{i}\hbar}{\tau_{\mathrm{imp}}} - \mathrm{i} \frac{v_{\mathrm{F}}^2\sigma_{\textsc{q}}B^2}{\epsilon+P}.
\end{equation}
If $\tau_{\mathrm{imp}}^{-1}=0$ and $\sigma_{\textsc{q}}=0$,   we observe the presence of poles on the real axis.   These are called cyclotron resonances -- their presence is guaranteed in a Galilean-invariant fluid by Kohn's theorem \cite{kohn}.  With Galilean invariance, the charge current is proportional to the momentum density, and so the cyclotron resonance simply denotes the rotation of the charge current in a uniform magnetic field, guaranteed by the Ward identity (\ref{eq:dTFJ}).
Although the derivation of these phenomena was not rigorous, we expect that they are qualitatively correct.  Can they be experimentally observed?  $\sigma_{\textsc{q}}$ is expected to be largest, relative to other coefficients, at the charge neutrality point (Section \ref{sec:kineticcoeff}), which will be accessible for $\mu/k_{\mathrm{B}} \lesssim 100$ K.   In this regime, we find that \begin{equation}
\frac{en B v_{\mathrm{F}}^2}{\epsilon+P}  \sim  10^{14} \;  \mathrm{Hz} \times \frac{B}{1\; \mathrm{T}}.
\end{equation}  
Cyclotron resonances will occur for $\mathrm{Re}(\omega) \sim 100$ GHz at $B= 1 $ mT, which is an extremely weak magnetic field (the Earth's magnetic field is $10^{-5}$ T).   Similarly, one finds \begin{equation}
\frac{v_{\mathrm{F}}^2\sigma_{\textsc{q}}B^2}{\hbar(\epsilon+P)} \sim 10^{14} \;  \mathrm{Hz} \times \left(\frac{B}{1\; \mathrm{T}}\right)^2,
\end{equation}
and for $B\sim1$ mT,  this suggests that $\mathrm{Im}(\omega) \sim 10^{-3} \mathrm{Re}(\omega)$.  This will likely be very hard to detect in graphene -- in particular the momentum relaxation rate will likely be significantly larger.

We also note that the ratio $\sigma_{xy}/\sigma_{xx}$, as computed from (\ref{eq:hallsigmaQ}), can exhibit novel scaling with temperature $T$ \cite{blakedonos}, which could shed light on experiments in the cuprates \cite{hallcuprate}.

Another effect of a finite magnetic field which is missed by the ``mean field" treatment of disorder is Hall viscosity \cite{avron, yarom}.  The Hall viscosity is a non-dissipative modification of $\widehat{T}^{\mu\nu}$, as given in (\ref{eq:widehatTmunu}): $\widehat{T}^{\mu\nu} \rightarrow\widehat{T}^{\mu\nu}  +  \widehat{T}^{\mu\nu}_{\mathrm{H}}$ where \begin{equation}
\widehat{T}^{\mu\nu}_{\mathrm{H}} = -\frac{\eta_{\mathrm{H}} }{2}\left(\epsilon^{\mu\alpha\rho} u_\alpha {\sigma_\rho}^\nu + \epsilon^{\nu\alpha\rho} u_\alpha {\sigma_\rho}^\mu \right),
\end{equation}
where $\sigma^{\mu\nu} = \mathcal{P}^{\alpha\mu} \mathcal{P}^{\beta \nu} (\partial_\alpha u_\beta + \partial_\beta u_\alpha - \eta_{\alpha \beta} \partial_\lambda u^\lambda)$.   Possible experimental signatures of this Hall viscosity have been proposed in \cite{scaffidi, delacretaz, polinihall}.   We also note that $\sigma_{\textsc{q}}$ can be generalized to contain an intrinsic Hall conductivity \cite{yarom}.

Magnetic fields also lead to qualitative changes to transport through inhomogeneous puddles.   Using the same long wavelength limit of Section \ref{sec:hydrotrans}, but now supposing that the magnetic field $B$ is not perturbatively small,  one finds that the momentum relaxation time becomes \cite{levchenko1, levchenko2} \begin{equation}
\frac{1}{\tau_{\mathrm{imp}}} \sim \frac{B^2}{\eta} \log \frac{L}{\xi},
\end{equation}
where $L$ is the sample size and $\xi$ is the size of the charge puddles.    This result is relatively insensitive to the precise details of the hydrodynamics (in contrast to (\ref{eq:hydrotautime})), and depends only on the fact that the electron fluid in graphene is two-dimensional. Strictly speaking, $\tau_{\mathrm{imp}}$ becomes so short as $L\rightarrow \infty$ that the momentum relaxation time approximation itself fails.  However, the prefactor of the logarithm is inversely proportional to $\eta$.   Is studying dissipative transport in magnetic fields  an effective way of measuring the viscosity of an electron fluid?  We caution that this effect is not present in three spatial dimensions \cite{baumgartner}, and may be less well suited for electron fluids other than graphene.

\section{Coulomb Drag}
\label{sec:drag}
So far we have looked for the signatures of electron-electron interactions directly from experiments on monolayer graphene.   One way to possibly probe electron interactions more directly is by separating two monolayers of graphene by a few layers of insulator, such as boron nitride.  Ensuring that the monolayers are not in electrical contact, we then ask for the currents/voltages in layer 1 due to a current/voltage induced in layer 2.  This cross layer signal is believed to be dominated by Coulomb interactions between electrons in the different layers, and so is coined ``Coulomb drag" \cite{dragreview}.       These experiments can also be performed on any effectively two-dimensional electron system.   In graphene, Coulomb drag is noticable for interlayer spacings $\lesssim 10$ nm, which can easily be achieved in heterostructures.

\subsection{Hydrodynamic Description}\label{sec:drag1}

In some respects, Coulomb drag physics could be expected to be quite similar to the imbalance mode physics described in Section \ref{sec:imbalance}.    For example, if there is a Fermi liquid in both monolayers, then (neglecting energy conservation) we expect only the total momentum of electrons in both layers to be conserved, together with the charge density of electrons in each layer separately.   Assuming linearized, time-dependent flows, one then writes down a set of coupled hydrodynamic equations \begin{subequations}\label{eq:8imb}\begin{align}
\partial_i \left(n^a v^i - \Sigma^{ab} \partial^i \mu_b\right) &= 0, \\
n^a \partial_i \mu^a - \partial^j \left(\eta \left(\partial_i v_j + \partial_j v_i\right)  + (\zeta-\eta) \mdelta_{ij}\partial_k v^k\right) &= n^a E^a_i,
\end{align}\end{subequations}
where $a=1,2$ denotes the electrical response in each monolayer, while $v^i$ denotes the collective fluid velocity.   The physics is then identical to the imbalance mode physics described previously, including the existence of an extra diffusive mode (see also \cite{hartnoll1704, hartnoll1705}).

However, it is more common in the literature to treat this set-up as consisting of two fluids with two long lived momenta.    We consider a theory of two coupled charged fluids with charge current $J^{\mu 1,2}$ and energy-momentum current $T^{\mu\nu 1,2}$:  \begin{subequations}\label{eq:2fluid}\begin{align}
\partial_\mu J^{\mu 1} &= 0, \\
\partial_\mu J^{\mu 2} &= 0, \\
\partial_\mu T^{\mu \nu 1} &= F^{\mu\nu1}J^1_\mu - \frac{T^{t\nu 1} - T^{t\nu 2}}{\tau_{\mathrm{d}}} , \\
\partial_\mu T^{\mu \nu 2} &= F^{\mu\nu2}J^2_\mu - \frac{T^{t\nu 1} - T^{t\nu 2}}{\tau_{\mathrm{d}}}.
\end{align}\end{subequations}
Analogous to our treatment of momentum relaxation in Section \ref{sec:dis4},  $\tau_{\mathrm{d}}$ is the relaxation rate of energy and momentum between the two layers, and will increase as the separation between the monolayers increases.  We will only rely on these equations to linear order in the velocity/momentum of each fluid. 

\subsection{Coulomb Drag and Transport}
\begin{figure}
\centering
\includegraphics[width=2.6in]{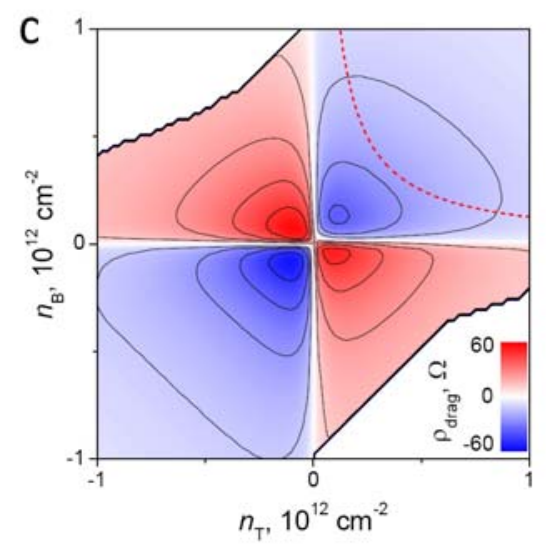} 
\hspace{0.5in}
\includegraphics[width=2.6in]{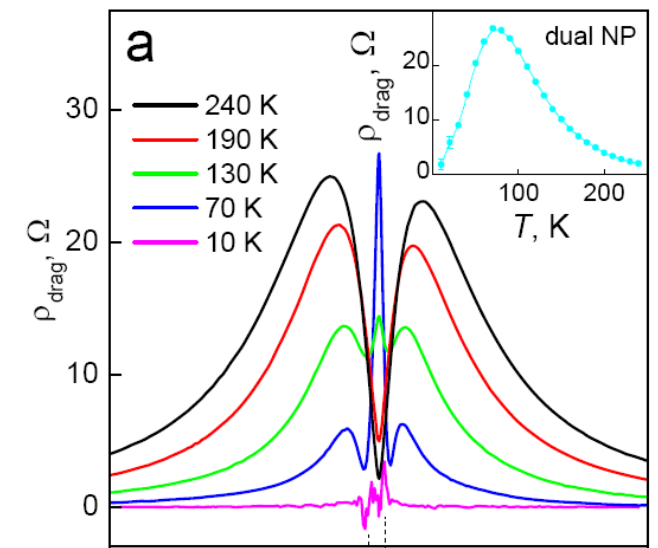}
\caption{Left: the Drag resistivity $\rho^{\mathrm{D}}$ at the charge neutrality point in graphene.  The sign reversals (denoted by the red vs. blue shading) can be understood simply from (\ref{eq:simpledrag}).   Right: a clearer view of the magnitude of $\rho_{\mathrm{d}}$ as a function of  the density $n$, at various temperatures.  At lower temperatures there is a very sharp enhancement of $\rho_{\mathrm{d}}$.   Figure adapted from \cite{gorbachev1206} with permission.}
\label{fig:dragexp}
\end{figure}

We now turn to the transport signatures of the fluid-like models above.   The most common measurement is of the drag resistivity \begin{equation}
E_i^2 = \rho_{\mathrm{d}} J^1_i.
\end{equation}
It tells us the electrical response of layer 2 due to a current flowing in layer 1.  We can easily generalize the model of Section \ref{sec:MFdisorder} to the theory (\ref{eq:2fluid}), to get a flavor for what happens.   The momentum balance equations in each layer read 
\begin{equation}
\left(\begin{array}{c} n_1 E_1 \\ n_2 E_2 \end{array}\right) = \left(\begin{array}{cc} \Gamma + \Gamma_{\mathrm{d}} &\ -\Gamma_{\mathrm{d}} \\ -\Gamma_{\mathrm{d}} &\ \Gamma + \Gamma_{\mathrm{d}} \end{array}\right) \left(\begin{array}{c} v_1 \\ v_2 \end{array}\right),
\end{equation}
 where $\Gamma = (\epsilon+P)/\tau_{\mathrm{imp}}$ is related to the momentum relaxation time, and the charge current in each layer takes the form of (\ref{eq:71charge}).
 For convenience and simplicity we have assumed that $\Gamma$ and $\Gamma_{\mathrm{d}}$ take the same, simple form above for each layer, even if the layers are at different charge densities, and we have also (in this two-fluid approximation) not allowed for any inter-layer cross-terms in $\sigma_{\textsc{q}}$, which could generically appear  as in (\ref{eq:8imb}).  After a short calculation we obtain \begin{equation}
\rho_{\mathrm{d}} = - \frac{n_1n_2\Gamma_{\mathrm{d}}}{(\sigma_{\textsc{q}}^2 \Gamma(\Gamma+2\Gamma_{\mathrm{d}}) + (n_1^2 + n_2^2)(\Gamma+\Gamma_{\mathrm{d}})\sigma_{\textsc{q}} + n_1^2n_2^2}.  \label{eq:simpledrag}
\end{equation}
Thus we predict that (as in a conventional two-dimensional double-layer system \cite{dragreview}), $\rho_{\mathrm{d}}$ will grow quite large as we approach neutrality (either $n_1=0$ or $n_2=0$) from  high density.  However, right at the neutrality point,  (\ref{eq:simpledrag}) predicts $\rho_{\mathrm{d}}$ vanishes due to charge conjugation symmetry.   We predict that the sign of $\rho_{\mathrm{d}}$ can also be flipped depending on the relative sign of $n$ in the two layers.   Many of these qualitative features are observed in experiments on graphene \cite{gorbachev1206}:  see Figure \ref{fig:dragexp}. 

However, there is one key experimental observation missed by (\ref{eq:simpledrag}).   At relatively low temperatures, there is an extremely large peak in $\rho_{\mathrm{d}}$ observable in Figure \ref{fig:dragexp}.  It has been argued \cite{levitov1205, dho1611} that this peak arises from the inhomogeneous puddle landscape of the Dirac fluid.   In particular, in the presence of an (approximate) diffusive mode  -- either the relative energy density of the two layers (\ref{eq:8imb}), or electron/hole imbalances in a single layer (Section \ref{sec:imbalance}) -- we can estimate a contribution to $\rho_{\mathrm{d}}$:  \cite{levitov1205} \begin{equation}
\rho_{\mathrm{d}} \propto  (\sigma_{\textsc{q}}^{-1})^{1a} \langle n_{\mathrm{dis}}^a(\mathbf{x}) n_{\mathrm{dis}}^b(\mathbf{x})  \rangle_{\mathrm{dis}} (\sigma_{\textsc{q}}^{-1})^{b2}.  \label{eq:imbpuddle}
\end{equation}
A similar result was found in  \cite{levitov1205}.    
(\ref{eq:imbpuddle}) suggests that even at the neutrality point, the drag resistivity will be finite, and the sign will be sensitive to whether charge puddles typically have the same or opposite sign in the two layers \cite{levitov1205}.  The experiment \cite{gorbachev1206} finds that this sign is positive.   The precise temperature dependence of (\ref{eq:imbpuddle}) depends on the dominant diffusive mode.
We also stress that the  derivation of (\ref{eq:imbpuddle}) makes certain assumptions about the leading order inhomogeneity-induced contributions to transport that may not be rigorous in general hydrodynamic models.  It would be interesting to carry out the analysis of \cite{lucas3} to second order in perturbation theory and to rigorously confirm or correct (\ref{eq:imbpuddle}) in the presence of inhomogeneity, in the hydrodynamic approximation.   For other approaches to this problem, see \cite{apostolov, dho1611}.

Magnetotransport drag phenomena have also been studied theoretically \cite{levitov1303} and experimentally \cite{titov1303}.   The most non-trivial effect is the presence of a non-vanishing Hall ($xy$) component to the drag resistivity tensor,  which would be vanishing in a Drude-like limit \cite{dragreview}.   See \cite{levitov1303, titov1303} for more details on this limit.

The experiments described above took place before the observation of viscous hydrodynamics in graphene.  It would be interesting to understand whether Coulomb drag experiments on viscous samples can help probe hydrodynamics more directly.

\section{Outlook}
\label{sec:outlook}
There is a great deal of emerging interest in the study of the hydrodynamics of electron fluids.   For concreteness, this review has focused on a specific material, graphene,  although given the universality of hydrodynamics, much of what we have said is far more general.   In this outlook, we look forward to some important open questions which remain unsolved, and to the possible promise of the field.

\subsection{Pinning Down Hydrodynamics}
Perhaps the most immediate question which must be addressed is a more ``absolute" probe of hydrodynamic flow, even in the Fermi liquid phase of graphene.   One of the most striking signatures of viscous electron flow in the Fermi liquid is the formation of vortices.  Such flows may be indirectly observed through a combination of nonlocal voltage probes and  classical magnetotransport \cite{levitovjuly}.  A more direct observation of vortices may be possible using a variety of techniques to image nanoscale current flow and magnetic fields \cite{grinolds, vasyukov, benlev}.     Another important open problem is to directly measure the electronic viscosity.  We discussed proposals for such a measurement in Section \ref{sec:viscometry}.  In particular, two independent measurements of the viscosity may shed light onto the quantitative reliability of simple hydrodynamic models for electron flow.

At the moment, observing the onset of electronic hydrodynamics in graphene (or any other metal) already requires some of the highest quality materials capable of being grown.   This can make it challenging to quantitatively determine the limiting factors in the development of hydrodynamics:  short-range impurities, umklapp (whether in electron-electron or electron-phonon scattering), other long lived dynamical modes, etc.  A better understanding of these issues will help to prepare optimal crystals for electronic hydrodynamic flow.   

An even more spectacular hydrodynamic phenomenon is turbulence.  However, as we discussed in Section \ref{sec:dis4turb}, this is not likely to be accessible experimentally for the near future.    Although the estimates made in Section \ref{sec:dis4turb} assume the textbook Navier-Stokes theory of turbulence, valid in the Fermi liquid limit of Section \ref{sec:FLhydro}, we would not be surprised if further complications spoil measures of turbulence in non-Fermi liquids.  As an example, the enhancement of the Reynolds number at strong coupling in the Dirac fluid of graphene will be offset by the complications of thermal modes in hydrodynamics:   at the charge neutrality point, electrical measurements probe diffusive dynamics (Section \ref{sec:DFtrans}) while turbulence is dominated by the energy-momentum sector, which is hard to measure.

Finally, there may be exotic signatures of hydrodynamics that we have not yet understood.  For example, recent work has suggested the effects of hydrodynamic flow on nonlinear optical conductivity \cite{fogler1, fogler2} or electromagnetic penetration depth \cite{zaanen}.

\subsection{Hydrodynamics of Complicated Materials}
Hydrodynamics is a common language -- it allows us to describe the motion of interacting electrons in graphene, as well as the air and water which flow around us every day.   Yet we have also focused in large part on the challenges of understanding electronic hydrodynamics in graphene, in this review.   In particular, our derivation of hydrodynamics focused on the relativistic gradient expansion, appropriate for interacting quasirelativistic fermions,  and our discussion of kinetic theory almost exclusively focused on graphene, along with our discussions of phonons and impurities.

One simple reason for the emphasis on graphene is that this is a material where many of the seminal experiments on hydrodynamic electron flow have been performed, and the sharpest signatures of hydrodynamics have been observed.   To understand why graphene -- out of thousands of other materials -- was such a promising candidate for these experiments, it is important to understand in some detail the material-specific obstacles to conducting regimes dominated by electron-electron interactions.  For example, in the context of graphene it was crucial to reduce the inhomogeneous ``charge puddle" landscape in order to see hydrodynamic flow -- but it was also possible to do this.   In other materials, such as quantum critical metals at ``optimal doping", the ``random" chemical composition may make this purification impossible.  Not all materials will be amenable to simple signatures of viscous hydrodynamic flow such as negative non-local resistance.

Similarly, graphene has a particularly simple band structure with approximate rotational invariance.   It can be well approximated, for most purposes, by a small single circular Fermi surface (in the Fermi liquid regime) or by a relativistic electron-hole plasma (in the Dirac fluid regime).    However, many other materials of interest will not have such simple band structures.  They may have larger Fermi surfaces, with disconnected pieces, or which badly break rotational invariance.   These more complicated Fermi surface geometries can have a significant impact on the ballistic-to-hydrodynamic crossover \cite{hartnoll1704, hartnoll1705}, which is the regime where most realistic experiments take place.  It has even been observed that the ballistic-to-hydrodynamic crossover is not trivial, even for a Fermi liquid with a circular Fermi surface \cite{ledwith1, ledwith2}.    
An important open problem in this field is to understand the extent to which these complications modify the experimental signatures of hydrodynamics.   For example, as we discussed in Section \ref{sec:FLtrans}, one of the key signatures of viscous transport in a Fermi liquid is a decreasing electrical resistivity at low temperatures:  $\partial \rho /\partial T < 0$, under rather generic circumstances.    This has not been directly observed in bulk transport measurements, even on high-quality samples of correlated electron materials.   \cite{hartnoll1704, hartnoll1705} noted that the non-trivial Fermi surface structure of many complicated materials can ``short circuit" this viscous effect, and lead to a more conventional $\partial \rho /\partial T > 0$.   Random magnetic fields can cause viscous effects themselves to lead to $\rho \propto T^2$ \cite{lucasRFB}.   To determine what mechanism destroys the conventional viscous $\rho \propto T^{-2}$ scaling, it is important to develop material-specific models of the ballistic-to-hydrodynamic crossover, and understand how to experimentally confirm the presence of a more complicated hydrodynamics.

\subsection{Thermalization and the Emergence of Classical Physics}
In this review, we have mostly focused on the practical challenges with viewing hydrodynamics of electrons in a metal.  As described in the introduction, there is also interest in understanding how such a classical, dissipative, description arises from unitary microscopic quantum mechanics.  Conjectures \cite{hartnoll1, kss} that hydrodynamics would be fundamentally limited by the constraints of quantum mecahnics have inspired a large body of theoretical work (see \cite{lucasrmp} for a review).  Recently, it has been noted \cite{hartman, lucasbound} that consistency with the theory of quantum chaos places fundamental constraints on the hydrodynamics of any many-body quantum system.  In particular, for a typical experimentally realized quantum many-body system, the diffusion constants are bounded: \begin{equation}
D \le v_{\mathrm{B}}^2\tau  \label{eq:diffbound}
\end{equation}
where $v_{\mathrm{B}}$ is the butterfly velocity -- the speed of quantum chaos \cite{localized} --  and $\tau$ is the time scale beyond which the diffusion equation breaks down \cite{lucasbound}.  The earliest evidence for (\ref{eq:diffbound}) comes from \cite{blakeB1, blakeB2}.   The theory of quantum chaos tells us about how quantum systems thermalize, and  so a measurement of $v_{\mathrm{B}}$ would directly inform us about the efficiency with which quantum information is scrambled and lost, and classical physics emerges.   Unfortuntately, direct probes of quantum chaos are extremely challenging \cite{monika, grover, yao}.   But through (\ref{eq:diffbound}), a simple measurement of diffusion and its breakdown provides exact constraints on $v_{\mathrm{B}}$.   Experimental detection of ``rapid" quantum chaos, perhaps along the lines suggested above, could serve as sharp signatures for whether certain electronic systems are as strongly interacting as has been suggested.   

As a theoretical question, the emergence of thermalization in closed quantum systems is a fascinating problem.   Perhaps hydrodynamic probes of experimental systems will be embarassingly useful in giving insight, or at least constraints, on this physics.

\subsection{Nanoscale Viscous Electronics}
Some of the recent interest in electronic hydrodynamics has also arisen from the possibility of practical applications.   For instance, we observed in Section \ref{sec:FLtrans} that in simple Fermi liquids, viscous electron flow enhances conductance,  which could be useful in nanoscale environments requiring minimal dissipation.   Another possible application of hydrodynamics is the creation of high quality thermoelectric devices, which efficiently convert charge into heat current -- such devices employ the breakdown of the Wiedemann-Franz law observed in Section \ref{sec:exptherm}.   The low specific heat and rapid thermalization time of the Dirac fluid have also recently proven useful in building a single photon detector \cite{karasik, fong17}.

As we discussed in Section \ref{sec:viscometry}, the Dyakonov-Shur instability arises in the flow of a hydrodynamic electron liquid in one dimension.   This instability has long been sought after experimentally for practical purposes:  it may provide a route to the generation of THz radiation, which has proven quite challenging with other techniques \cite{kitaeva}.    A robust source of THz radiation could lead to breakthroughs in nanoscale imaging technology, with immediate medical, military and industrial applications.  Experimental evidence for the presence of this instability in ultra-clean, weakly interacting two-dimensional electron liquids such as silicon is rather lacking \cite{giliberti, tauk}, and we expect this is in no small part due to the challenge in observing the hydrodynamic limit of electron flow.

We caution the reader that the industrial applications of viscous electronic flow appear will likely not arise for some time.  Indeed, the extremely small scales necessary to observe hydrodynamic effects could very well spoil any practical application, or limit it to a very narrow set of devices.   Nevertheless, we believe that understanding of the hydrodynamic flow of correlated electrons remains a beautiful (and, perhaps, simple) outstanding problem in theoretical and experimental physics.       Given the convergence of a broad range of theoretical and experimental techniques and ideas, we predict the rapid advance of this field over the next few years.  Thus, we hope that this review serves as an invitation to an emerging area of physics, and not as a summary of a well-understood problem.  We will not be surprised if, in two decades, it is common knowledge that a hydrodynamic limit of electronic dynamics explains quite a few of the present day puzzles in electronic transport.

\addcontentsline{toc}{section}{Acknowledgements}
\section*{Acknowledgements}

We thank Sankar Das Sarma, Zhiyuan Sun and Dmitrii Svintsov   for comments on a draft of this review.   AL is supported by the Gordon and Betty Moore Foundation's EPiQS Initiative through Grant GBMF4302.
KCF is supported by Raytheon BBN Technologies.   The authors also acknowledge the hospitality of the Simons Center for Geometry and Physics during the writing of this review.

\bibliographystyle{unsrt}
\addcontentsline{toc}{section}{References}
\bibliography{soundbib}

\begin{thebibliography}{100}

\bibitem{ashcroft}
N.~W. Ashcroft and N.~D. Mermin.
\newblock \emph{Solid-State Physics}
  \href{http://www.amazon.com/Solid-State-Physics-Neil-Ashcroft/dp/0030839939/ref=sr_1_1?ie=UTF8&qid=1439625478&sr=8-1&keywords=ashcroft+and+mermin}{(Brooks
  Cole, 1976)}.

\bibitem{wallace}
P.~R. Wallace.
\newblock ``The band theory of graphite",
  \href{http://journals.aps.org/pr/abstract/10.1103/PhysRev.71.622}{\textsl{Physical
  Review} \textbf{71} 622 (1947)}.

\bibitem{geim2005}
K.~S. Novoselov, A.~K. Geim, S.~V. Morozov, D.~Jiang, M.~I. Katsnelson, I.~V.
  Grigorieva, S.~V. Dubonos, and A.~A. Firsov.
\newblock ``Two-dimensional gas of massless Dirac fermions in graphene",
  \href{http://www.nature.com/nature/journal/v438/n7065/full/nature04233.html}{\textsl{Nature}
  \textbf{438} 197 (2005)}.

\bibitem{kim2005}
Y.~Zhang, Y.~W. Tan, H.~L. Stormer, and P.~Kim.
\newblock ``Experimental observation of the quantum Hall effect and Berry's
  phase in graphene",
  \href{http://www.nature.com/nature/journal/v438/n7065/full/nature04235.html}{\textsl{Nature}
  \textbf{438} 201 (2005)}.

\bibitem{landau}
L.D. Landau and E.M. Lifshitz.
\newblock \emph{Fluid Mechanics}
  \href{https://www.amazon.com/Fluid-Mechanics-Second-Theoretical-Physics/dp/0750627670/ref=sr_1_1?ie=UTF8&qid=1478217116&sr=8-1&keywords=fluid+mechanics+landau}{(Butterworth
  Heinemann, $2^{\mathrm{nd}}$ ed., 1987)}.

\bibitem{kadanoff}
L.~P. Kadanoff and P.~C. Martin.
\newblock ``Hydrodynamic equations and correlation functions",
  \href{http://www.sciencedirect.com/science/article/pii/0003491663900782}{\textsl{Annals
  of Physics} \textbf{24} 419 (1963)}.

\bibitem{forster}
D.~Forster.
\newblock \emph{Hydrodynamic Fluctuations, Broken Symmetry and Correlation
  Functions}
  \href{http://www.amazon.com/Hydrodynamic-Fluctuations-Symmetry-Correlation-Functions/dp/0201410494/ref=sr_1_1?ie=UTF8&qid=1418231940&sr=8-1&keywords=hydrodynamic+fluctuations+broken+symmetry+and+correlation+functions&pebp=1418231941724}{(Perseus
  Books, 1975)}.

\bibitem{putterman}
S.~J. Putterman.
\newblock \emph{Superfluid Hydrodynamics},
  \href{https://www.amazon.com/Superfluid-hydrodynamics-North-Holland-temperature-physics/dp/0444106812}{(Elsevier,
  1974)}.

\bibitem{kamenev}
A.~Kamenev.
\newblock \emph{Field Theory of Non-Equilibrium Systems}
  \href{https://www.amazon.com/Field-Theory-Non-Equilibrium-Systems-Kamenev/dp/0521760828/ref=sr_1_1?ie=UTF8&qid=1478548419&sr=8-1&keywords=field+theory+of+non-equilibrium+systems}{(Cambridge
  University Press, 2011)}.

\bibitem{cao}
C.~Cao, E.~Elliott, J.~Joseph, H.~Wu, J.~Petricka, T.~Sch\"afer, and J.~E.
  Thomas.
\newblock ``Universal quantum viscosity in a unitary Fermi gas",
  \href{http://www.sciencemag.org/content/331/6013/58.full}{\textsl{Science}
  \textbf{331} 58 (2011)},
  \href{http://arxiv.org/abs/1007.2625}{\texttt{arXiv:1007.2625}}.

\bibitem{shuryak}
E.~Shuryak.
\newblock ``Physics of strongly coupled quark-gluon plasma",
  \href{http://www.sciencedirect.com/science/article/pii/S0146641008000732}{\textsl{Progress
  in Particle and Nuclear Physics} \textbf{62} 48 (2009)},
  \href{http://arxiv.org/abs/0807.3033}{\texttt{arXiv:0807.3033}}.

\bibitem{molenkamp2}
L.~W. Molenkamp and M.~J.~M. de~Jong.
\newblock ``Observation of Knudsen and Gurzhi transport regimes in a
  two-dimensional wire",
  \href{https://doi.org/10.1016/0038-1101(94)90244-5}{\textsl{Solid State
  Electronics} \textbf{37} 551 (1994)}.

\bibitem{molenkamp}
M.~J.~M. de~Jong and L.~W. Molenkamp.
\newblock ``Hydrodynamic electron flow in high-mobility wires",
  \href{http://journals.aps.org/prb/abstract/10.1103/PhysRevB.51.13389}{\textsl{Physical
  Review} \textbf{B51} 11389 (1995)},
  \href{http://arxiv.org/abs/cond-mat/9411067}{\texttt{arXiv:cond-mat/9411067}}.

\bibitem{bandurin}
D.~A.~Bandurin \emph{et al.}
\newblock ``Negative local resistance due to viscous electron backflow in
  graphene",
  \href{http://science.sciencemag.org/content/351/6277/1055}{\textsl{Science}
  \textbf{351} 1055 (2016)},
  \href{http://arxiv.org/abs/1509.04165}{\texttt{arXiv:1509.04165}}.

\bibitem{crossno}
J.~Crossno \emph{et al.}
\newblock ``Observation of the Dirac fluid and the breakdown of the
  Wiedemann-Franz law in graphene",
  \href{http://science.sciencemag.org/content/351/6277/1058}{\textsl{Science}
  \textbf{351} 1058 (2016)},
  \href{http://arxiv.org/abs/1509.04713}{\texttt{arXiv:1509.04713}}.

\bibitem{levitov1703}
R.~Krishna~Kumar \emph{et al}.
\newblock ``Super-ballistic flow of viscous electron fluid through graphene
  constrictions",
  \href{http://arxiv.org/abs/1703.06672}{\texttt{arXiv:1703.06672}}.

\bibitem{mackenzie}
P.~J.~W. Moll, P.~Kushwaha, N.~Nandi, B.~Schmidt, and A.~P. Mackenzie.
\newblock ``Evidence for hydrodynamic electron flow in $\mathrm{PdCoO}_2$",
  \href{http://science.sciencemag.org/content/351/6277/1061}{\textsl{Science}
  \textbf{351} 1061 (2016)},
  \href{http://arxiv.org/abs/1509.05691}{\texttt{arXiv:1509.05691}}.

\bibitem{felser}
J.~Gooth, F.~Menges, C.~Shekhar, V.~S\"uss, N.~Kumar, Y.~Sun, U.~Drechsler,
  R.~Zierold, C.~Felser, and B.~Gotsmann.
\newblock ``Electrical and thermal transport at the Planckian bound of
  dissipation in the hydrodynamic electron fluid of $\mathrm{WP}_2$",
  \href{http://arxiv.org/abs/1706.05925}{\texttt{arXiv:1706.05925}}.

\bibitem{gurzhi}
R.~N. Gurzhi.
\newblock ``Minimum of resistance in impurity-free conductors",
  \href{http://www.jetp.ac.ru/cgi-bin/e/index/e/17/2/p521?a=list}{\textsl{Journal
  of Experimental and Theoretical Physics} \textbf{17} 521 (1963)}.

\bibitem{SSBK11}
S.~Sachdev and B.~Keimer.
\newblock ``Quantum criticality",
  \href{http://dx.doi.org/10.1063/1.3554314}{\textsl{Physics Today}
  \textbf{64(2)} 29 (2011)},
  \href{http://arxiv.org/abs/1102.4628}{\texttt{arXiv:1102.4628}}.

\bibitem{lucasrmp}
S.~A. Hartnoll, A.~Lucas, and S.~Sachdev.
\newblock ``Holographic quantum matter",
  \href{http://arxiv.org/abs/1612.07324}{\texttt{arXiv:1612.07324}}.

\bibitem{mackenzie2013}
J.~A.~N. Bruin, H.~Sakai, R.~S. Perry, and A.~P. Mackenzie.
\newblock ``Similarity of scattering rates in metals showing $T$-linear
  resistivity",
  \href{http://science.sciencemag.org/content/339/6121/804}{\textsl{Science}
  \textbf{339} 804 (2013)}.

\bibitem{dsz}
R.~A. Davison, K.~Schalm, and J.~Zaanen.
\newblock ``Holographic duality and the resistivity of strange metals",
  \href{http://journals.aps.org/prb/abstract/10.1103/PhysRevB.89.245116}{\textsl{Physical
  Review} \textbf{B89} \texttt{245116} (2014)},
  \href{http://arxiv.org/abs/1311.2451}{\texttt{arXiv:1311.2451}}.

\bibitem{hartnoll1}
S.~A. Hartnoll.
\newblock ``Theory of universal incoherent metallic transport",
  \href{http://www.nature.com/nphys/journal/v11/n1/full/nphys3174.html}{\textsl{Nature
  Physics} \textbf{11} 54 (2015)},
  \href{http://arxiv.org/abs/1405.3651}{\texttt{arXiv:1405.3651}}.

\bibitem{hartnoll1704}
A.~Lucas and S.~A. Hartnoll.
\newblock ``Resistivity bound for hydrodynamic bad metals",
  \href{http://arxiv.org/abs/1704.07384}{\texttt{arXiv:1704.07384}}.

\bibitem{hartnoll1705}
A.~Lucas and S.~A. Hartnoll.
\newblock ``Kinetic theory of transport for inhomogeneous electron fluids",
  \href{http://arxiv.org/abs/1706.04621}{\texttt{arXiv:1706.04621}}.

\bibitem{geimrmp}
A.~H.~Castro Neto, F.~Guinea, N.~M.~R. Peres, K.~S. Novoselov, and A.~K. Geim.
\newblock ``The electronic properties of graphene",
  \href{http://journals.aps.org/rmp/abstract/10.1103/RevModPhys.81.109}{\textsl{Reviews
  of Modern Physics} \textbf{81} 109 (2009)},
  \href{http://arxiv.org/abs/0709.1163}{\texttt{arXiv:0709.1163}}.

\bibitem{Sarma:2011br}
S.~Das Sarma, S.~Adam, E.~H. Hwang, and E.~Rossi.
\newblock ``Electronic transport in two-dimensional graphene",
  \href{https://doi.org/10.1103/RevModPhys.83.407}{\textsl{Reviews of Modern
  Physics} \textbf{83} 407 (2011)},
  \href{http://arxiv.org/abs/1003.4731}{\texttt{arXiv:1003.4731}}.

\bibitem{sachdevhubbard}
S.~Sachdev.
\newblock ``The landscape of the Hubbard model",
  \href{http://arxiv.org/abs/1012.0299}{\texttt{arXiv:1012.0299}}.

\bibitem{zhni}
Z.~H. Ni, T.~Yu, Y.~H. Lu, Y.~Y. Wang, Y.~P. Feng, and Z.~X. Shen.
\newblock ``Uniaxial strain on graphene: Raman spectroscopy study and bandgap
  opening", \href{http://pubs.acs.org/doi/abs/10.1021/nn800459e}{\textsl{ACS
  Nano} \textbf{2} 2301 (2009)},
  \href{http://arxiv.org/abs/0810.3476}{\texttt{arXiv:0810.3476}}.

\bibitem{Nomura:2007ed}
K.~Nomura and A.~H. MacDonald.
\newblock ``Quantum transport of massless Dirac fermions in graphene",
  \href{https://doi.org/10.1103/PhysRevLett.98.076602}{\textsl{Physical Review
  Letters} \textbf{98} \texttt{076602} (2007)},
  \href{http://arxiv.org/abs/cond-mat/0606589}{\texttt{arXiv:cond-mat/0606589}}.

\bibitem{sarma1}
S.~Adam, E.~H. Hwang, V.~Galitski, and S.~Das Sarma.
\newblock ``A self-consistent theory for graphene transport",
  \href{http://www.pnas.org/content/104/47/18392}{\textsl{Proceedings of the
  National Academy of Sciences} \textbf{104} 18392 (2007)},
  \href{http://arxiv.org/abs/0705.1540}{\texttt{arXiv:0705.1540}}.

\bibitem{Chen:2008hp}
J.~H. Chen, C.~Jang, M.~S. Fuhrer, E.~D. Williams, and M.~Ishigami.
\newblock ``Charged impurity scattering in graphene",
  \href{https://doi.org/10.1038/nphys935}{\textsl{Nature Physics} \textbf{4}
  377 (2008)}, \href{http://arxiv.org/abs/0708.2408}{\texttt{arXiv:0708.2408}}.

\bibitem{adamchargepuddles}
S.~Samaddar, I.~Yudhistira, S.~Adam, H.~Courtois, and C.~B. Winkelmann.
\newblock ``Charge puddles in graphene near the Dirac point",
  \href{http://journals.aps.org/prl/abstract/10.1103/PhysRevLett.116.126804}{\textsl{Physical
  Review Letters} \textbf{116} \texttt{126804} (2016)},
  \href{http://arxiv.org/abs/1512.05304}{\texttt{arXiv:1512.05304}}.

\bibitem{xue}
J.~Xue, J.~Sanchez-Yamagishi, D.~Bulmash, P.~Jacquod, A.~Deshpande,
  K.~Watanabe, T.~Taniguchi, P.~Jarillo-Herrero, and B.~J. LeRoy.
\newblock ``Scanning tunnelling microscopy and spectroscopy of ultra-flat
  graphene on hexagonal boron nitride",
  \href{http://www.nature.com/nmat/journal/v10/n4/full/nmat2968.html}{\textsl{Nature
  Materials} \textbf{10} 282 (2011)},
  \href{http://arxiv.org/abs/1102.2642}{\texttt{arXiv:1102.2642}}.

\bibitem{crommie}
Y.~Zhang, V.~W. Brar, C.~Girit, A.~Zettl, and M.~F. Crommie.
\newblock ``Origin of spatial charge inhomogeneity in graphene",
  \href{http://www.nature.com/nphys/journal/v5/n10/abs/nphys1365.html}{\textsl{Nature
  Physics} \textbf{5} 722 (2009)},
  \href{http://arxiv.org/abs/0902.4793}{\texttt{arXiv:0902.4793}}.

\bibitem{yacoby2007}
J.~Martin, N.~Akerman, G.~Ulbricht, T.~Lohmann, J.~H. Smet, K.~von Klitzing,
  and A.~Yacoby.
\newblock ``Observation of electron-hole puddles in graphene using a scanning
  single electron transistor",
  \href{http://www.nature.com/nphys/journal/v4/n2/full/nphys781.html}{\textsl{Nature
  Physics} \textbf{4} 144 (2008)},
  \href{http://arxiv.org/abs/0705.2180}{\texttt{arXiv:0705.2180}}.

\bibitem{bolotin}
K.~I. Bolotin, K.~J. Sikes, Z.~Jiang, M.~Klima, G.~Fudenberg, J.~Hone, P.~Kim,
  and H.~L. Stormer.
\newblock ``Ultrahigh electron mobility in suspended graphene",
  \href{http://www.sciencedirect.com/science/article/pii/S0038109808001178}{\textsl{Solid
  State Communications} \textbf{146} 351 (2008)},
  \href{http://arxiv.org/abs/0802.2389}{\texttt{arXiv:0802.2389}}.

\bibitem{Dean:2010jy}
C.~R.~Dean \emph{et al.}
\newblock ``Boron nitride substrates for high-quality graphene electronics",
  \href{https://doi.org/10.1038/nnano.2010.172}{\textsl{Nature Nanotechnology}
  \textbf{5} 722 (2010)},
  \href{http://arxiv.org/abs/1005.4917}{\texttt{arXiv:1005.4917}}.

\bibitem{ortix}
C.~Ortix, L.~Yang, and J.~van~den Brink.
\newblock ``Graphene on incommensurate substrates: trigonal warping and
  emerging Dirac cone replicas with halved group velocity",
  \href{https://doi.org/10.1103/PhysRevB.86.081405}{\textsl{Physical Review}
  \textbf{B86} \texttt{081405} (2012)},
  \href{http://arxiv.org/abs/1111.0399}{\texttt{arXiv:1111.0399}}.

\bibitem{yankowitz}
M.~Yankowitz, J.~Xue, D.~Cormode, J.~D. Sanchez-Yamagishi, K.~Watanabe,
  T.~Taniguchi, P.~Jarillo-Herrero, P.~Jacquod, and B.~J. LeRoy.
\newblock ``Emergence of superlattice Dirac points in graphene on hexagonal
  boron nitride", \href{https://doi.org/10.1038/nphys2272}{\textsl{Nature
  Physics} \textbf{8} 382 (2012)},
  \href{http://arxiv.org/abs/1202.2870}{\texttt{arXiv:1202.2870}}.

\bibitem{hwang07}
E.~H. Hwang and S.~Das Sarma.
\newblock ``Acoustic phonon scattering limited carrier mobility in 2D extrinsic
  graphene",
  \href{http://journals.aps.org/prb/abstract/10.1103/PhysRevB.77.115449}{\textsl{Physical
  Review} \textbf{B77} \texttt{115449} (2008)},
  \href{http://arxiv.org/abs/0711.0754}{\texttt{arXiv:0711.0754}}.

\bibitem{Chen:2008by}
J-H. Chen, C.~Jang, S.~Xiao, M.~Ishigami, and M.~S. Fuhrer.
\newblock ``Intrinsic and extrinsic performance limits of graphene devices on
  $\mathrm{SiO}_2$",
  \href{https://doi.org/10.1038/nnano.2008.58}{\textsl{Nature Nanotechnology}
  \textbf{3} 206 (2008)},
  \href{http://arxiv.org/abs/0711.3646}{\texttt{arXiv:0711.3646}}.

\bibitem{Morozov:2008uf}
S.~V. Mozorov, K.~S. Novoselov, M.~I. Katsnelson, F.~Schedin, D.~C. Elias,
  J.~A. Jaszczak, and A.~K. Geim.
\newblock ``Giant intrinsic carrier mobilities in graphene and its bilayer",
  \href{https://doi.org/10.1103/PhysRevLett.100.016602}{\textsl{Physical Review
  Letters} \textbf{100} \texttt{016602} (2008)},
  \href{http://arxiv.org/abs/0710.5304}{\texttt{arXiv:0710.5304}}.

\bibitem{wang13}
L.~Wang \emph{et al.}
\newblock ``One-dimensional electrical contact to a two-dimensional material",
  \href{http://www.sciencemag.org/content/342/6158/614}{\textsl{Science}
  \textbf{342} 614 (2013)}.

\bibitem{sarma13}
Q.~Li and S.~Das Sarma.
\newblock ``Finite temperature inelastic mean free path and quasiparticle
  lifetime in graphene",
  \href{https://doi.org/10.1103/PhysRevB.87.085406}{\textsl{Physical Review}
  \textbf{B87} \texttt{085406} (2013)},
  \href{http://arxiv.org/abs/1211.6430}{\texttt{arXiv:1211.6430}}.

\bibitem{ancona}
B.~R. Bennett, R.~Magno, J.~B. Boos, W.~Kruppa, and M.~G. Ancona.
\newblock ``Antimonide-based compound semiconductors for electronic devices: a
  review", \href{https://doi.org/10.1016/j.sse.2005.09.008}{\textsl{Solid-State
  Electronics} \textbf{49} 1875 (2005)}.

\bibitem{ashley}
J.~M.~S. Orr, A.~M. Gilbertson, M.~Fearn, O.~W. Croad, C.~J. Storey, L.~Buckle,
  M.~T. Emeny, P.~D. Buckle, and T.~Ashley.
\newblock ``Electronic transport in modulation-doped InSb quantum well
  heterostructures",
  \href{https://doi.org/10.1103/PhysRevB.77.165334}{\textsl{Physical Review}
  \textbf{B77} \texttt{165334} (2008)}.

\bibitem{Efetov:2010fu}
D.~K. Efetov and P.~Kim.
\newblock ``Controlling electron-phonon interactions in graphene at ultrahigh
  carrier densities",
  \href{https://doi.org/10.1103/PhysRevLett.105.256805}{\textsl{Physical Review
  Letters} \textbf{105} \texttt{256805} (2010)},
  \href{http://arxiv.org/abs/1009.2988}{\texttt{arXiv:1009.2988}}.

\bibitem{Yan:2007ie}
J.~Yan, Y.~Zhang, P.~Kim, and A.~Pinczuk.
\newblock ``Electric field effect tuning of electron-phonon coupling in
  graphene",
  \href{https://doi.org/10.1103/PhysRevLett.98.166802}{\textsl{Physical Review
  Letters} \textbf{98} \texttt{166802} (2007)},
  \href{http://arxiv.org/abs/cond-mat/0612634}{\texttt{arXiv:cond-mat/0612634}}.

\bibitem{Ferrari:2007fba}
A.~C. Ferrari.
\newblock ``Raman spectroscopy of graphene and graphite: Disorder,
  electron?phonon coupling, doping and nonadiabatic effects",
  \href{https://doi.org/10.1016/j.ssc.2007.03.052}{\textsl{Solid State
  Communications} \textbf{143} 37 (2007)}.

\bibitem{Tse:2009il}
W-K. Tse and S.~Das Sarma.
\newblock ``Energy relaxation of hot Dirac fermions in graphene",
  \href{https://doi.org/10.1103/PhysRevB.79.235406}{\textsl{Physical Review}
  \textbf{B79} \texttt{235406} (2009)},
  \href{http://arxiv.org/abs/0812.1008}{\texttt{arXiv:0812.1008}}.

\bibitem{Bistritzer:2009ht}
R.~Bistritzer and A.~H. MacDonald.
\newblock ``Electronic cooling in graphene",
  \href{https://doi.org/10.1103/PhysRevLett.102.206410}{\textsl{Physical Review
  Letters} \textbf{102} \texttt{206410} (2009)},
  \href{http://arxiv.org/abs/0901.4159}{\texttt{arXiv:0901.4159}}.

\bibitem{Kubakaddi:2009br}
S.~S. Kubakaddi.
\newblock ``Interaction of massless Dirac electrons with acoustic phonons in
  graphene at low temperatures",
  \href{https://doi.org/10.1103/PhysRevB.79.075417}{\textsl{Physical Review}
  \textbf{B79} \texttt{075417} (2009)}.

\bibitem{viljas}
J.~K. Viljas and T.~T. Heikkil\"a.
\newblock ``Electron-phonon heat transfer in monolayer and bilayer graphene",
  \href{http://journals.aps.org/prl/abstract/10.1103/PhysRevB.81.245404}{\textsl{Physical
  Review} \textbf{B81} \texttt{245404} (2010)},
  \href{http://arxiv.org/abs/1002.3502}{\texttt{arXiv:1002.3502}}.

\bibitem{Chen:2012et}
W.~Chen and A.~A. Clerk.
\newblock ``Electron-phonon mediated heat flow in disordered graphene",
  \href{https://doi.org/10.1103/PhysRevB.86.125443}{\textsl{Physical Review}
  \textbf{B86} \texttt{125443} (2012)},
  \href{http://arxiv.org/abs/1207.2730}{\texttt{arXiv:1207.2730}}.

\bibitem{Graham:2013vl}
M.~W. Graham, S-F. Shi, D.~C. Ralph, J.~Park, and P.~L. McEuen.
\newblock ``Photocurrent measurements of supercollision cooling in graphene",
  \href{https://doi.org/10.1038/nphys2493}{\textsl{Nature Physics} \textbf{9}
  103 (2013)}, \href{http://arxiv.org/abs/1207.1249}{\texttt{arXiv:1207.1249}}.

\bibitem{Betz:2012wya}
A.~C.~Betz \emph{et al}.
\newblock ``Hot electron cooling by acoustic phonons in graphene",
  \href{http://journals.aps.org/prl/abstract/10.1103/PhysRevLett.109.056805}{\textsl{Physical
  Review Letters} \textbf{109} \texttt{056805} (2012)},
  \href{http://arxiv.org/abs/1203.2753}{\texttt{arXiv:1203.2753}}.

\bibitem{Fong:2012ut}
K.~C. Fong and K.~C. Schwab.
\newblock ``Ultrasensitive and wide-bandwidth thermal measurements of graphene
  at low temperatures",
  \href{http://journals.aps.org/prb/abstract/10.1103/PhysRevX.2.031006}{\textsl{Physical
  Review} \textbf{X2} \texttt{031006} (2012)},
  \href{http://arxiv.org/abs/1202.5737}{\texttt{arXiv:1202.5737}}.

\bibitem{Betz:2013up}
A.~C. Betz, S.~Jhang, E.~Pallecchi, R.~Ferreira, G.~F{\`e}ve, J-M. Berroir, and
  B.~Pla{\c c}ais.
\newblock ``Supercollision cooling in undoped graphene",
  \href{http://www.nature.com/nphys/journal/v9/n2/full/nphys2494.html}{\textsl{Nature
  Physics} \textbf{9} 109 (2013)},
  \href{http://arxiv.org/abs/1210.6894}{\texttt{arXiv:1210.6894}}.

\bibitem{Fong:2013hl}
K.~C. Fong, E.~Wollman, H.~Ravi, W.~Chen, A.~Clerk, M.~D. Shaw, H.~D. LeDuc,
  and K.~C. Schwab.
\newblock ``Measurement of the electronic thermal conductance channels and heat
  capacity of graphene at low temperature",
  \href{http://journals.aps.org/prx/abstract/10.1103/PhysRevX.3.041008}{\textsl{Physical
  Review} \textbf{X3} \texttt{041008} (2013)},
  \href{http://arxiv.org/abs/1308.2265}{\texttt{arXiv:1308.2265}}.

\bibitem{song2011}
J.~C.~W. Song, M.~Y. Reizer, and L.~S. Levitov.
\newblock ``Disorder-assisted electron-phonon scattering and cooling pathways
  in graphene",
  \href{https://doi.org/10.1103/PhysRevLett.109.106602}{\textsl{Physical Review
  Letters} \textbf{109} \texttt{106602} (2012)},
  \href{http://arxiv.org/abs/1111.4678}{\texttt{arXiv:1111.4678}}.

\bibitem{crossno2}
J.~Crossno, X.~Liu, T.~A. Ohki, P.~Kim, and K.~C. Fong.
\newblock ``Development of high frequency and wide bandwidth Johnson noise
  thermometry",
  \href{http://scitation.aip.org/content/aip/journal/apl/106/2/10.1063/1.4905926}{\textsl{Applied
  Physics Letters} \textbf{106} \texttt{023121} (2015)},
  \href{http://arxiv.org/abs/1411.4596}{\texttt{arXiv:1411.4596}}.

\bibitem{ma1}
N.~M. Gabor, J.~C.~W. Song, Q.~Ma, N.~L. Nair, T.~Taychatanapat, K.~Watanabe,
  T.~Taniguchi, L.~S. Levitov, and P.~Jarillo-Herrero.
\newblock ``Hot carrier assisted intrinsic photoresponse in graphene",
  \href{http://science.sciencemag.org/content/334/6056/648}{\textsl{Science}
  \textbf{334} 648 (2011)},
  \href{http://arxiv.org/abs/1108.3826}{\texttt{arXiv:1108.3826}}.

\bibitem{song2}
J.~C.~W. Song, M.~S. Rudner, C.~M. Marcus, and L.~S. Levitov.
\newblock ``Hot carrier transport and photocurrent response in graphene",
  \href{http://pubs.acs.org/doi/abs/10.1021/nl202318u}{\textsl{Nano Letters}
  \textbf{11} 4688 (2011)},
  \href{http://arxiv.org/abs/1105.1142}{\texttt{arXiv:1105.1142}}.

\bibitem{song}
K.~J. Tielrooij, J.~C.~W. Song, S.~A. Jensen, A.~Centeno, A.~Pesquera,
  A.~Zurutuza Elorza, M.~Bonn, L.~S. Levitov, and F.~H.~L. Koppens.
\newblock ``Photoexcitation cascade and multiple hot-carrier generation in
  graphene",
  \href{http://www.nature.com/nphys/journal/v9/n4/abs/nphys2564.html}{\textsl{Nature
  Physics} \textbf{9} 248 (2013)},
  \href{http://arxiv.org/abs/1210.1205}{\texttt{arXiv:1210.1205}}.

\bibitem{johannsen}
J.~C.~Johannsen \emph{et al}.
\newblock ``Direct view of hot carrier dynamics in graphene",
  \href{http://journals.aps.org/prl/abstract/10.1103/PhysRevLett.111.027403}{\textsl{Physical
  Review Letters} \textbf{111} \texttt{027403} (2013)},
  \href{http://arxiv.org/abs/1304.2615}{\texttt{arXiv:1304.2615}}.

\bibitem{ma2}
Q.~Ma, N.~M. Gabor, T.~I. Andersen, N.~L. Nair, K.~Watanabe, T.~Taniguchi, and
  P.~Jarillo-Herrero.
\newblock ``Competing channels for hot-electron cooling in graphene",
  \href{http://journals.aps.org/prl/abstract/10.1103/PhysRevLett.112.247401}{\textsl{Physical
  Review Letters} \textbf{112} \texttt{247401} (2014)},
  \href{http://arxiv.org/abs/1403.8152}{\texttt{arXiv:1403.8152}}.

\bibitem{hubbard}
J.~Hubbard.
\newblock ``Electron correlations in narrow energy bands",
  \href{https://doi.org/10.1098/rspa.1963.0204}{\textsl{Proceedings of the
  Royal Society} \textbf{A276} 238 (1963)}.

\bibitem{herbut1}
I.~F. Herbut.
\newblock ``Interactions and phase transitions on graphene{'}s honeycomb
  lattice",
  \href{http://journals.aps.org/prl/abstract/10.1103/PhysRevLett.97.146401}{\textsl{Physical
  Review Letters} \textbf{97} \texttt{146401} (2006)},
  \href{http://arxiv.org/abs/cond-mat/0606195}{\texttt{arXiv:cond-mat/0606195}}.

\bibitem{herbut2}
I.~F. Herbut, V.~Juricic, and B.~Roy.
\newblock ``Theory of interacting electrons on the honeycomb lattice",
  \href{http://journals.aps.org/prb/abstract/10.1103/PhysRevB.79.085116}{\textsl{Physical
  Review} \textbf{B79} \texttt{085116} (2009)},
  \href{http://arxiv.org/abs/0811.0610}{\texttt{arXiv:0811.0610}}.

\bibitem{herbut3}
I.~F. Herbut, V.~Juricic, and O.~Vafek.
\newblock ``Relativistic Mott criticality in graphene",
  \href{http://journals.aps.org/prb/abstract/10.1103/PhysRevB.80.075432}{\textsl{Physical
  Review} \textbf{B80} \texttt{075432} (2009)},
  \href{http://arxiv.org/abs/0904.1019}{\texttt{arXiv:0904.1019}}.

\bibitem{adam1505}
H-K. Tang, E.~Laksono, J.~N.~B. Rodrigues, P.~Sengupta, F.~F. Assaad, and
  S.~Adam.
\newblock ``Interaction driven metal-insulator transition in strained
  graphene",
  \href{http://journals.aps.org/prl/abstract/10.1103/PhysRevLett.115.186602}{\textsl{Physical
  Review Letters} \textbf{115} \texttt{186602} (2015)},
  \href{http://arxiv.org/abs/1505.04188}{\texttt{arXiv:1505.04188}}.

\bibitem{Sarma:2007ej}
S.~Das Sarma, E.~H. Hwang, and W-K. Tse.
\newblock ``Many-body interaction effects in doped and undoped graphene: Fermi
  liquid versus non-Fermi liquid",
  \href{https://doi.org/10.1103/PhysRevB.75.121406}{\textsl{Physical Review}
  \textbf{B75} \texttt{121406} (2007)},
  \href{http://arxiv.org/abs/cond-mat/0610581}{\texttt{arXiv:cond-mat/0610581}}.

\bibitem{pines}
D.~Pines and P.~Nozi\`eres.
\newblock \emph{The Theory of Quantum Liquids, Volume I}
  \href{http://www.amazon.com/Theory-Of-Quantum-Liquids-Advanced/dp/0201407744}{(W.
  A. Benjamin, 1966)}.

\bibitem{shankarRMP}
R.~Shankar.
\newblock ``Renormalization group approach to interacting fermions",
  \href{https://doi.org/10.1103/RevModPhys.66.129}{\textsl{Reviews of Modern
  Physics} \textbf{66} 129 (1994)},
  \href{http://arxiv.org/abs/cond-mat/9307009}{\texttt{arXiv:cond-mat/9307009}}.

\bibitem{uchoa}
B.~Uchoa and A.~H.~Castro Neto.
\newblock ``Superconducting states of pure and doped graphene",
  \href{https://doi.org/10.1103/PhysRevLett.98.146801}{\textsl{Physical Review
  Letters} \textbf{98} \texttt{146801} (2007)},
  \href{http://arxiv.org/abs/cond-mat/0608515}{\texttt{arXiv:cond-mat/0608515}}.

\bibitem{mcchesney}
J.~L. McChesney, A.~Bostwick, T.~Ohta, T.~Seyller, K.~Horn, J.~Gonz\'alez, and
  E.~Rotenberg.
\newblock ``Extended van Hove singularity and superconducting instability in
  graphene",
  \href{https://journals.aps.org/prl/abstract/10.1103/PhysRevLett.104.136803}{\textsl{Physical
  Review Letters} \textbf{104} \texttt{136803} (2010)}.

\bibitem{nandkishore2011}
R.~Nandkishore, L.~S. Levitov, and A.~V. Chubukov.
\newblock ``Chiral superconductivity from repulsive interactions in doped
  graphene",
  \href{http://www.nature.com/nphys/journal/v8/n2/full/nphys2208.html}{\textsl{Nature
  Physics} \textbf{8} 158 (2012)},
  \href{http://arxiv.org/abs/1107.1903}{\texttt{arXiv:1107.1903}}.

\bibitem{eegraphenermp}
V.~N. Kotov, B.~Uchoa, V.~M. Pereira, F.~Guinea, and A.~H.~Castro Neto.
\newblock ``Electron-electron interactions in graphene: current status and
  perspectives",
  \href{https://doi.org/10.1103/RevModPhys.84.1067}{\textsl{Reviews of Modern
  Physics} \textbf{84} 1067 (2012)},
  \href{http://arxiv.org/abs/1012.3484}{\texttt{arXiv:1012.3484}}.

\bibitem{wilsonRG}
K.~G. Wilson.
\newblock ``The renormalization group: Critical phenomena and the Kondo
  problem",
  \href{https://journals.aps.org/rmp/abstract/10.1103/RevModPhys.47.773}{\textsl{Reviews
  of Modern Physics} \textbf{47} 773 (1975)}.

\bibitem{kardar2}
M.~Kardar.
\newblock \emph{Statistical Physics of Fields},
  \href{https://www.amazon.com/Statistical-Physics-Fields-Mehran-Kardar/dp/052187341X/ref=sr_1_1?ie=UTF8&qid=1512005644&sr=8-1&keywords=statistical+physics+of+fields}{(Cambridge
  University Press, 2007)}.

\bibitem{vozmediano1}
J.~Gonz\'alez, F.~Guinea, and M.~A.~H. Vozmediano.
\newblock ``Non-Fermi liquid behaviour of electrons in the half-filled
  honeycomb lattice (A renormalization group approach)",
  \href{https://doi.org/10.1016/0550-3213(94)90410-3}{\textsl{Nuclear Physics}
  \textbf{B424} 595 (1994)},
  \href{http://arxiv.org/abs/hep-th/9311105}{\texttt{arXiv:hep-th/9311105}}.

\bibitem{vozmediano2}
J.~Gonz\'alez, F.~Guinea, and M.~A.~H. Vozmediano.
\newblock ``Marginal-Fermi-liquid behavior from two-dimensional Coulomb
  interaction",
  \href{https://journals.aps.org/prb/abstract/10.1103/PhysRevB.59.R2474}{\textsl{Physical
  Review} \textbf{B59} 2474 (1999)},
  \href{http://arxiv.org/abs/cond-mat/9807130}{\texttt{arXiv:cond-mat/9807130}}.

\bibitem{vafek}
O.~Vafek.
\newblock ``Anomalous thermodynamics of Coulomb-interacting massless Dirac
  fermions in two spatial dimensions",
  \href{http://journals.aps.org/prl/abstract/10.1103/PhysRevLett.98.216401}{\textsl{Physical
  Review Letters} \textbf{98} \texttt{216401} (2007)},
  \href{http://arxiv.org/abs/cond-mat/0701145}{\texttt{arXiv:cond-mat/0701145}}.

\bibitem{schmalian}
D.~E. Sheehy and J.~Schmalian.
\newblock ``Quantum critical scaling in graphene",
  \href{http://journals.aps.org/prl/abstract/10.1103/PhysRevLett.99.226803}{\textsl{Physical
  Review Letters} \textbf{99} \texttt{226803} (2007)},
  \href{http://arxiv.org/abs/0707.2945}{\texttt{arXiv:0707.2945}}.

\bibitem{vafek2}
O.~Vafek and M.~J. Case.
\newblock ``Renormalization group approach to 2D Coulomb interacting Dirac
  fermions with random gauge potential",
  \href{https://doi.org/10.1103/PhysRevB.77.033410}{\textsl{Physical Review}
  \textbf{B77} \texttt{033410} (2008)},
  \href{http://arxiv.org/abs/0710.2907}{\texttt{arXiv:0710.2907}}.

\bibitem{sarma14}
E.~Barnes, E.~H. Hwang, R.~E. Throckmorton, and S.~Das Sarma.
\newblock ``Effective field theory, three-loop perturbative expansion, and
  their experimental implications in graphene many-body effects",
  \href{https://doi.org/10.1103/PhysRevB.89.235431}{\textsl{Physical Review}
  \textbf{B89} \texttt{235431} (2014)},
  \href{http://arxiv.org/abs/1401.7011}{\texttt{arXiv:1401.7011}}.

\bibitem{drut1}
J.~E. Drut and T.~A. L\"ahde.
\newblock ``Is graphene in vacuum an insulator?",
  \href{https://journals.aps.org/prl/abstract/10.1103/PhysRevLett.102.026802}{\textsl{Physical
  Review Letters} \textbf{102} \texttt{026802} (2009)},
  \href{http://arxiv.org/abs/0807.0834}{\texttt{arXiv:0807.0834}}.

\bibitem{drut2}
J.~E. Drut and T.~A. L\"ahde.
\newblock ``Critical exponents of the semimetal-insulator transition in
  graphene: A Monte Carlo study",
  \href{https://journals.aps.org/prb/abstract/10.1103/PhysRevB.79.241405}{\textsl{Physical
  Review} \textbf{B79} \texttt{241405} (2009)},
  \href{http://arxiv.org/abs/0905.1320}{\texttt{arXiv:0905.1320}}.

\bibitem{sarmaRPA}
J.~Hofmann, E.~Barnes, and S.~Das Sarma.
\newblock ``Why does graphene behave as a weakly interacting system?",
  \href{https://doi.org/10.1103/PhysRevLett.113.105502}{\textsl{Physical Review
  Letters} \textbf{113} \texttt{105502} (2014)},
  \href{http://arxiv.org/abs/1405.7036}{\texttt{arXiv:1405.7036}}.

\bibitem{kopietz}
A.~Sharma and P.~Kopietz.
\newblock ``Multilogarithmic velocity renormalization in graphene",
  \href{https://doi.org/10.1103/PhysRevB.93.235425}{\textsl{Physical Review}
  \textbf{B93} \texttt{235425} (2016)},
  \href{http://arxiv.org/abs/1603.01188}{\texttt{arXiv:1603.01188}}.

\bibitem{polikarpov}
M.~V. Ulybyshev, P.~V. Buividovich, M.~I. Kastnelson, and M.~I. Polikarpov.
\newblock ``Monte-Carlo study of the semimetal-insulator phase transition in
  monolayer graphene with realistic inter-electron interaction potential",
  \href{https://journals.aps.org/prl/abstract/10.1103/PhysRevLett.111.056801}{\textsl{Physical
  Review Letters} \textbf{111} \texttt{056801} (2013)},
  \href{http://arxiv.org/abs/1304.3360}{\texttt{arXiv:1304.3360}}.

\bibitem{tupitsyn}
I.~Tupitsyn and N.~Prokof'ev.
\newblock ``Stability of Dirac liquids with strong Coulomb interaction",
  \href{http://journals.aps.org/prl/abstract/10.1103/PhysRevLett.118.026403}{\textsl{Physical
  Review Letters} \textbf{118} \texttt{026403} (2017)},
  \href{http://arxiv.org/abs/1608.00133}{\texttt{arXiv:1608.00133}}.

\bibitem{elias}
D.~C.~Elias \emph{et al.}
\newblock ``Dirac cones reshaped by interaction effects in suspended graphene",
  \href{http://www.nature.com/nphys/journal/v7/n9/full/nphys2049.html}{\textsl{Nature
  Physics} \textbf{7} 701 (2011)},
  \href{http://arxiv.org/abs/1104.1396}{\texttt{arXiv:1104.1396}}.

\bibitem{sharapov}
S.~G. Sharapov, V.~P. Gusynin, and H.~Beck.
\newblock ``Magnetic oscillations in planar systems with the Dirac-like
  spectrum", \href{https://doi.org/10.1103/PhysRevB.69.075104}{\textsl{Physical
  Review} \textbf{B69} \texttt{075104} (2004)},
  \href{http://arxiv.org/abs/cond-mat/0308216}{\texttt{arXiv:cond-mat/0308216}}.

\bibitem{siegel2011}
D.~A. Siegel, C-H. Park, C.~Hwang, J.~Deslippe, A.~V. Fedorov, S.~G. Louie, and
  A.~Lanzara.
\newblock ``Many-body interactions in quasi-freestanding graphene",
  \href{https://doi.org/10.1073/pnas.1100242108}{\textsl{Proceedings of the
  National Academy of Sciences} \textbf{108} 11365 (2011)},
  \href{http://arxiv.org/abs/1106.5822}{\texttt{arXiv:1106.5822}}.

\bibitem{hkms}
S.~A. Hartnoll, P.~K. Kovtun, M.~M\"uller, and S.~Sachdev.
\newblock ``Theory of the Nernst effect near quantum phase transitions in
  condensed matter, and in dyonic black holes",
  \href{http://journals.aps.org/prb/abstract/10.1103/PhysRevB.76.144502}{\textsl{Physical
  Review} \textbf{B76} \texttt{144502} (2007)},
  \href{http://arxiv.org/abs/0706.3215}{\texttt{arXiv:0706.3215}}.

\bibitem{lucas3}
A.~Lucas, J.~Crossno, K.~C. Fong, P.~Kim, and S.~Sachdev.
\newblock ``Transport in inhomogeneous quantum critical fluids and in the Dirac
  fluid in graphene",
  \href{http://journals.aps.org/prb/abstract/10.1103/PhysRevB.93.075426}{\textsl{Physical
  Review} \textbf{B93} \texttt{075426} (2016)},
  \href{http://arxiv.org/abs/1510.01738}{\texttt{arXiv:1510.01738}}.

\bibitem{rangamani}
F.~M. Haehl, R.~Loganayagam, and M.~Rangamani.
\newblock ``The fluid manifesto: Emergent symmetries, hydrodynamics and black
  holes",
  \href{http://link.springer.com/article/10.1007%2FJHEP01%282016%29184}{\textsl{Journal
  of High Energy Physics} \textbf{01} \texttt{186} (\textbf{2016})},
  \href{http://arxiv.org/abs/1510.02494}{\texttt{arXiv:1510.02494}}.

\bibitem{glorioso}
M.~Crossley, P.~Glorioso, and H.~Liu.
\newblock ``Effective field theory of dissipative fluids",
  \href{http://arxiv.org/abs/1511.03646}{\texttt{arXiv:1511.03646}}.

\bibitem{levitovsound}
T.~V. Phan, J.~C.~W. Song, and L.~S. Levitov.
\newblock ``Ballistic heat transfer and energy waves in an electron system",
  \href{http://arxiv.org/abs/1306.4972}{\texttt{arXiv:1306.4972}}.

\bibitem{kovtun}
P.~Kovtun.
\newblock ``Lectures on hydrodynamic fluctuations in relativistic theories",
  \href{http://m.iopscience.iop.org/1751-8121/45/47/473001/}{\textsl{Journal of
  Physics} \textbf{A45} \texttt{473001} (2012)},
  \href{http://arxiv.org/abs/1205.5040}{\texttt{arXiv:1205.5040}}.

\bibitem{lucasplasma}
A.~Lucas.
\newblock ``Sound waves and resonances in electron-hole plasma",
  \href{http://journals.aps.org/prb/abstract/10.1103/PhysRevB.93.245153}{\textsl{Physical
  Review} \textbf{B93} \texttt{245153} (2016)},
  \href{http://arxiv.org/abs/1604.03955}{\texttt{arXiv:1604.03955}}.

\bibitem{polini}
I.~Torre, A.~Tomadin, A.~K. Geim, and M.~Polini.
\newblock ``Non-local transport and the hydrodynamic shear viscosity in
  graphene",
  \href{http://journals.aps.org/prb/abstract/10.1103/PhysRevB.92.165433}{\textsl{Physical
  Review} \textbf{B92} \texttt{165433} (2016)},
  \href{http://arxiv.org/abs/1508.00363}{\texttt{arXiv:1508.00363}}.

\bibitem{kimflow}
I.~Meric, M.~Y. Han, A.~F. Young, B.~Oyzilmaz, P.~Kim, and K.~L. Shepard.
\newblock ``Current saturation in zero-bandgap, top-gated graphene field-effect
  transistors",
  \href{http://www.nature.com/nnano/journal/v3/n11/full/nnano.2008.268.html}{\textsl{Nature
  Nanotechnology} \textbf{3} 654 (2008)}.

\bibitem{dorgan}
V.~E. Dorgan, M-H. Bae, and E.~Pop.
\newblock ``Mobility and saturation velocity in graphene on $\mathrm{SiO}_2$",
  \href{https://doi.org/10.1063/1.3483130}{\textsl{Applied Physics Letters}
  \textbf{97} \texttt{082112} (2010)},
  \href{http://arxiv.org/abs/1005.2711}{\texttt{arXiv:1005.2711}}.

\bibitem{succiturb}
M.~Mendoza, H.~J. Herrmann, and S.~Succi.
\newblock ``Preturbulent regimes in graphene flow",
  \href{http://journals.aps.org/prl/abstract/10.1103/PhysRevLett.106.156601}{\textsl{Physical
  Review Letters} \textbf{106} \texttt{156601} (2011)},
  \href{http://arxiv.org/abs/1201.6590}{\texttt{arXiv:1201.6590}}.

\bibitem{muller2}
M.~M\"uller, L.~Fritz, and S.~Sachdev.
\newblock ``Quantum-critical relativistic magnetotransport in graphene",
  \href{http://journals.aps.org/prb/abstract/10.1103/PhysRevB.78.115406}{\textsl{Physical
  Review} \textbf{B78} \texttt{115406} (2008)},
  \href{http://arxiv.org/abs/0805.1413}{\texttt{arXiv:0805.1413}}.

\bibitem{muller4}
L.~Fritz, J.~Schmalian, M.~M\"uller, and S.~Sachdev.
\newblock ``Quantum critical transport in clean graphene",
  \href{http://journals.aps.org/prb/abstract/10.1103/PhysRevB.78.085416}{\textsl{Physical
  Review} \textbf{B78} \texttt{085416} (2008)},
  \href{http://arxiv.org/abs/0802.4289}{\texttt{arXiv:0802.4289}}.

\bibitem{grozdanov1610}
S.~Grozdanov, D.~Hofman, and N.~Iqbal.
\newblock ``Generalized global symmetries and dissipative
  magnetohydrodynamics",
  \href{https://journals.aps.org/prd/abstract/10.1103/PhysRevD.95.096003}{\textsl{Physical
  Review} \textbf{D95} \texttt{096003} (2017)},
  \href{http://arxiv.org/abs/1610.07392}{\texttt{arXiv:1610.07392}}.

\bibitem{kovtun1703}
J.~Hernandez and P.~Kovtun.
\newblock ``Relativistic magnetohydrodynamics",
  \href{http://arxiv.org/abs/1703.08757}{\texttt{arXiv:1703.08757}}.

\bibitem{muller1}
M.~M\"uller and S.~Sachdev.
\newblock ``Collective cyclotron motion of the relativistic plasma in
  graphene",
  \href{http://journals.aps.org/prb/abstract/10.1103/PhysRevB.78.115419}{\textsl{Physical
  Review} \textbf{B78} \texttt{115419} (2008)},
  \href{http://arxiv.org/abs/0801.2970}{\texttt{arXiv:0801.2970}}.

\bibitem{marston}
A.~Houghton, H-J. Kwon, J.~B. Marston, and R.~Shankr.
\newblock ``Coulomb interaction and the Fermi liquid state: solution by
  bosonization",
  \href{https://doi.org/10.1088/0953-8984/6/26/012}{\textsl{Journal of Physics:
  Condensed Matter} \textbf{6} 4909 (1994)},
  \href{http://arxiv.org/abs/cond-mat/9312067}{\texttt{arXiv:cond-mat/9312067}}.

\bibitem{hwangplasmon}
E.~H. Hwang and S.~Das Sarma.
\newblock ``Dielectric function, screening and plasmons in two-dimensional
  graphene",
  \href{http://journals.aps.org/prb/abstract/10.1103/PhysRevB.75.205418}{\textsl{Physical
  Review} \textbf{B75} \texttt{205418} (2007)},
  \href{http://arxiv.org/abs/cond-mat/0610561}{\texttt{arXiv:cond-mat/0610561}}.

\bibitem{sarmaplasmon}
S.~Das Sarma and E.~H. Hwang.
\newblock ``Collective modes of the massless Dirac plasma",
  \href{https://doi.org/10.1103/PhysRevLett.102.206412}{\textsl{Physical Review
  Letters} \textbf{102} \texttt{206412} (2009)},
  \href{http://arxiv.org/abs/0902.3822}{\texttt{arXiv:0902.3822}}.

\bibitem{soljacic2}
M.~Jablan, M.~{Solja\v{c}i\'c}, and H.~Buljan.
\newblock ``Plasmons in graphene: fundamental properties and potential
  applications",
  \href{http://ieeexplore.ieee.org/abstract/document/6519306/}{\textsl{Proceedings
  of the IEEE} \textbf{101} 1689 (2013)}.

\bibitem{chen12}
J.~Chen \emph{et al.}
\newblock ``Optical nano-imaging of gate-tunable graphene plasmons",
  \href{http://www.nature.com/nature/journal/v487/n7405/full/nature11254.html}{\textsl{Nature}
  \textbf{487} 77 (2012)}.

\bibitem{koppens12}
Z.~Fei \emph{et al.}
\newblock ``Gate tuning of graphene plasmons revealed by infrared
  nano-imaging",
  \href{http://www.nature.com/nature/journal/v487/n7405/full/nature11253.html}{\textsl{Nature}
  \textbf{487} 82 (2012)}.

\bibitem{landaukinetic}
E.~M. Lifshitz and L.~P. Pitaevskii.
\newblock \emph{Physical Kinetics},
  \href{https://www.amazon.com/Physical-Kinetics-Course-Theoretical-Physics/dp/0750626356/ref=sr_1_1?ie=UTF8&qid=1509042515&sr=8-1&keywords=physical+kinetics}{(Butterworth
  Heinemann, 1981)}.

\bibitem{fogler3}
Z.~Sun, D.~N. Basov, and M.~M. Fogler.
\newblock ``Adiabatic amplification of plasmons and demons in 2d systems",
  \href{https://doi.org/10.1103/PhysRevLett.117.076805}{\textsl{Physical Review
  Letters} \textbf{117} \texttt{076805} (2016)},
  \href{http://arxiv.org/abs/1601.02722}{\texttt{arXiv:1601.02722}}.

\bibitem{fogler1}
Z.~Sun, D.~N. Basov, and M.~M. Fogler.
\newblock ``Universal linear and nonlinear electrodynamics of the Dirac fluid",
  \href{http://arxiv.org/abs/1704.07334}{\texttt{arXiv:1704.07334}}.

\bibitem{sodemann}
I.~Sodemann.
\newblock ``Current induced and interaction driven Dirac-point drag of massless
  quasi-relativistic fermions",
  \href{http://journals.aps.org/prb/abstract/10.1103/PhysRevB.93.235154}{\textsl{Physical
  Review} \textbf{B93} \texttt{235154} (2016)},
  \href{http://arxiv.org/abs/1601.02619}{\texttt{arXiv:1601.02619}}.

\bibitem{sybesma}
J.~de~Boer, J.~Hartong, N.~A. Obers, W.~Sybesma, and S.~Vandoren.
\newblock ``Perfect fluids",
  \href{http://arxiv.org/abs/1710.04708}{\texttt{arXiv:1710.04708}}.

\bibitem{lucas1612}
A.~Lucas.
\newblock ``Stokes paradox in electronic Fermi liquids",
  \href{http://link.aps.org/doi/10.1103/PhysRevB.95.115425}{\textsl{Physical
  Review} \textbf{B95} \texttt{115425} (2017)},
  \href{http://arxiv.org/abs/1612.00856}{\texttt{arXiv:1612.00856}}.

\bibitem{boffetta}
G.~Boffetta and R.~E. Ecke.
\newblock ``Two-dimensional turbulence",
  \href{https://doi.org/10.1146/annurev-fluid-120710-101240}{\textsl{Annual
  Review of Fluid Mechanics} \textbf{44} 427 (2012)}.

\bibitem{lehner}
F.~Carrasco, L.~Lehner, R.~C. Myers, O.~Reula, and A.~Singh.
\newblock ``Turbulent flows for relativistic conformal fluids in 2+1
  dimensions",
  \href{https://doi.org/10.1103/PhysRevD.86.126006}{\textsl{Physical Review}
  \textbf{D86} \texttt{126006} (2012)},
  \href{http://arxiv.org/abs/1210.6702}{\texttt{arXiv:1210.6702}}.

\bibitem{schmalian2}
B.~N. Narozhny, I.~V. Gornyi, A.~D. Mirlin, and J.~Schmalian.
\newblock ``Hydrodynamic approach to electronic transport in graphene",
  \href{http://arxiv.org/abs/1704.03494}{\texttt{arXiv:1704.03494}}.

\bibitem{kardar}
M.~Kardar.
\newblock \emph{Statistical Physics of Particles},
  \href{https://www.amazon.com/Statistical-Physics-Particles-Mehran-Kardar/dp/0521873428/ref=sr_1_1?ie=UTF8&qid=1507915016&sr=8-1&keywords=statistical+physics+of+particles}{(Cambridge
  University Press, 2007)}.

\bibitem{bgk}
P.~L. Bhatnagar, E.~P. Gross, and M.~Krook.
\newblock ``A model for collision processes in gases. I. Small amplitude
  processes in charged and neutral one-component systems",
  \href{https://journals.aps.org/pr/abstract/10.1103/PhysRev.94.511}{\textsl{Physical
  Review} \textbf{94} 511 (1954)}.

\bibitem{foster}
M.~S. Foster and I.~L. Aleiner.
\newblock ``Slow imbalance relaxation and thermoelectric transport in
  graphene",
  \href{http://journals.aps.org/prb/abstract/10.1103/PhysRevB.79.085415}{\textsl{Physical
  Review} \textbf{B79} \texttt{085415} (2009)},
  \href{http://arxiv.org/abs/0810.4342}{\texttt{arXiv:0810.4342}}.

\bibitem{mirlin11}
M.~Sch\"utt, P.~M. Ostrovky, I.~V. Gornyi, and A.~D. Mirlin.
\newblock ``Coulomb interaction in graphene: relaxation rates and transport",
  \href{https://doi.org/10.1103/PhysRevB.83.155441}{\textsl{Physical Review}
  \textbf{B83} \texttt{155441} (2011)},
  \href{http://arxiv.org/abs/1011.5217}{\texttt{arXiv:1011.5217}}.

\bibitem{svinstov}
D.~Svinstov, V.~Vyurkov, S.~Yurchenko, T.~Otsuji, and V.~Ryzhii.
\newblock ``Hydrodynamic model for electron-hole plasma in graphene",
  \href{http://aip.scitation.org/doi/10.1063/1.4705382}{\textsl{Journal of
  Applied Physics} \textbf{111} \texttt{083715} (2012)},
  \href{http://arxiv.org/abs/1201.0592}{\texttt{arXiv:1201.0592}}.

\bibitem{narozhny}
B.~D. Narozhny, I.~V. Gornyi, M.~Titov, M.~Sch\"utt, and A.~D. Mirlin.
\newblock ``Hydrodynamics in graphene: linear-response transport",
  \href{http://journals.aps.org/prb/abstract/10.1103/PhysRevB.91.035414}{\textsl{Physical
  Review} \textbf{B91} \texttt{035414} (2015)},
  \href{http://arxiv.org/abs/1411.0819}{\texttt{arXiv:1411.0819}}.

\bibitem{vafek2008}
I.~F. Herbut, V.~Juricic, and O.~Vafek.
\newblock ``Coulomb interaction, ripples, and the minimal conductivity of
  graphene",
  \href{https://doi.org/10.1103/PhysRevLett.100.046403}{\textsl{Physical Review
  Letters} \textbf{100} \texttt{046403} (2008)},
  \href{http://arxiv.org/abs/0707.4171}{\texttt{arXiv:0707.4171}}.

\bibitem{trushin}
M.~Trushin.
\newblock ``Collinear scattering of photoexcited carriers in graphene",
  \href{https://doi.org/10.1103/PhysRevB.94.205306}{\textsl{Physical Review}
  \textbf{B94} \texttt{205306} (2016)},
  \href{http://arxiv.org/abs/1606.07064}{\texttt{arXiv:1606.07064}}.

\bibitem{brida}
D.~Brida \emph{et al}.
\newblock ``Ultrafast collinear scattering and carrier multiplication in
  graphene", \href{https://doi.org/10.1038/ncomms2987}{\textsl{Nature
  Communications} \textbf{4} \texttt{1987} (2013)},
  \href{http://arxiv.org/abs/1209.5729}{\texttt{arXiv:1209.5729}}.

\bibitem{svintsov1710}
D.~Svintsov.
\newblock ``Hydrodynamic-to-ballistic crossover in Dirac fluid",
  \href{http://arxiv.org/abs/1710.05054}{\texttt{arXiv:1710.05054}}.

\bibitem{mirlin2015}
U.~Briskot, M.~Sch\"utt, I.~V. Gornyi, M.~Titov, B.~N. Narozhny, and A.~D.
  Mirlin.
\newblock ``Collision-dominated nonlinear hydrodynamics in graphene",
  \href{https://doi.org/10.1103/PhysRevB.92.115426}{\textsl{Physical Review}
  \textbf{B92} \texttt{115426} (2015)},
  \href{http://arxiv.org/abs/1507.08946}{\texttt{arXiv:1507.08946}}.

\bibitem{svintsov1709}
G.~Alymov, D.~Svintsov, V.~Vyurkov, V.~Ryzhii, and A.~Satou.
\newblock ``Auger recombination in two-dimensional Dirac materials",
  \href{http://arxiv.org/abs/1709.09015}{\texttt{arXiv:1709.09015}}.

\bibitem{sarma0703}
E.~H. Hwang, B.~Y-K. Hu, and S.~Das Sarma.
\newblock ``Density dependent exchange contribution to $\partial \mu/\partial
  n$ in extrinsic graphene",
  \href{http://journals.aps.org/prl/abstract/10.1103/PhysRevLett.99.226801}{\textsl{Physical
  Review Letters} \textbf{99} \texttt{226801} (2007)},
  \href{http://arxiv.org/abs/cond-mat/0703499}{\texttt{arXiv:cond-mat/0703499}}.

\bibitem{teitel}
P.~Olsson and S.~Teitel.
\newblock ``Critical scaling of shear viscosity at the jamming transition",
  \href{https://doi.org/10.1103/PhysRevLett.99.178001}{\textsl{Physical Review
  Letters} \textbf{99} \texttt{178001} (2007)},
  \href{http://arxiv.org/abs/0704.1806}{\texttt{arXiv:0704.1806}}.

\bibitem{polini1506}
A.~Principi, G.~Vignale, M.~Carrega, and M.~Polini.
\newblock ``Bulk and shear viscosities of the 2D electron liquid in a doped
  graphene sheet",
  \href{http://journals.aps.org/prb/abstract/10.1103/PhysRevB.93.125410}{\textsl{Physical
  Review} \textbf{B93} \texttt{125410} (2016)},
  \href{http://arxiv.org/abs/1506.06030}{\texttt{arXiv:1506.06030}}.

\bibitem{wilkins}
C.~Hodges, H.~Smith, and J.~W. Wilkins.
\newblock ``Effect of Fermi surface geometry on electron-electron scattering",
  \href{https://journals.aps.org/pr/abstract/10.1103/PhysRevB.4.302}{\textsl{Physical
  Review} \textbf{B4} 302 (1971)}.

\bibitem{quinn}
G.~F. Giuliani and J.~J. Quinn.
\newblock ``Lifetime of a quasiparticle in a two-dimensional electron gas",
  \href{https://journals.aps.org/pr/abstract/10.1103/PhysRevB.26.4421}{\textsl{Physical
  Review} \textbf{B26} 4421 (1982)}.

\bibitem{jungwirth}
T.~Jungwirth and A.~H. MacDonald.
\newblock ``Electron-electron interactions and two-dimensional-two-dimensional
  tunneling", \href{https://doi.org/10.1103/PhysRevB.53.7403}{\textsl{Physical
  Review} \textbf{B53} 7403 (1996)},
  \href{http://arxiv.org/abs/cond-mat/9603001}{\texttt{arXiv:cond-mat/9603001}}.

\bibitem{zala}
B.~N. Narozhny, G.~Zala, and I.~L. Aleiner.
\newblock ``Interaction corrections at intermediate temperatures: dephasing
  time", \href{https://doi.org/10.1103/PhysRevB.65.180202}{\textsl{Physical
  Review} \textbf{B65} \texttt{180202} (2002)},
  \href{http://arxiv.org/abs/cond-mat/0201379}{\texttt{arXiv:cond-mat/0201379}}.

\bibitem{ledwith1}
P.~Ledwith, H.~Guo, and L.~Levitov.
\newblock ``Fermion collisions in two dimensions",
  \href{http://arxiv.org/abs/1708.01915}{\texttt{arXiv:1708.01915}}.

\bibitem{ledwith2}
P.~Ledwith, H.~Guo, A.~V. Shytov, and L.~Levitov.
\newblock ``Head-on collisions and scale-dependent viscosity in two-dimensional
  electron systems",
  \href{http://arxiv.org/abs/1708.02376}{\texttt{arXiv:1708.02376}}.

\bibitem{novikov}
D.~S. Novikov.
\newblock ``Viscosity of a two-dimensional Fermi liquid",
  \href{http://arxiv.org/abs/cond-mat/0603184}{\texttt{arXiv:cond-mat/0603184}}.

\bibitem{aleiner06}
I.~L. Aleiner and K.~B. Efetov.
\newblock ``Supersymmetric low-energy theory and renormalization group for a
  clean Fermi gas with a repulsion in arbitrary dimensions",
  \href{https://doi.org/10.1103/PhysRevB.74.075102}{\textsl{Physical Review}
  \textbf{B74} \texttt{075102} (2006)},
  \href{http://arxiv.org/abs/cond-mat/0602309}{\texttt{arXiv:cond-mat/0602309}}.

\bibitem{abrikosov}
A.~A. Abrikosov and I.~M. Khalatnikov.
\newblock ``The theory of a fermi liquid (the properties of liquid $^3$He at
  low temperatures)",
  \href{http://iopscience.iop.org/article/10.1088/0034-4885/22/1/310}{\textsl{Reports
  on Progress in Physics} \textbf{22} 329 (1959)}.

\bibitem{muller3}
M.~M\"uller, J.~Schmalian, and L.~Fritz.
\newblock ``Graphene -- a nearly perfect fluid",
  \href{http://journals.aps.org/prl/abstract/10.1103/PhysRevLett.103.025301}{\textsl{Physical
  Review Letters} \textbf{103} \texttt{025301} (2009)},
  \href{http://arxiv.org/abs/0903.4178}{\texttt{arXiv:0903.4178}}.

\bibitem{kss}
P.~Kovtun, D.~T. Son, and A.~O. Starinets.
\newblock ``Viscosity in strongly interacting quantum field theories from black
  hole physics",
  \href{https://doi.org/10.1103/PhysRevLett.94.111601}{\textsl{Physical Review
  Letters} \textbf{94} \texttt{111601} (2005)},
  \href{http://arxiv.org/abs/hep-th/0405231}{\texttt{arXiv:hep-th/0405231}}.

\bibitem{lock}
R.~C. Lock.
\newblock ``The stability of a flow of an electrically conducting fluid between
  parallel planes under a transverse magnetic field",
  \href{https://doi.org/10.1098/rspa.1955.0249}{\textsl{Proceedings of the
  Royal Society} \textbf{A233} 105 (1955)}.

\bibitem{ziman}
J.~Ziman.
\newblock \emph{Electrons and Phonons}
  \href{http://www.amazon.com/Electrons-Phonons-Transport-Phenomena-Physical/dp/0198507798/ref=sr_1_1?ie=UTF8&qid=1433301493&sr=8-1&keywords=electrons+and+phonons&pebp=1433301494471&perid=0S820PYT3GQQ1X5Q7Z92}{(Oxford
  University Press, 1960)}.

\bibitem{dgg}
M.~Lee, J.~R. Wallbank, P.~Gallagher, K.~Watanabe, T.~Taniguchi, V.~I. Fal'ko,
  and D.~Goldhaber-Gordon.
\newblock ``Ballistic miniband conduction in a graphene superlattice",
  \href{http://science.sciencemag.org/content/353/6307/1526}{\textsl{Science}
  \textbf{353} 1526 (2016)},
  \href{http://arxiv.org/abs/1603.01260}{\texttt{arXiv:1603.01260}}.

\bibitem{zettledge}
K.~Kim, S.~Coh, C.~Kisielowski, M.~F. Crommie, S.~G. Louie, M.~L. Cohen, and
  A.~Zettl.
\newblock ``Atomically perfectly torn graphene edges and their reversible
  reconstruction",
  \href{https://www.nature.com/articles/ncomms3723}{\textsl{Nature
  Communications} \textbf{4} \texttt{2723} (2013)}.

\bibitem{levitov1607}
H.~Guo, E.~Ilseven, G.~Falkovich, and L.~Levitov.
\newblock ``Higher-than-ballistic conduction of viscous electron flows",
  \href{http://www.pnas.org/content/114/12/3068.abstract}{\textsl{Proceedings
  of the National Academy of Sciences} \textbf{114} 3068 (2017)},
  \href{http://arxiv.org/abs/1607.07269}{\texttt{arXiv:1607.07269}}.

\bibitem{spivak02}
M.~Hruska and B.~Spivak.
\newblock ``Conductivity of the classical two-dimensional electron gas",
  \href{http://journals.aps.org/prb/abstract/10.1103/PhysRevB.65.033315}{\textsl{Physical
  Review} \textbf{B65} \texttt{033315} (2002)},
  \href{http://arxiv.org/abs/cond-mat/0102219}{\texttt{arXiv:cond-mat/0102219}}.

\bibitem{spivak06}
B.~Spivak and S.~A. Kivelson.
\newblock ``Transport in two-dimensional electronic micro-emulsions",
  \href{https://doi.org/10.1103/10.1016/j.aop.2005.12.002}{\textsl{Annals of
  Physics} \textbf{321} 2071 (2006)}.

\bibitem{levitov1612}
H.~Guo, E.~Ilseven, G.~Falkovich, and L.~Levitov.
\newblock ``Stokes paradox, back reflections and interaction-enhanced
  conductance",
  \href{http://arxiv.org/abs/1612.09239}{\texttt{arXiv:1612.09239}}.

\bibitem{levitovhydro}
L.~Levitov and G.~Falkovich.
\newblock ``Electron viscosity, current vortices and negative nonlocal
  resistance in graphene",
  \href{http://www.nature.com/nphys/journal/v12/n7/full/nphys3667.html}{\textsl{Nature
  Physics} \textbf{12} 672 (2016)},
  \href{http://arxiv.org/abs/1508.00836}{\texttt{arXiv:1508.00836}}.

\bibitem{levitovjuly}
G.~Falkovich and L.~Levitov.
\newblock ``Linking spatial distributions of potential and current in viscous
  electronics",
  \href{https://doi.org/10.1103/PhysRevLett.119.066601}{\textsl{Physical Review
  Letters} \textbf{119} \texttt{066601} (2017)},
  \href{http://arxiv.org/abs/1607.00986}{\texttt{arXiv:1607.00986}}.

\bibitem{torre}
F.~M.~D. Pellegrino, I.~Torre, A.~K. Geim, and M.~Polini.
\newblock ``Electron hydrodynamics dilemma: whirlpools or no whirlpools",
  \href{http://journals.aps.org/prb/abstract/10.1103/PhysRevB.94.155414}{\textsl{Physical
  Review} \textbf{B94} \texttt{155414} (2016)},
  \href{http://arxiv.org/abs/1607.03726}{\texttt{arXiv:1607.03726}}.

\bibitem{DS}
M.~Dyakonov and M.~Shur.
\newblock ``Shallow water analogy for a ballistic field effect transistor: New
  mechanism of plasma wave generation by dc current",
  \href{http://journals.aps.org/prl/abstract/10.1103/PhysRevLett.71.2465}{\textsl{Physical
  Review Letters} \textbf{71} 2465 (1993)}.

\bibitem{DSkin}
V.~Yu.~Kachorovskii A.~P.~Dmitriev and M.~S. Shur.
\newblock ``Plasma wave instability in gated collisionless two-dimensional
  electron gas", \href{http://dx.doi.org/10.1063/1.1391395}{\textsl{Applied
  Physics Letters} \textbf{79} 922 (2001)}.

\bibitem{giliberti}
V.~Giliberti, A.~Di Gaspare, E.~Giovine, M.~Ortolani, L.~Sorba, G.~Biasiol,
  V.~V. Popov, D.~V. Fateev, and F.~Evangelisti.
\newblock ``Downconversion of terahertz radiation due to intrinsic hydrodynamic
  nonlinearity of a two-dimensional electron plasma",
  \href{http://dx.doi.org/10.1103/PhysRevB.91.165313}{\textsl{Physical Review}
  \textbf{B91} \texttt{165313} (2015)}.

\bibitem{svintsov13}
D.~Svintsov, V.~Vyurkov, V.~Ryzhii, and T.~Otsuji.
\newblock ``Hydrodynamic electron transport and nonlinear waves in graphene",
  \href{https://doi.org/10.1103/PhysRevB.88.245444}{\textsl{Physical Review}
  \textbf{B88} \texttt{245444} (2013)},
  \href{http://arxiv.org/abs/1310.3963}{\texttt{arXiv:1310.3963}}.

\bibitem{tomadin}
A.~Tomadin, G.~Vignale, and M.~Polini.
\newblock ``A Corbino disk viscometer for 2d quantum electron liquids",
  \href{http://journals.aps.org/prl/abstract/10.1103/PhysRevLett.113.235901}{\textsl{Physical
  Review Letters} \textbf{113} \texttt{235901} (2014)},
  \href{http://arxiv.org/abs/1401.0938}{\texttt{arXiv:1401.0938}}.

\bibitem{surowka}
R.~Moessner, P.~Sur\'owka, and P.~Witkowski.
\newblock ``Floquet hydrodynamics in a two-dimensional electronic fluid",
  \href{http://arxiv.org/abs/1710.00354}{\texttt{arXiv:1710.00354}}.

\bibitem{alekseev}
P.~S. Alekseev.
\newblock ``Negative magnetoresistance in viscous flow of two-dimensional
  electrons",
  \href{http://journals.aps.org/prl/abstract/10.1103/PhysRevLett.117.166601}{\textsl{Physical
  Review Letters} \textbf{117} \texttt{166601} (2016)}.

\bibitem{jerome1}
E.~Banks, A.~Donos, J.~P. Gauntlett, T.~Griffin, and L.~Melgar.
\newblock ``Thermal backflow in CFTs",
  \href{https://doi.org/10.1103/PhysRevD.95.025022}{\textsl{Physical Review}
  \textbf{D95} \texttt{025022} (2017)},
  \href{http://arxiv.org/abs/1610.00392}{\texttt{arXiv:1610.00392}}.

\bibitem{hofman}
S.~A. Hartnoll and D.~M. Hofman.
\newblock ``Locally critical umklapp scattering and holography",
  \href{https://doi.org/10.1103/PhysRevLett.108.241601}{\textsl{Physical Review
  Letters} \textbf{108} \texttt{241601} (2012)},
  \href{http://arxiv.org/abs/1201.3917}{\texttt{arXiv:1201.3917}}.

\bibitem{lucasMM}
A.~Lucas and S.~Sachdev.
\newblock ``Memory matrix theory of magnetotransport in strange metals",
  \href{http://journals.aps.org/prb/abstract/10.1103/PhysRevB.91.195122}{\textsl{Physical
  Review} \textbf{B91} \texttt{195122} (2015)},
  \href{http://arxiv.org/abs/1502.04704}{\texttt{arXiv:1502.04704}}.

\bibitem{scheffler}
M.~Scheffler, M.~Dressel, M.~Jourdan, and H.~Adrian.
\newblock ``Extremely slow Drude relaxation of correlated electrons",
  \href{https://doi.org/10.1038/nature04232}{\textsl{Nature} \textbf{438} 1135
  (2005)}.

\bibitem{vignale}
A.~Principi and G.~Vignale.
\newblock ``Violation of the Wiedemann-Franz law in hydrodynamic electron
  liquids",
  \href{http://journals.aps.org/prl/abstract/10.1103/PhysRevLett.115.056603}{\textsl{Physical
  Review Letters} \textbf{115} \texttt{056603} (2015)}.

\bibitem{mahajan}
R.~Mahajan, M.~Barkeshli, and S.~A. Hartnoll.
\newblock ``Non-Fermi liquids and the Wiedemann-Franz law",
  \href{https://doi.org/10.1103/PhysRevB.88.125107}{\textsl{Physical Review}
  \textbf{B88} \texttt{125107} (2013)},
  \href{http://arxiv.org/abs/1304.4249}{\texttt{arXiv:1304.4249}}.

\bibitem{andreev}
A.~V. Andreev, S.~A. Kivelson, and B.~Spivak.
\newblock ``Hydrodynamic description of transport in strongly correlated
  electron systems",
  \href{http://journals.aps.org/prl/abstract/10.1103/PhysRevLett.106.256804}{\textsl{Physical
  Review Letters} \textbf{106} \texttt{256804} (2011)},
  \href{http://arxiv.org/abs/1011.3068}{\texttt{arXiv:1011.3068}}.

\bibitem{lucas}
A.~Lucas.
\newblock ``Hydrodynamic transport in strongly coupled disordered quantum field
  theories",
  \href{http://iopscience.iop.org/article/10.1088/1367-2630/17/11/113007/meta}{\textsl{New
  Journal of Physics} \textbf{17} \texttt{113007} (2015)},
  \href{http://arxiv.org/abs/1506.02662}{\texttt{arXiv:1506.02662}}.

\bibitem{scopelliti}
V.~Scopelliti, K.~Schalm, and A.~Lucas.
\newblock ``Hydrodynamic charge and heat transport on inhomogeneous curved
  spaces", \href{https://doi.org/10.1103/PhysRevB.96.075150}{\textsl{Physical
  Review} \textbf{B96} \texttt{075150} (2017)},
  \href{http://arxiv.org/abs/1705.04325}{\texttt{arXiv:1705.04325}}.

\bibitem{jerome2}
A.~Donos, J.~P. Gauntlett, and V.~Ziogas.
\newblock ``Diffusion in inhomogeneous media",
  \href{http://arxiv.org/abs/1708.05412}{\texttt{arXiv:1708.05412}}.

\bibitem{camiola}
V.~D. Camiola and V.~Romano.
\newblock ``Hydrodynamical model for charge transport in graphene",
  \href{https://doi.org/10.1007/s10955-014-1102-z}{\textsl{Journal of
  Statistical Physics} \textbf{157} 1114 (2014)}.

\bibitem{barletti}
L.~Barletti.
\newblock ``Hydrodynamic equations for electrons in graphene obtained from the
  maximum entropy principle",
  \href{https://doi.org/10.1063/1.4886698}{\textsl{Journal of Mathematical
  Physics} \textbf{55} \texttt{083303} (2014)},
  \href{http://arxiv.org/abs/1311.5392}{\texttt{arXiv:1311.5392}}.

\bibitem{yigen}
S.~Yigen, V.~Tayari, J.~O. Island, J.~M. Porter, and A.~M. Champagne.
\newblock ``Electronic thermal conductivity measurements in intrinsic
  graphene", \href{https://doi.org/10.1103/PhysRevB.87.241411}{\textsl{Physical
  Review} \textbf{B87} \texttt{241411} (2013)},
  \href{http://arxiv.org/abs/1303.2390}{\texttt{arXiv:1303.2390}}.

\bibitem{dicke}
R.~H. Dicke.
\newblock ``The measurement of thermal radiation at microwave frequencies",
  \href{https://doi.org/10.1063/1.1770483}{\textsl{Review of Scientific
  Instruments} \textbf{17} 268 (1946)}.

\bibitem{johnsonnoise}
J.~B. Johnson.
\newblock ``Thermal agitation of electricity in conductors",
  \href{https://doi.org/10.1103/PhysRev.32.97}{\textsl{Physical Review}
  \textbf{32} 97 (1928)}.

\bibitem{foster2}
H-Y. Xie and M.~S. Foster.
\newblock ``Transport coefficients of graphene: Interplay of impurity
  scattering, Coulomb interaction, and optical phonons",
  \href{http://journals.aps.org/prb/abstract/10.1103/PhysRevB.93.195103}{\textsl{Physical
  Review} \textbf{B93} \texttt{195103} (2016)},
  \href{http://arxiv.org/abs/1601.05862}{\texttt{arXiv:1601.05862}}.

\bibitem{ghahari}
F.~Ghahari, H-Y. Xie, T.~Taniguchi, K.~Watanabe, M.~S. Foster, and P.~Kim.
\newblock ``Enhanced thermoelectric power in graphene: violation of the Mott
  relation by inelastic scattering",
  \href{http://journals.aps.org/prl/abstract/10.1103/PhysRevLett.116.136802}{\textsl{Physical
  Review Letters} \textbf{116} \texttt{136802} (2016)},
  \href{http://arxiv.org/abs/1601.05859}{\texttt{arXiv:1601.05859}}.

\bibitem{seebeck1}
Y.~M. Zuev, W.~Chang, and P.~Kim.
\newblock ``Thermoelectric and magnetothermoelectric transport measurements of
  graphene",
  \href{https://doi.org/10.1103/PhysRevLett.102.096807}{\textsl{Physical Review
  Letters} \textbf{102} \texttt{096807} (2009)},
  \href{http://arxiv.org/abs/0812.1393}{\texttt{arXiv:0812.1393}}.

\bibitem{seebeck2}
P.~Wei, W.~Bao, Y.~Pu, C.~N. Lau, and J.~Shi.
\newblock ``Anomalous thermoelectric transport of Dirac particles in graphene",
  \href{https://doi.org/10.1103/PhysRevLett.102.166808}{\textsl{Physical Review
  Letters} \textbf{102} \texttt{166808} (2009)},
  \href{http://arxiv.org/abs/0812.1411}{\texttt{arXiv:0812.1411}}.

\bibitem{seebeck3}
J.~G. Checkelsky and N.~P. Ong.
\newblock ``The thermopower and Nernst effect in graphene in a magnetic field",
  \href{https://doi.org/10.1103/PhysRevB.80.081413}{\textsl{Physical Review
  Letters} \textbf{B80} \texttt{081413} (2009)},
  \href{http://arxiv.org/abs/0812.2866}{\texttt{arXiv:0812.2866}}.

\bibitem{sangjin}
Y.~Seo, G.~Song, P.~Kim, S.~Sachdev, and S-J. Sin.
\newblock ``Holography of the Dirac fluid in graphene with two currents",
  \href{https://doi.org/10.1103/PhysRevLett.118.036601}{\textsl{Physical Review
  Letters} \textbf{118} \texttt{036601} (2017)},
  \href{http://arxiv.org/abs/1609.03582}{\texttt{arXiv:1609.03582}}.

\bibitem{goldsmid}
G.~S. Nolas and H.~J. Goldsmid.
\newblock ``Thermal conductivity of semiconductors", in \emph{Thermal
  Conductivity: Theory, Properties and Applications} (ed. T. M. Tritt), 105,
  \href{http://www.amazon.com/Quantum-Phase-Transitions-Subir-Sachdev/dp/0521514681/ref=sr_1_1?ie=UTF8&qid=1433031213&sr=8-1&keywords=quantum+phase+transitions}{(Kluwer
  Academic, 2004)}.

\bibitem{yoshino}
H.~Yoshino and K.~Murata.
\newblock ``Significant enhancement of electronic thermal conductivity of
  two-dimensional zero-gap systems by bipolar-diffusion effect",
  \href{http://journals.jps.jp/doi/abs/10.7566/JPSJ.84.024601}{\textsl{Journal
  of the Physical Society of Japan} \textbf{84} \texttt{024601} (2015)}.

\bibitem{kohn}
W.~Kohn.
\newblock ``Cyclotron resonance and de Haas-van Alphen oscillations of an
  interacting electron gas",
  \href{https://doi.org/10.1103/PhysRev.123.1242}{\textsl{Physical Review}
  \textbf{123} 1242 (1961)}.

\bibitem{blakedonos}
M.~Blake and A.~Donos.
\newblock ``Quantum critical transport and the Hall angle in holographic
  models",
  \href{https://doi.org/10.1103/PhysRevLett.114.021601}{\textsl{Physical Review
  Letters} \textbf{114} \texttt{021601} (2015)},
  \href{http://arxiv.org/abs/1406.1659}{\texttt{arXiv:1406.1659}}.

\bibitem{hallcuprate}
T.~R. Chien, Z.~Z. Wang, and N.~P. Ong.
\newblock ``Effect of Zn impurities on the normal-state Hall angle in
  single-crystal $\mathrm{YBa}_2\mathrm{Cu}_{3-x} \mathrm{Zn}_x
  \mathrm{O}_{7-x}$",
  \href{https://doi.org/10.1103/PhysRevLett.67.2088}{\textsl{Physical Review
  Letters} \textbf{67} 2088 (1991)}.

\bibitem{avron}
J.~E. Avron.
\newblock ``Odd viscosity",
  \href{https://link.springer.com/article/10.1023/A:1023084404080}{\textsl{Journal
  of Statistical Physics} \textbf{92} 543 (1998)},
  \href{http://arxiv.org/abs/physics/9712050}{\texttt{arXiv:physics/9712050}}.

\bibitem{yarom}
K.~Jensen, M.~Kaminski, P.~Kovtun, R.~Meyer, A.~Ritz, and A.~Yarom.
\newblock ``Parity-violating hydrodynamics in 2+1 dimensions",
  \href{https://doi.org/10.1007/JHEP05(2012)102}{\textsl{Journal of High Energy
  Physics} \textbf{05} \texttt{102} (\textbf{2012})},
  \href{http://arxiv.org/abs/1112.4498}{\texttt{arXiv:1112.4498}}.

\bibitem{scaffidi}
T.~Scaffidi, N.~Nandi, B.~Schmidt, A.~P. Mackenzie, and J.~E. Moore.
\newblock ``Hydrodynamic electron flow and Hall viscosity",
  \href{https://doi.org/10.1103/PhysRevLett.118.226601}{\textsl{Physical Review
  Letters} \textbf{118} \texttt{226601} (2017)},
  \href{http://arxiv.org/abs/1703.07325}{\texttt{arXiv:1703.07325}}.

\bibitem{delacretaz}
L.~V. Delacretaz and A.~Gromov.
\newblock ``Transport signatures of Hall viscosity",
  \href{http://arxiv.org/abs/1706.03773}{\texttt{arXiv:1706.03773}}.

\bibitem{polinihall}
F.~M.~D. Pellegrino, I.~Torre, and M.~Polini.
\newblock ``Non-local transport and the Hall viscosity of 2D hydrodynamic
  electron liquids",
  \href{https://doi.org/10.1103/PhysRevB.96.195401}{\textsl{Physical Review}
  \textbf{B96} \texttt{195401} (2016)},
  \href{http://arxiv.org/abs/1706.08363}{\texttt{arXiv:1706.08363}}.

\bibitem{levchenko1}
A.~Levchenko, H-Y. Xie, and A.~V. Andreev.
\newblock ``Viscous magnetoresistance of correlated electron liquids",
  \href{https://journals.aps.org/prd/abstract/10.1103/PhysRevB.95.121301}{\textsl{Physical
  Review} \textbf{B95} \texttt{121301} (2017)},
  \href{http://arxiv.org/abs/1612.09275}{\texttt{arXiv:1612.09275}}.

\bibitem{levchenko2}
A.~A. Patel, R.~A. Davison, and A.~Levchenko.
\newblock ``Hydrodynamic flows of non-Fermi liquids: magnetotransport and
  bilayer drag",
  \href{http://arxiv.org/abs/1706.03775}{\texttt{arXiv:1706.03775}}.

\bibitem{baumgartner}
A.~Baumgartner, A.~Karch, and A.~Lucas.
\newblock ``Magnetoresistance in relativistic hydrodynamics without anomalies",
  \href{http://link.springer.com/article/10.1007/JHEP10%282017%29054}{\textsl{Journal
  of High Energy Physics} \textbf{06} \texttt{054} (\textbf{2017})},
  \href{http://arxiv.org/abs/1704.01592}{\texttt{arXiv:1704.01592}}.

\bibitem{dragreview}
B.~N. Narozhny and A.~Levchenko.
\newblock ``Coulomb drag",
  \href{http://journals.aps.org/rmp/abstract/10.1103/RevModPhys.88.025003}{\textsl{Reviews
  of Modern Physics} \textbf{88} \texttt{025003} (2016)},
  \href{http://arxiv.org/abs/1505.07468}{\texttt{arXiv:1505.07468}}.

\bibitem{gorbachev1206}
R.~V. Gorbachev, A.~K. Geim, M.~I. Katsnelson, K.~S. Novoselov, T.~Tudorovskiy,
  I.~V. Grigorieva, A.~H. MacDonald, K.~Watanabe, T.~Taniguchi, and L.~A.
  Ponomarenko.
\newblock ``Strong Coulomb drag and broken symmetry in double-layer graphene",
  \href{https://doi.org/10.1038/nphys2441}{\textsl{Nature Physics} \textbf{8}
  896 (2012)}, \href{http://arxiv.org/abs/1206.6626}{\texttt{arXiv:1206.6626}}.

\bibitem{levitov1205}
J.~C.~W. Song and L.~S. Levitov.
\newblock ``Energy-driven drag at charge neutrality in graphene",
  \href{https://doi.org/10.1103/PhysRevLett.109.236602}{\textsl{Physical Review
  Letters} \textbf{109} \texttt{236602} (2012)},
  \href{http://arxiv.org/abs/1205.5257}{\texttt{arXiv:1205.5257}}.

\bibitem{dho1611}
D.~Y.~H. Ho, I.~Yudhistira, B.~Y-K. Hu, and S.~Adam.
\newblock ``Non-monotonic temperature dependence of Coulomb drag peaks in
  graphene", \href{http://arxiv.org/abs/1611.03089}{\texttt{arXiv:1611.03089}}.

\bibitem{apostolov}
S.~S. Apostolov, A.~Levchenko, and A.~V. Andreev.
\newblock ``Hydrodynamic Coulomb drag of strongly correlated electron liquids",
  \href{https://doi.org/10.1103/PhysRevB.89.121104}{\textsl{Physical Review}
  \textbf{B89} \texttt{121104} (2014)},
  \href{http://arxiv.org/abs/1312.6890}{\texttt{arXiv:1312.6890}}.

\bibitem{levitov1303}
J.~C.~W. Song and L.~S. Levitov.
\newblock ``Hall drag and magnetodrag in graphene",
  \href{https://doi.org/10.1103/PhysRevLett.111.126601}{\textsl{Physical Review
  Letters} \textbf{111} \texttt{126601} (2013)},
  \href{http://arxiv.org/abs/1303.3529}{\texttt{arXiv:1303.3529}}.

\bibitem{titov1303}
M.~Titov \emph{et al.}
\newblock ``Giant magneto-drag in graphene at charge neutrality ",
  \href{https://doi.org/10.1103/PhysRevLett.111.166601}{\textsl{Physical Review
  Letters} \textbf{111} \texttt{166601} (2013)},
  \href{http://arxiv.org/abs/1303.6264}{\texttt{arXiv:1303.6264}}.

\bibitem{grinolds}
M.~S. Grinolds, S.~Hong, P.~Maletinsky, L.~Luan, M.~D. Lukin, R.~L. Walsworth,
  and A.~Yacoby.
\newblock ``Nanoscale magnetic imaging of a single electron spin under ambient
  conditions", \href{https://doi.org/10.1038/nphys2543}{\textsl{Nature Physics}
  \textbf{9} 215 (2013)},
  \href{http://arxiv.org/abs/1209.0203}{\texttt{arXiv:1209.0203}}.

\bibitem{vasyukov}
D.~Vasyukov \emph{et al.}
\newblock ``Scanning nano-SQUID with single electron spin sensitivity",
  \href{https://doi.org/10.1038/nnano.2013.169}{\textsl{Nature Nanotechnology}
  \textbf{8} 169 (2013)},
  \href{http://arxiv.org/abs/1308.0694}{\texttt{arXiv:1308.0694}}.

\bibitem{benlev}
F.~Yang, A.~J. Koll\'ar, S.~F. Taylor, R.~W. Turner, and B.~L. Lev.
\newblock ``Scanning quantum cryogenic atom microscope",
  \href{https://doi.org/10.1103/PhysRevApplied.7.034026}{\textsl{Physical
  Review Applied} \textbf{7} \texttt{034026} (2017)},
  \href{http://arxiv.org/abs/1608.06922}{\texttt{arXiv:1608.06922}}.

\bibitem{fogler2}
Z.~Sun, D.~N. Basov, and M.~M. Fogler.
\newblock ``The third-order optical conductivity of an electron fluid",
  \href{http://arxiv.org/abs/1704.07334}{\texttt{arXiv:1710.02297}}.

\bibitem{zaanen}
D.~Forcella, J.~Zaanen, D.~Valentinis, and D.~van~der Marel.
\newblock ``Electromagnetic properties of viscous charged fluids",
  \href{http://journals.aps.org/prl/abstract/10.1103/PhysRevB.90.035143}{\textsl{Physical
  Review} \textbf{B90} \texttt{035143} (2014)},
  \href{http://arxiv.org/abs/1406.1356}{\texttt{arXiv:1406.1356}}.

\bibitem{lucasRFB}
A.~Lucas.
\newblock ``Kinetic theory of electronic transport in random magnetic fields",
  \href{http://arxiv.org/abs/1710.11141}{\texttt{arXiv:1710.11141}}.

\bibitem{hartman}
T.~Hartman, S.~A. Hartnoll, and R.~Mahajan.
\newblock ``An upper bound on transport",
  \href{http://arxiv.org/abs/1706.00019}{\texttt{arXiv:1706.00019}}.

\bibitem{lucasbound}
A.~Lucas.
\newblock ``Constraints on hydrodynamics from many-body quantum chaos",
  \href{http://arxiv.org/abs/1710.01005}{\texttt{arXiv:1710.01005}}.

\bibitem{localized}
D.~Roberts, S.~Shenker, and D.~Stanford.
\newblock ``Localized shocks ",
  \href{http://link.springer.com/article/10.1007%2FJHEP03%282015%29051}{\textsl{Journal
  of High Energy Physics} \textbf{03} \texttt{051} (\textbf{2015})},
  \href{http://arxiv.org/abs/1409.8180}{\texttt{arXiv:1409.8180}}.

\bibitem{blakeB1}
M.~Blake.
\newblock ``Universal charge diffusion and the butterfly effect in holographic
  theories",
  \href{http://journals.aps.org/prl/abstract/10.1103/PhysRevLett.117.091601}{\textsl{Physical
  Review Letters} \textbf{117} \texttt{091601} (2016)},
  \href{http://arxiv.org/abs/1603.08510}{\texttt{arXiv:1603.08510}}.

\bibitem{blakeB2}
M.~Blake.
\newblock ``Universal diffusion in incoherent black holes",
  \href{http://journals.aps.org/prd/abstract/10.1103/PhysRevD.94.086014}{\textsl{Physical
  Review} \textbf{D94} \texttt{086014} (2016)},
  \href{http://arxiv.org/abs/1604.01754}{\texttt{arXiv:1604.01754}}.

\bibitem{monika}
B.~Swingle, G.~Bentsen, M.~Schleier-Smith, and P.~Hayden.
\newblock ``Measuring the scrambling of quantum information",
  \href{https://doi.org/10.1103/PhysRevA.94.040302}{\textsl{Physical Review}
  \textbf{A94} \texttt{040302} (2016)},
  \href{http://arxiv.org/abs/1602.06271}{\texttt{arXiv:1602.06271}}.

\bibitem{grover}
G.~Zhu, M.~Hafezi, and T.~Grover.
\newblock ``Measurement of many-body chaos using a quantum clock",
  \href{https://doi.org/10.1103/PhysRevA.94.062329}{\textsl{Physical Review}
  \textbf{A94} \texttt{062329} (2016)},
  \href{http://arxiv.org/abs/1607.00079}{\texttt{arXiv:1607.00079}}.

\bibitem{yao}
N.~Y. Yao, F.~Grusdt, B.~Swingle, M.~D. Lukin, D.~M. Stamper-Kurn, J.~E. Moore,
  and E.~A. Demler.
\newblock ``Interferometric approach to probing fast scrambling",
  \href{http://arxiv.org/abs/1607.01801}{\texttt{arXiv:1607.01801}}.

\bibitem{karasik}
C.~B. McKitterick, D.~E. Prober, and B.~S. Karasik.
\newblock ``Performance of graphene photon detectors",
  \href{https://doi.org/10.1063/1.4789360}{\textsl{Journal of Applied Physics}
  \textbf{113} \texttt{044512} (2013)},
  \href{http://arxiv.org/abs/1210.5495}{\texttt{arXiv:1210.5495}}.

\bibitem{fong17}
E.~D. Walsh, D.~K. Efetov, G-H. Lee, M.~Heuck, J.~Crossno, T.~A. Ohki, P.~Kim,
  D.~Englund, and K.~C. Fong.
\newblock ``Graphene-based Josephson-junction single-photon detector",
  \href{https://doi.org/10.1103/PhysRevApplied.8.024022}{\textsl{Physical
  Review Applied} \textbf{8} \texttt{024022} (2017)},
  \href{http://arxiv.org/abs/1703.09736}{\texttt{arXiv:1703.09736}}.

\bibitem{kitaeva}
G.~Kh. Kitaeva.
\newblock ``Terahertz generation by means of optical lasers",
  \href{https://doi.org/10.1002/lapl.200810039}{\textsl{Laser Physics Letters}
  \textbf{5} 559 (2008)}.

\bibitem{tauk}
R.~Tauk \emph{et al.}
\newblock ``Plasma wave detection of terahertz radiation by silicon field
  effects transistors: responsivity and noise equivalent power",
  \href{http://scitation.aip.org/content/aip/journal/apl/89/25/10.1063/1.2410215}{\textsl{Applied
  Physics Letters} \textbf{89} \texttt{253511} (2006)}.

\end{thebibliography}

\end{document}